\documentclass[11pt,a4paper,epsf,elsart,showkeys]{article}
\usepackage{amssymb,a4wide}
\usepackage{graphicx}
\usepackage[figuresright]{rotating}
\usepackage[small]{caption}
\title{{\hfill\small IEKP-KA/2003-17}\\[1cm]\Large Indirect
Evidence for the Supersymmetric Nature of Dark Matter from the
Combined Data on Galactic Positrons, Antiprotons and Gamma Rays}

\author{W. de Boer, M. Herold, C. Sander, V. Zhukov \\
{\it Institut f\"ur Experimentelle Kernphysik},\\
Universit\"at Karlsruhe (TH),\\
P.O. Box 6980, 76128 Karlsruhe, Germany}

\newlength{\dslashwidth}

\newcommand{\bsg}{\ensuremath{b\to X_s\gamma}}

\newcommand{\tb}{\ensuremath{\tan\beta}}

\newcommand{\bq}{\begin{equation}}
\newcommand{\eq}{\end{equation}}
\newcommand{\ba}{\begin{array}}
\newcommand{\ea}{\end{array}}
\newcommand{\bqa}{\begin{eqnarray}}
\newcommand{\eqa}{\end{eqnarray}}

\newcommand{\ra}{\rightarrow}
\newcommand{\lnf}{{\ifmmode \Lambda^{(N_f)} \else $\Lambda^{(N_f)}$\fi}}
\newcommand{\ms}{{\ifmmode \overline{MS} \else $\overline{MS}$\fi}}
\newcommand{\dr}{{\ifmmode \overline{DR} \else $\overline{DR}$\fi}}
\newcommand{\lms}{{\ifmmode \Lambda^{(5)}_{\overline{MS}} \else $\Lambda^{(5)}_{\overline{MS}}$\fi}}
\newcommand{\lam}{{\ifmmode \Lambda \else $\Lambda$\fi}}
\newcommand{\gev}{{\ifmmode {\rm GeV} \else ${\rm GeV}$\fi}}
\newcommand{\gevc}{{\ifmmode {\rm GeV/c^2} \else ${\rm GeV/c^2}$\fi}}
\newcommand{\tev}{{\ifmmode {\rm TeV} \else ${\rm TeV}$\fi}}
\newcommand{\tevc}{{\ifmmode {\rm TeV/c^2} \else ${\rm TeV/c^2}$\fi}}
\newcommand{\lp}{{\ifmmode L^+  \else $L^+$\fi}}
\newcommand{\lm}{{\ifmmode L^-  \else $L^-$\fi}}
\newcommand{\mlp}{{\ifmmode M(L^-) \else $M(L^-)$\fi}}
\newcommand{\mlz}{{\ifmmode M(L^0) \else $M(L^0)$\fi}}
\newcommand{\lz}{{\ifmmode L^0 \else $L^0$\fi}}
\newcommand{\ev}{{\ifmmode GeV/c^2 \else $GeV/c^2$\fi}}
\newcommand{\tri}{{\ifmmode \triangleup \else $\triangleup$\fi}}
\newcommand{\unl}{{\ifmmode U_{lL^0} \else $U_{lL^0}$\fi}}\newcommand{\gL}{{\ifmmode g_L \else $g_{L}$\fi}}
\newcommand{\gR}{{\ifmmode g_R  \else $g_{R}$\fi}}
\newcommand{\gumu}{{\ifmmode \gamma^{\mu} \else $\gamma^{\mu}$\fi}}
\newcommand{\gunu}{{\ifmmode \gamma^{\nu} \else $\gamma^{\nu}$\fi}}
\newcommand{\gdmu}{{\ifmmode \gamma_{\mu} \else $\gamma_{\mu}$\fi}}
\newcommand{\gdnu}{{\ifmmode \gamma_{\nu} \else $\gamma_{\nu}$\fi}}
\newcommand{\stw}{{\ifmmode\sin^2\theta_W \else $\sin^{2}\theta_{W}$ \fi}}
\newcommand{\sws}{{\ifmmode \;\sin^2\theta_W  \else $\;\sin^{2}\theta_{W}$ \fi}}
\newcommand{\cws}{{\ifmmode \;\cos^2\theta_W  \else $\;\cos^{2}\theta_{W}$ \fi}}
\newcommand{\sw}{{\ifmmode \;\sin\theta_W  \else $\sin\theta_{W}$ \fi}}
\newcommand{\cw}{{\ifmmode \;\cos\theta_W  \else $\;\cos\theta_{W}$ \fi}}
\newcommand{\tw}{{\ifmmode \;\tan\theta_W  \else $\;\tan\theta_{W}$ \fi}}
\newcommand{\qq}{{\ifmmode q\overline{q} \else $q\overline{q}$\fi}}
\newcommand{\lR}{{\ifmmode l_R \else $l_R$\fi}}
\newcommand{\lL}{{\ifmmode l_L \else $l_L$\fi}}
\newcommand{\nt}{{\ifmmode \nu_{\tau} \else $\nu_{\tau}$\fi}}
\newcommand{\nuR}{{\ifmmode \nu_R  \else $\nu_R$\fi}}
\newcommand{\nuL}{{\ifmmode \nu_L  \else $\nu_L$\fi}}
\newcommand{\qR}{{\ifmmode g_R  \else $q_R$\fi}}
\newcommand{\qL}{{\ifmmode q_L  \else $q_L$\fi}}
\newcommand{\qRp}{{\ifmmode q_R'  \else $q_{R}$'\fi}}
\newcommand{\qLp}{{\ifmmode q_L'  \else $q_{L}$'\fi}}
\newcommand{\est}{{\ifmmode e^{\bf \ast} \else $e^{\bf \ast}$\fi}}
\newcommand{\lst}{{\ifmmode l^{\bf \ast} \else $l^{\bf \ast}$\fi}}
\newcommand{\must}{{\ifmmode \mu^{\bf \ast} \else $\mu^{\bf \ast}$\fi}}
\newcommand{\taust}{{\ifmmode \tau^{\bf \ast} \else $\tau^{\bf \ast}$ \fi}}
\newcommand{\pperp}{{\ifmmode p_t  \else $p_t$\fi}}
\newcommand{\et}{{\ifmmode E_t  \else $E_t$\fi}}
\newcommand{\xt}{{\ifmmode x_t  \else $x_t$\fi}}
\newcommand{\smumu}{{\ifmmode \sigma_{\mu\mu}  \else $\sigma_{\mu\mu}$ \fi}}
\newcommand{\eg}{{\ifmmode e\gamma  \else $e\gamma$\fi}}
\newcommand{\epem}{{\ifmmode e^+e^-  \else $e^+e^-$\fi}}
\newcommand{\lplm}{{\ifmmode L^+L^-  \else $L^+L^-$\fi}}
\newcommand{\pp}{{\ifmmode p\overline p  \else $p\overline p$\fi}}
\newcommand{\llz}{{\ifmmode L^0\overline{L}^0 \else $L^0\overline{L}^0$\fi}}
\newcommand{\epemt}{{\ifmmode e^+e^- \to  \else $e^+e^- \to$\fi}}
\newcommand{\eb}{{\ifmmode E_{beam}  \else $E_{beam}$\fi}}
\newcommand{\ip}{{\ifmmode pb^{-1}  \else $pb^{-1}$\fi}}
\newcommand{\upm}{{\ifmmode ^{\pm}  \else $^{\pm}$\fi}}
\newcommand{\de}{{\ifmmode ^{\circ}  \else $^{\circ}$ \fi}}
\newcommand{\appr}{{\ifmmode \sim \else $\sim$ \fi}}
\newcommand{\corresp}{{\ifmmode \stackrel{\wedge}{=} \else $\stackrel{\wedge}{=}$ \fi}}
\newcommand{\sqrts}{{\ifmmode \sqrt{s} \else $\sqrt{s}$\fi}}
\newcommand{\zz}{{\ifmmode Z^0  \else $Z^0$\fi}}
\newcommand{\mz}{{\ifmmode M_{Z}  \else $M_{Z}$\fi}}
\newcommand{\mzs}{{\ifmmode M_{Z}^2  \else $M_{Z}^2$\fi}}
\newcommand{\mw}{{\ifmmode M_{W}  \else $M_{W}$\fi}}
\newcommand{\mws}{{\ifmmode M_{W}^2  \else $M_{W}^2$\fi}}
\newcommand{\mh}{{\ifmmode M_{Higgs}  \else $M_{Higgs}$\fi}}
\newcommand{\gt}{{\ifmmode \Gamma_{tot} \else $\Gamma_{tot}$\fi}}
\newcommand{\msusy}{{\ifmmode M_{SUSY}  \else $M_{SUSY}$\fi}}
\newcommand{\msusys}{{\ifmmode M_{SUSY}^2  \else $M_{SUSY}^2$\fi}}
\newcommand{\su}{{\ifmmode SU(3)_C\otimes\- SU(2)_L\otimes\- U(1)_Y \else $SU(3)_C\otimes SU(2)_L\otimes U(1)_Y$\fi}}
\newcommand{\suthree}{{\ifmmode SU(3)_C  \else $SU(3)_C$\fi}}
\newcommand{\sutwo}{{\ifmmode  SU(2)_L\otimes U(1)_Y \else $SU(2)_L\otimes U(1)_Y$\fi}}
\newcommand{\taup} {{\ifmmode \tau_{proton} \else $\tau_{proton}$\fi}}
\newcommand{\as}{{\ifmmode \alpha_{s}  \else $\alpha_{s}$\fi}}
\newcommand{\mgut}{{\ifmmode M_{GUT}  \else $M_{GUT}$\fi}}
\newcommand{\mguts}{{\ifmmode M_{GUT}^2  \else $M_{GUT}^2$\fi}}
\newcommand{\mze} {{\ifmmode m_0        \else $m_0$\fi}}
\newcommand{\mha}{{\ifmmode m_{1/2}    \else $m_{1/2}$\fi}}
\newcommand{\mb} {{\ifmmode m_{b}    \else $m_{b}$\fi}}
\newcommand{\mt} {{\ifmmode m_{t}    \else $m_{t}$\fi}}
\newcommand{\mts} {{\ifmmode m_{t}^2    \else $m_{t}^2$\fi}}

\newcommand{\mtau}{{\ifmmode m_{\tau}  \else $m_{\tau}$\fi}}
\newcommand{\dpp}{{\ifmmode \delta_{pert} \else $\delta_{pert}$\fi}}
\newcommand{\dnp}{{\ifmmode\delta_{non-pert}\else$\delta_{non-pert}$\fi}}
\newcommand{\dew}{{\ifmmode \delta_{\rm EW}\else $\delta_{\rm EW}$\fi}}
\newcommand{\rt}{{\ifmmode R_{\tau}  \else $R_{\tau} $\fi}}
\newcommand{\rz}{{\ifmmode R_{Z}  \else $R_{Z} $\fi}}

\newcommand{\swb}{{\ifmmode \sin^2\theta_{\overline{MS}} \else $\sin^2\theta_{\overline{MS}}$\fi}}
\newcommand{\cwb}{{\ifmmode \cos^2\theta_{\overline{MS}} \else $\cos^2\theta_{\overline{MS}}$\fi}}

\newcommand{\AmS}{{\protect\the\textfont2 A\kern-.1667em\lower.5ex\hbox{M}\kern-.125emS}}

\begin{document}
\maketitle
\begin{abstract}

Two new observations have strengthened the case for the
supersymmetric nature of the Cold Dark Matter component in our
universe: First, it was shown that new data on the nuclear
abundance, B/C - and $^{10}$Be/$^9$Be ratios constrain the
diffusion parameters in Galactic Models so strongly, that they
lead to a clear deficiency in the production of diffuse hard gamma
rays, antiprotons and hard positrons, if no anomalous sources or
anomalous energy dependence of the diffusion coefficients are
postulated. Second, from the precise relic density measurement by
WMAP the WIMP annihilation cross section can be determined in a
model independent way. If the WIMPS are postulated to be the
neutralinos of Supersymmetry, then only a limited region of
parameter space matches this annihilation cross section. It is
shown that the resulting positrons, antiprotons and gamma rays
from the neutralino annihilation (mainly into $b\overline{b}$
quark pairs) provide the correct shape and order of magnitude for
the missing fluxes in the Galactic Models. Here the shape is the
most discriminating feature, since for  relatively heavy
neutralinos  the positron and gamma spectra from neutralino
annihilation are significantly harder than   the spectra produced
by secondary interactions. For the antiprotons the shapes of
background and signal are similar, but they still provide a strong
constraint by the requirement of not overproducing antiprotons.
The fitted normalization factors for antiprotons, positrons and
gammas from neutralino annihilation come all out to be similar and
below 10 for a standard NRW halo profile. It should be pointed
out, that it is absolutely non-trivial to solve all three
deficiencies {\it simultaneously} with a common halo profile,
because the mean free paths of photons, antiprotons and positrons
are quite different, so the observed fluxes come from different
regions of the galaxy.
 The probability of a global fit to the galactic spectra of diffuse
gamma rays, positrons and antiprotons improves from about
$10^{-8}$ to 0.5, if Dark Matter, as predicted by Supersymmetry,
is taken into account.
This corresponds to about 6$\sigma$ evidence for the supersymmetric
nature of Dark Matter in case of Gaussian errors.

\end{abstract}

\section{Introduction}
Cold Dark Matter (CDM) makes up 23\% of the energy of the
universe, as deduced from the temperature anisotropies in the
Cosmic Microwave Background (CMB) in combination with data on the
Hubble expansion and the density fluctuations in the
universe~\cite{wmap}-\cite{verde}.
The nature of the CDM is unknown, but one of the most popular explanation
for it is the neutralino, a stable neutral particle predicted by
Supersymmetry~\cite{lspdm,jungman}. The neutralinos are spin 1/2
Majorana particles, which can annihilate into pairs of Standard Model (SM)
particles.
The stable decay and fragmentation products are neutrinos, photons, protons,
antiprotons, electrons and positrons. From these, the protons and
electrons are drown in the many matter particles in the universe,
but the others may be detectable above the background from nuclear
interactions, especially because of the much harder spectra
expected from neutralino annihilation. The background is strongly
constrained by the recent, precise measurements of the fluxes of all
nuclei in the galaxy, especially the ratios of secondary/primary
nuclei and radioactive/non-radioactive isotopes. The diffusion and
reacceleration parameters deduced from these data describe the
fluxes of {\it all} matter particles in the galaxy, but they
predict {\it too few} antiprotons, hard gammas, and
positrons~\cite{ms_problems,ms_gamma}. Several ad hoc proposals
have been made to solve these deficiencies, like unphysical
breaks in the diffusion coefficient or postulating ''unprocessed''
fresh components in a ``Local Bubble'', which is different from
the rest of the galaxy~\cite{ms_nucl}, or in order to obtain more
high energy gammas one has either to postulate a local hard
nuclear component generating hard $\pi_0$'s or a local hard
electron spectrum to create hard gammas by inverse compton
scattering~\cite{ms_gamma}.

However, none of these proposals solves these deficiencies
simultaneously. In this paper we consider the annihilation of Dark
Matter particles as a source for the missing gamma rays,
antiprotons and positrons. A statistical analysis of all three
species simultaneously provides a rather stringent test, since the
mean free path is quite different for gamma rays, antiprotons and
positrons. If one can solve the deficiencies of all three
simultaneously for a single halo density profile and a single set
of SUSY parameters, then this can be considered as a 'smoking gun'
signature for the supersymmetric nature of the Cold Dark Matter,
at least if the statistical significance is sufficient.
Fortunately, the recent improvement in the cosmic parameters
severely constrains the background calculation, while the precise
WMAP data on the relic density implies a precise and
model-independent measurement of the annihilation cross section: a
larger annihilation cross section would imply a lower relic
density and vice versa. Such an annihilation cross section can
only be obtained for rather restricted regions of the SUSY
parameter space. It will be shown that these SUSY parameters
combined with the best known Galactic Model improve the
probability of the fit to the measured galactic fluxes of diffuse
gamma rays, antiprotons and positrons by 8 orders of magnitude,
which corresponds to more than 6$\sigma$ evidence for the
supersymmetric nature of Dark Matter in case of Gaussian errors.

Indirect detection of Dark Matter has been discussed much before
~\cite{bergstrom}-\cite{bergstrom1}. Our results differ from these
previous results by performing a statistical analysis to
positrons, antiprotons and gamma rays {\it simultaneously} and
taking into account the best known propagation models.
Furthermore, previous results never considered the shape of the
gamma spectrum, which turns out to be a discriminating feature of
the present analysis, since for the relatively heavy neutralinos
the diffuse gamma spectrum from $\pi_0$ decays is considerably
harder than the spectrum from secondary interactions. Also most of
the previous analysis were done before the WMAP data became
available and before the new satellite data on nuclear fluxes
constrained the galactic background processes.

We do not consider neutrino detection, since for neutrinos no
precise data are available. We restrict us to neutralino
annihilation in the Minimal Supersymetric Model (MSSM) with
gravity inspired supersymmetry breaking and radiative electroweak
symmetry breaking, since this simplest model already gives a good
description of the considered fluxes. Actually all we need from
the model is stable neutralinos in the mass range of a few 100
GeV, which annihilate predominantly into hadronic final states.
The cross sections turn out to be correct for  values of the ratio
of neutral Higgs vacuum expectation values around 50, as expected
in SO(10) type of models\cite{susyrev}. In this case the
neutralino annihilation into b-quarks pairs via pseudoscalar Higgs
exchange dominates and the neutralino is predominantly
photino-like, or more popular, Dark Matter is the supersymmetric
partner of the Cosmic Microwave Background (CMB).

In the following we first describe the model independent
determination of the annihilation cross section from the relic
density. In the next section the predictions from Supersymmetry
concerning neutralino annihilation are discussed, while in the
following section the deficiencies in the gamma rays, antiprotons
and positrons, as predicted by the Galprop
model~\cite{ms_1998_nucl,ms_1998_pos} are discussed. In the last
section the global fits are discussed. They are performed within
the frame work of the DarkSusy~\cite{darksusy} program, after
modifying it to obtain the background and propagation model from
Galprop in DarkSusy in order to have consistent propagation and
diffusion for antiprotons and positrons from the two sources:
nuclear interactions and neutralino annihilation. For the SUSY
particle spectrum we use Suspect~\cite{suspect}, for the Higgs
masses Feynhiggsfast~\cite{fhf} and as a cross check for the cross
sections and relic density we use Micromegas~\cite{micro}.
Finally, some expectations for direct detection and indirect
searches for Dark Matter from solar neutrinos will be given.
\begin{figure}[t]
\begin{center}
 \includegraphics[width=0.4\textwidth,angle=0]{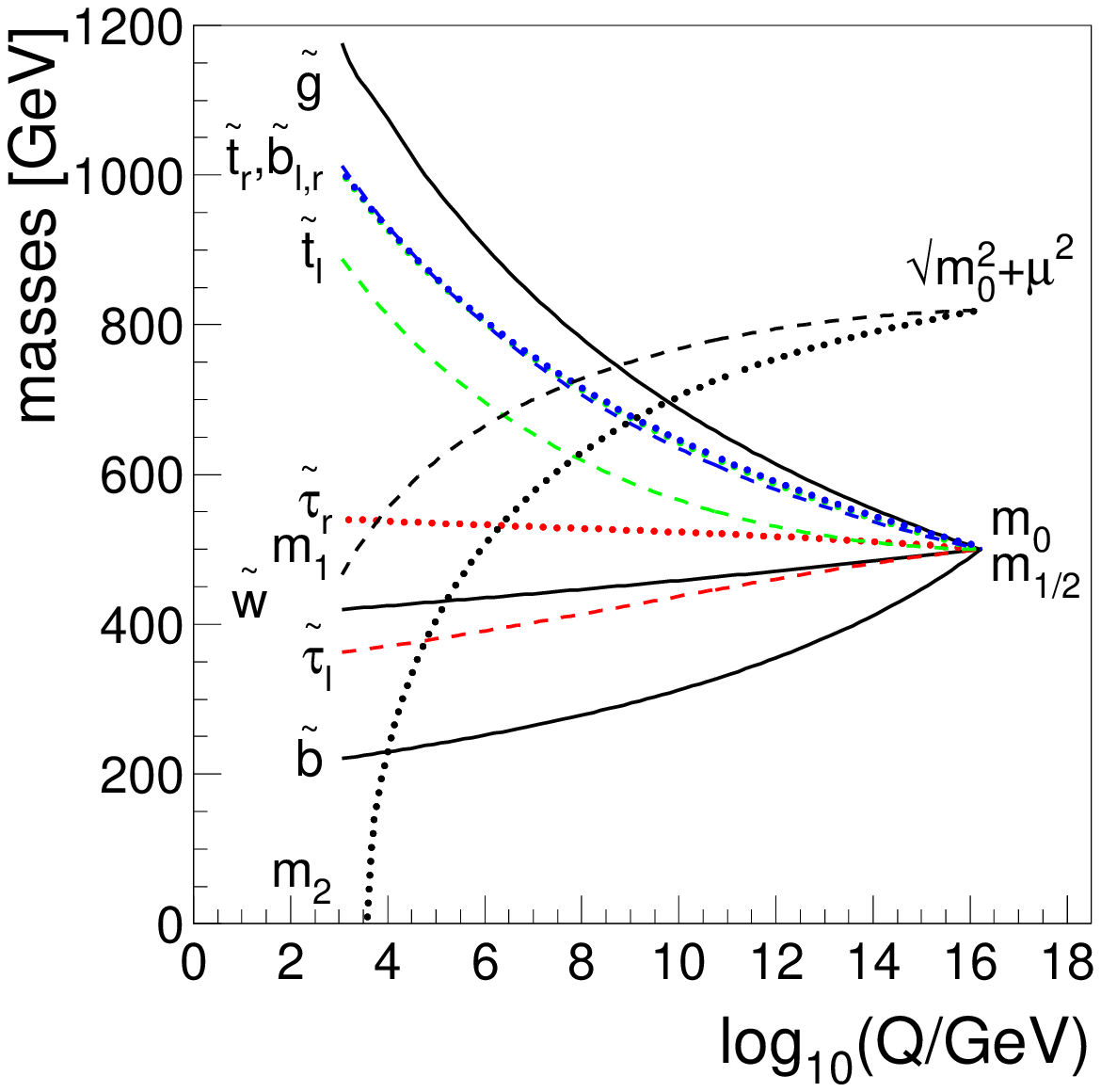}
 \includegraphics[width=0.4\textwidth,angle=0]{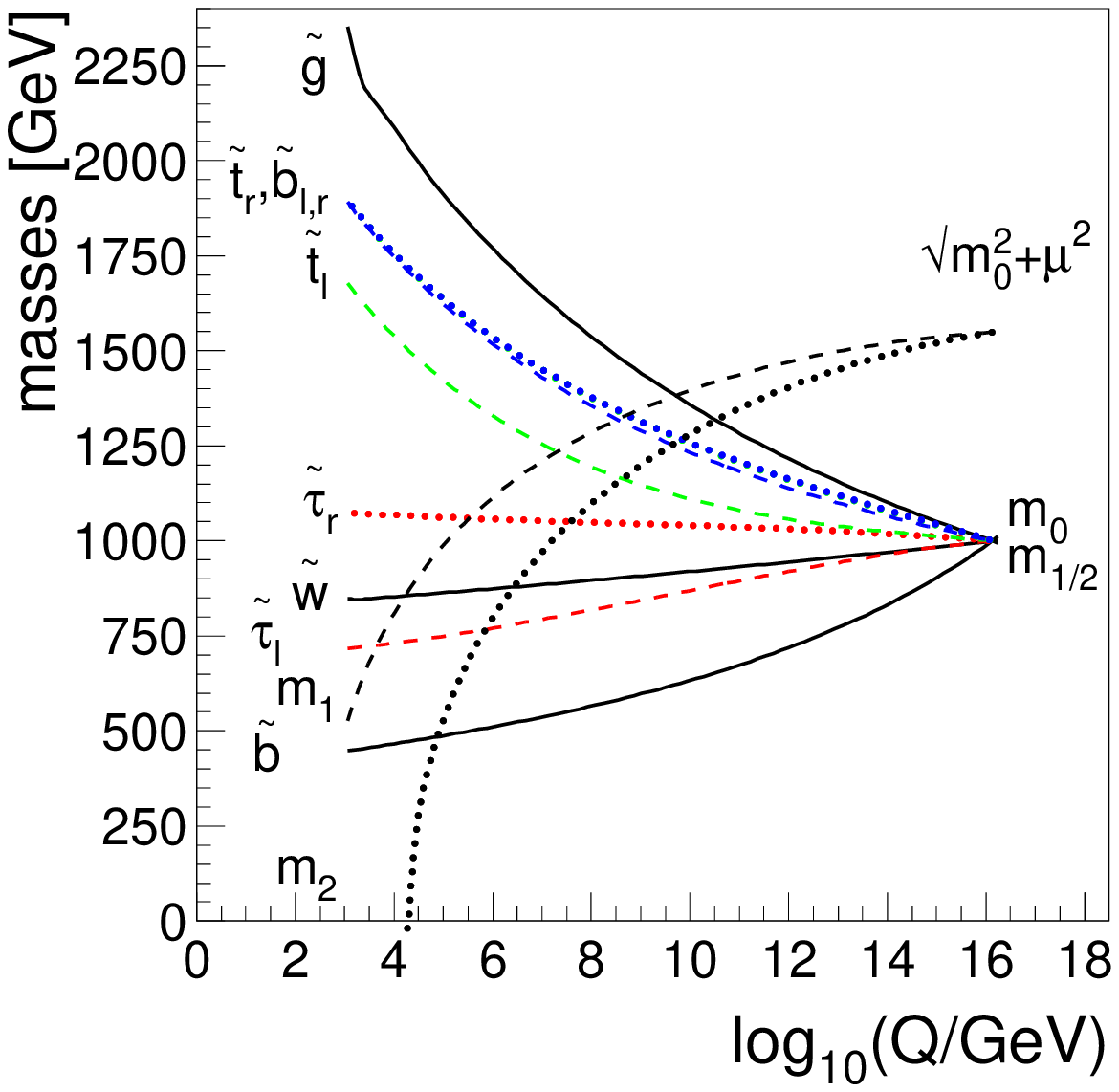}
 \caption{\it
 The running of the squark - and
 slepton masses starting at $m_0$, gaugino masses starting at $m_{1/2}$ and
 Higgs mass parameters starting at $\sqrt{m_0^2+\mu^2}$
 for $m_0=m_{1/2}$ = 500 GeV and \tb=51 (left) and
 for $m_0=m_{1/2}$ = 1000 GeV and \tb=53 (right), which are the parameters
 of interest for the present analysis.}
 \label{evol}
\end{center}
\end{figure}

\begin{figure}
\begin{center}
 \includegraphics[width=0.45\textwidth,angle=0]{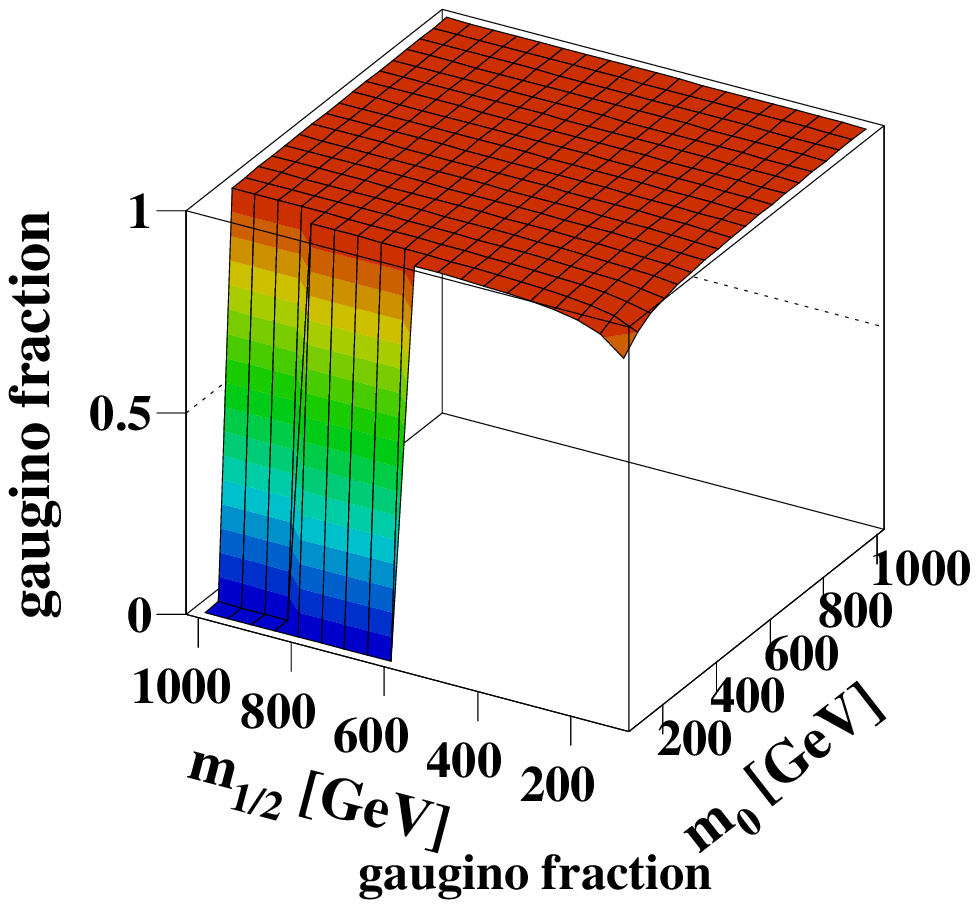}
 \includegraphics[width=0.45\textwidth,angle=0]{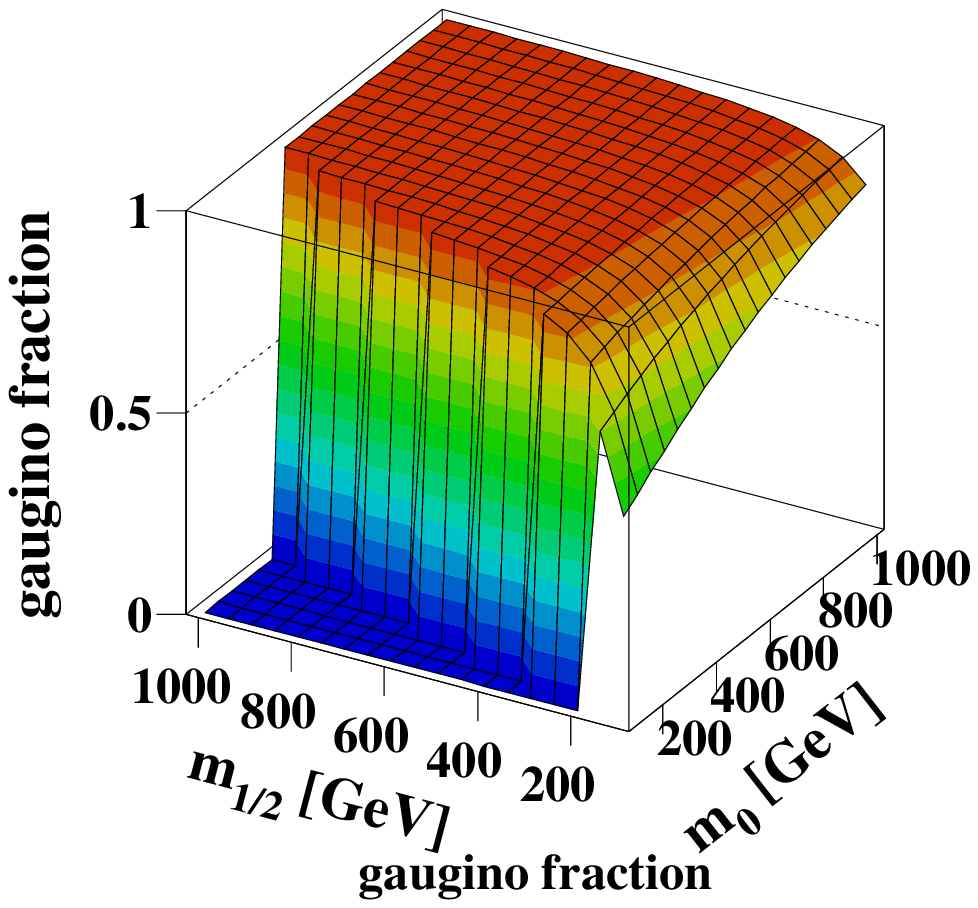}
 \caption{\it
 The gaugino fraction of the lightest neutralino as
 function of $m_0$ and $m_{1/2}$ for \tb=10 (left) and \tb=50 (right).}
 \label{gaugino}
\end{center}
\end{figure}
\begin{figure}
\begin{center}
  \includegraphics[width=\textwidth,clip]{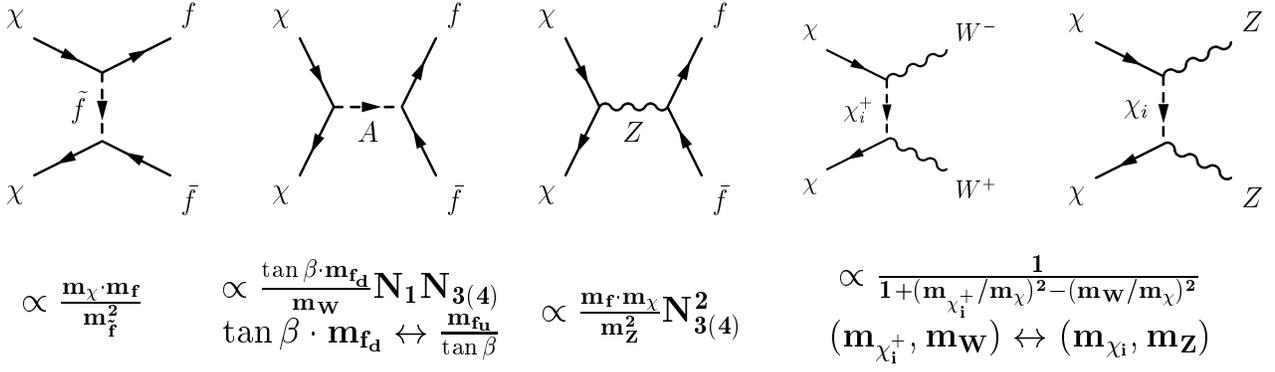}
  \caption[]{\it
  Dominant tree diagrams for Dark Matter annihilation.
  Note that the amplitudes of the graphs shown at the top are proportional
  to the mass of the final state fermion, while the Higgs exchange
  is proportional to \tb~for d-type quarks and 1/\tb~ for up-type
  quarks. This implies that light fermion final states can be neglected
  and at large \tb~ the bottom final states dominate.}
  \label{anni}
\end{center}
\end{figure}
\begin{figure}
\begin{center}
 \includegraphics[width=0.45\textwidth,clip]{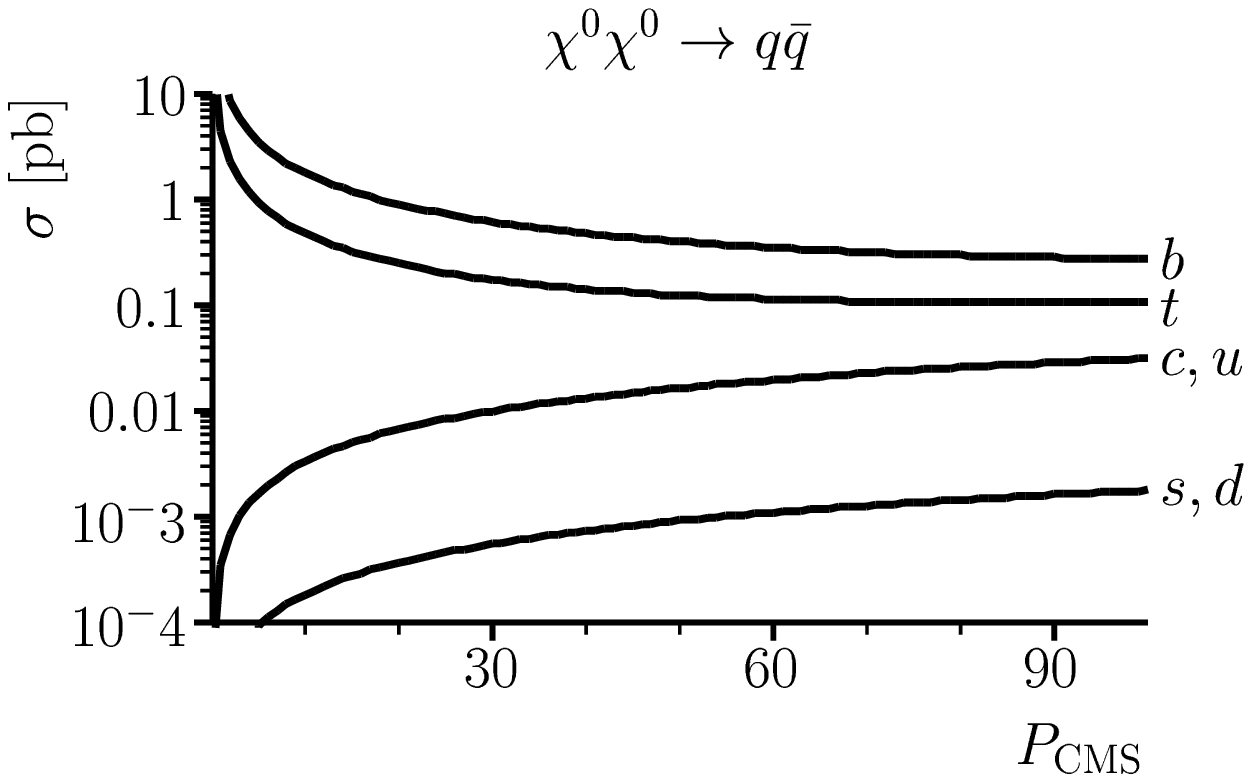}
 \includegraphics[width=0.45\textwidth,clip]{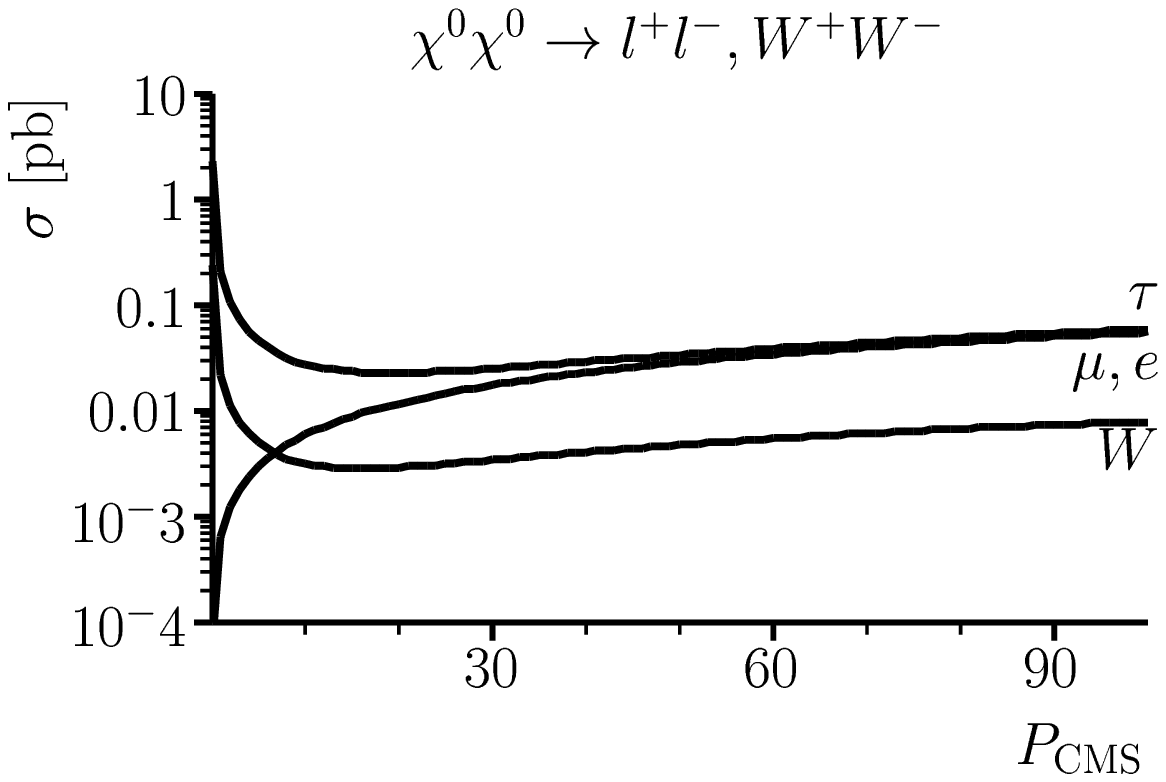}
 \caption[]{\it
 The neutralino annihilation total cross section for \tb=35
 as function of the center of mass momenta in GeV of the neutralinos for quark -,
 lepton - and $W^+W^-$ final states, as calculated with CalcHEP~\cite{calchep}.
 Note the helicity suppression at low momenta for light fermions.}
 \label{sigmap}
\end{center}
\end{figure}
\begin{figure}
\begin{center}
 \includegraphics [width=0.45\textwidth,clip]{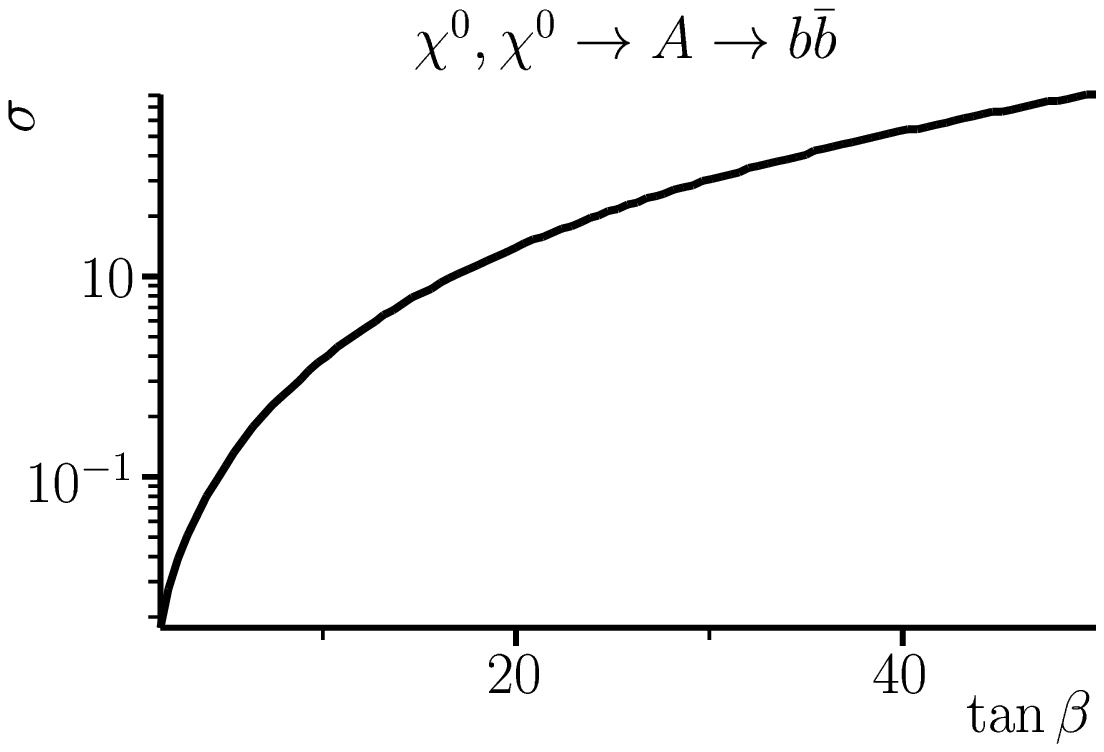}
 \includegraphics [width=0.45\textwidth,clip]{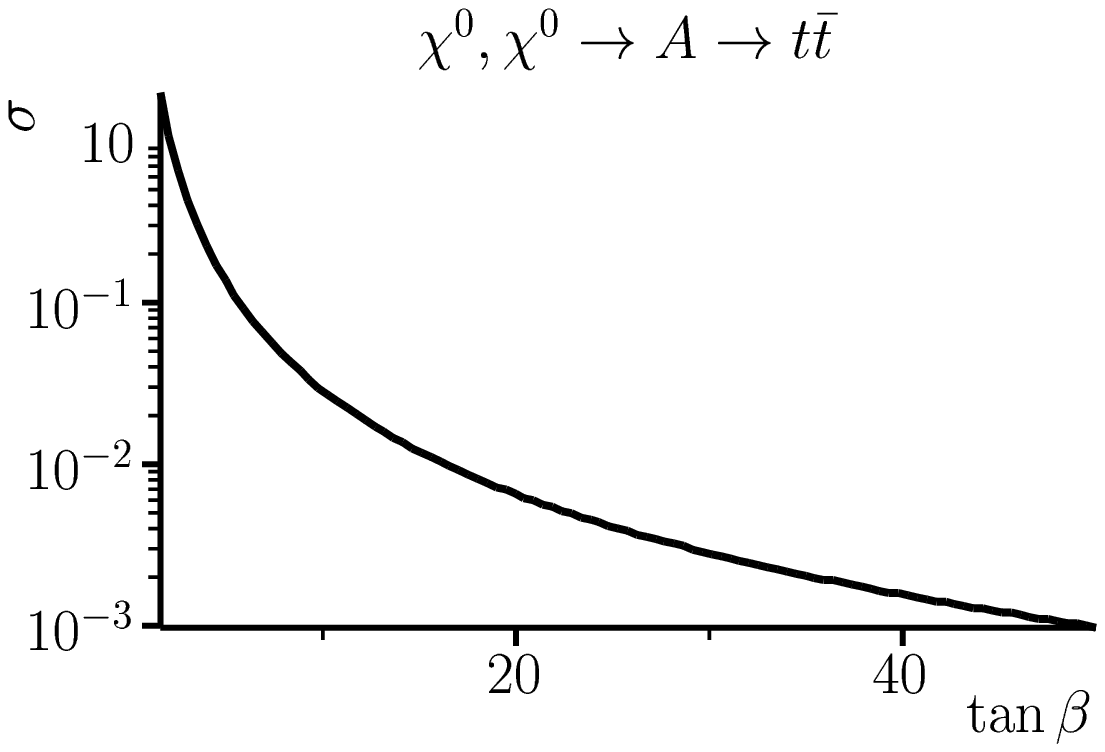}
 \caption[]{\label{sigmaA} \it
 The neutralino annihilation cross
 section for pseudoscalar Higgs exchange for bottom and top final
 states as function of \tb~, as calculated
 with CalcHEP~\cite{calchep}.}
\end{center}
\end{figure}

\begin{figure}
\begin{center}
 \includegraphics [width=0.34\textwidth,clip]{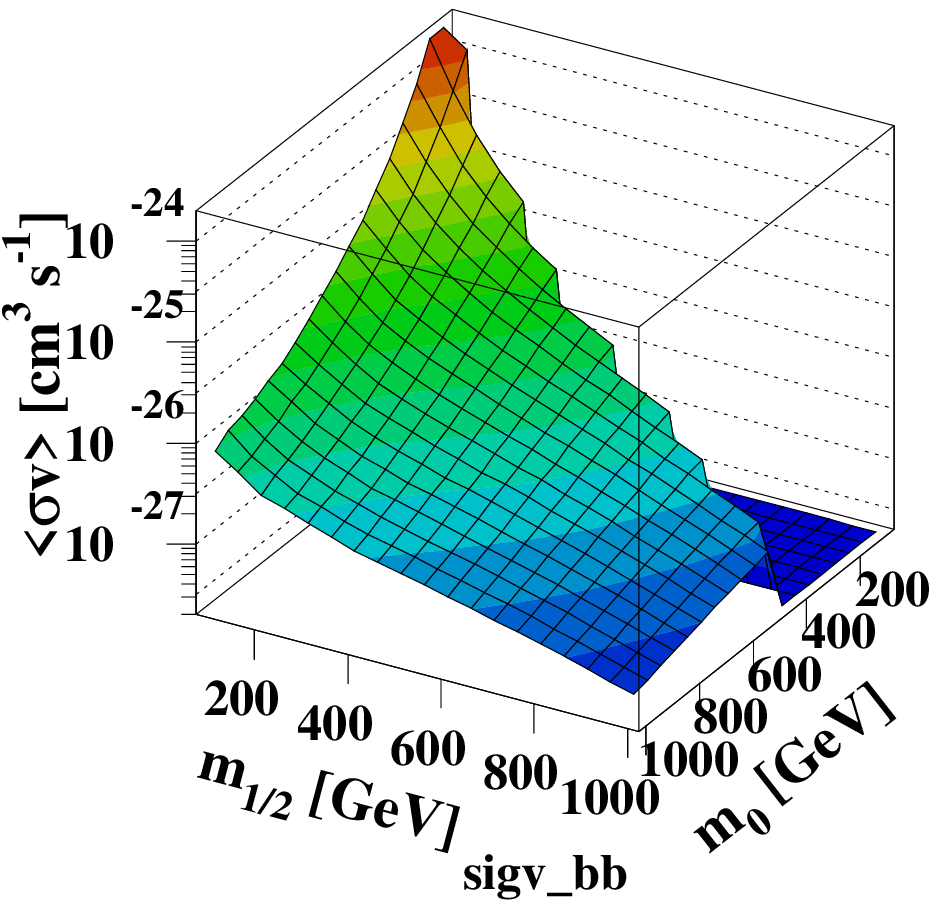}
 \includegraphics [width=0.34\textwidth,clip]{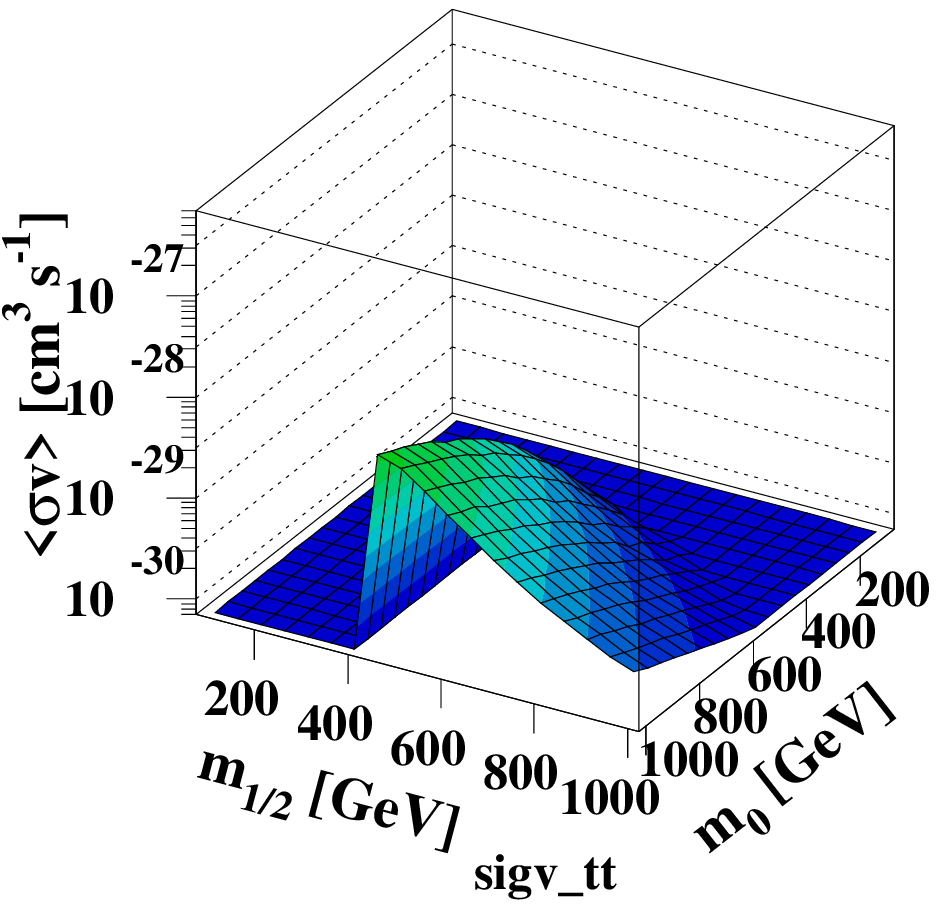}
 \includegraphics [width=0.34\textwidth,clip]{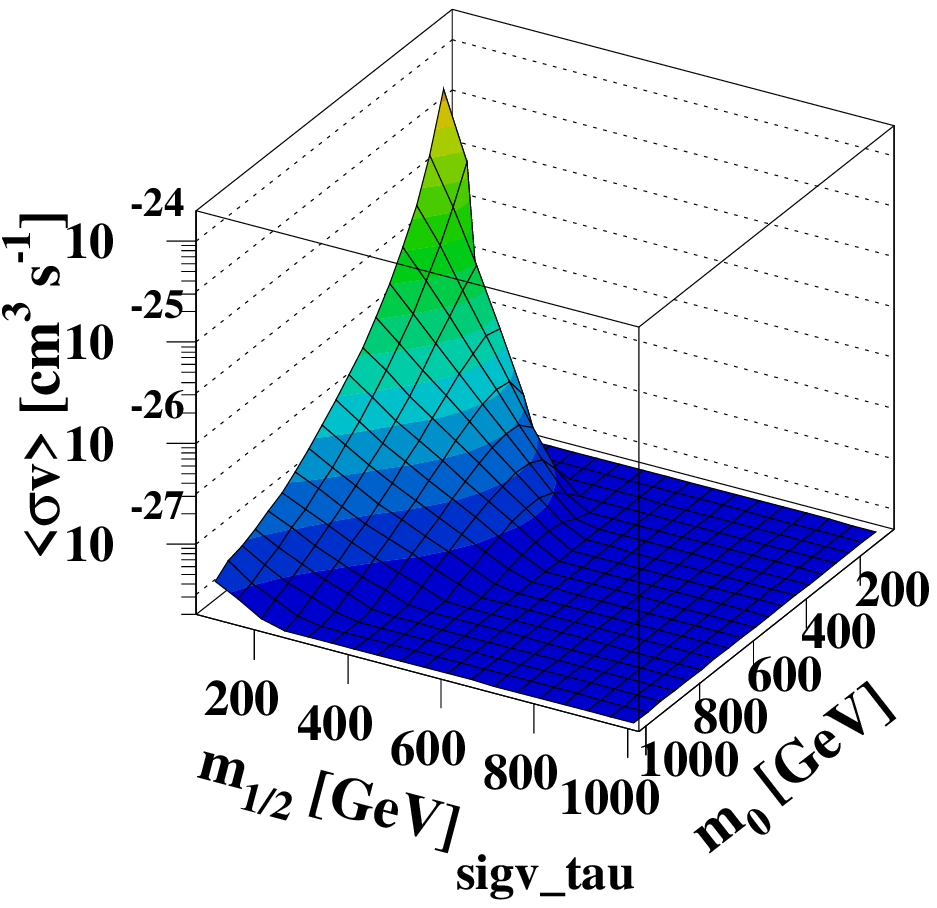}
 \includegraphics [width=0.34\textwidth,clip]{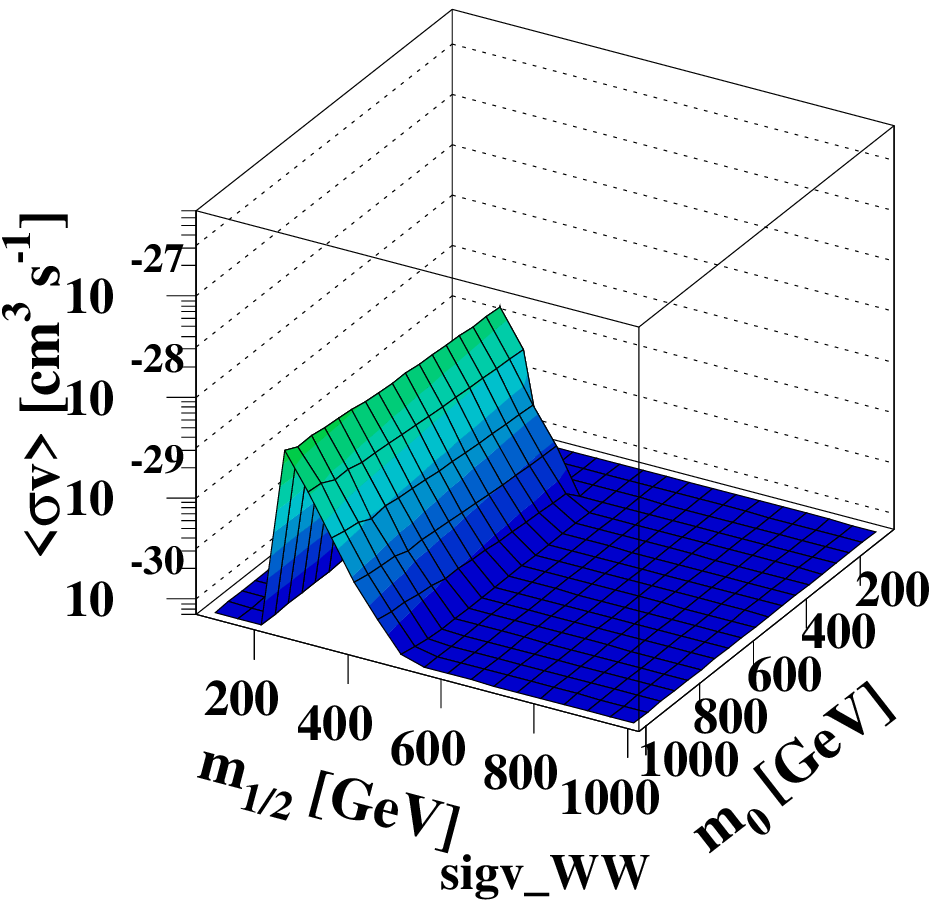}
 \caption[]{\label{sigmavall} \it
 The thermally averaged annihilation cross section times velocity for
 neutralino annihilation as function of $m_0$ and $m_{1/2}$ for
  $\tan\beta$= 50 and $b\overline{b}$, $t\overline{t}$,
  $W^+W^-$, and $\tau\overline{\tau}$ final states (clockwise from top left).
  The results were calculated with DarkSusy~\cite{darksusy}.}
\end{center}
\end{figure}

\begin{figure}
\begin{center}
 \includegraphics [width=0.34\textwidth,clip]{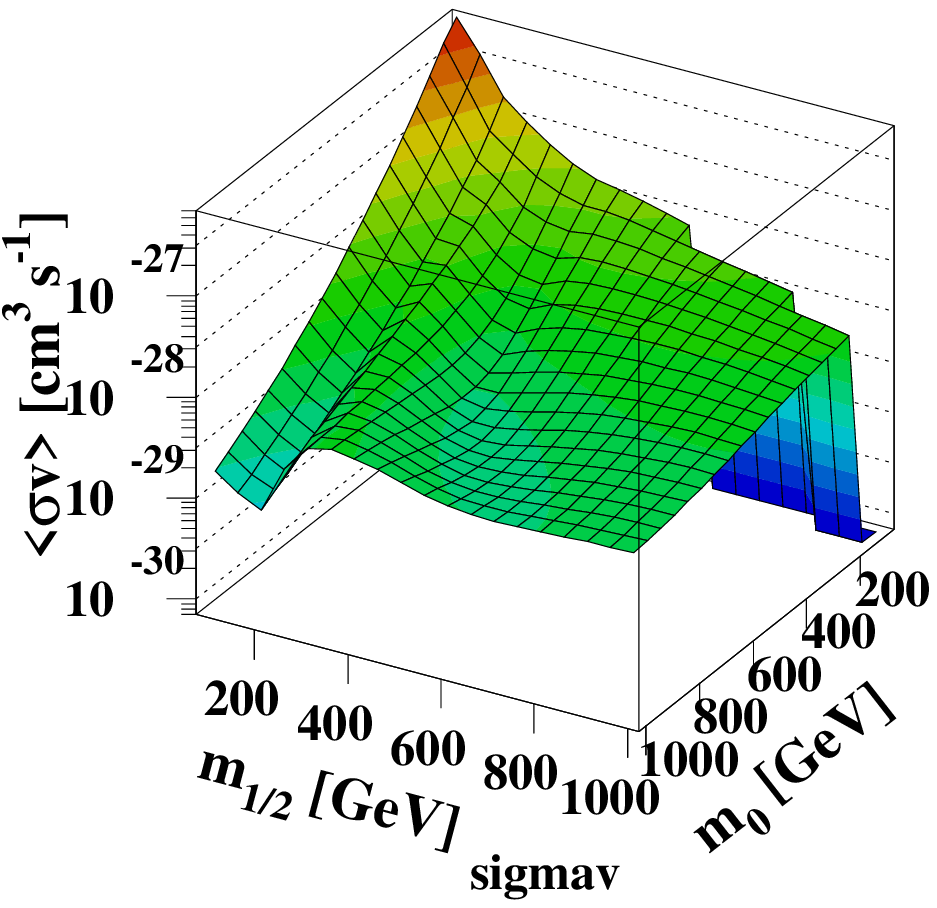}
 \includegraphics [width=0.34\textwidth,clip]{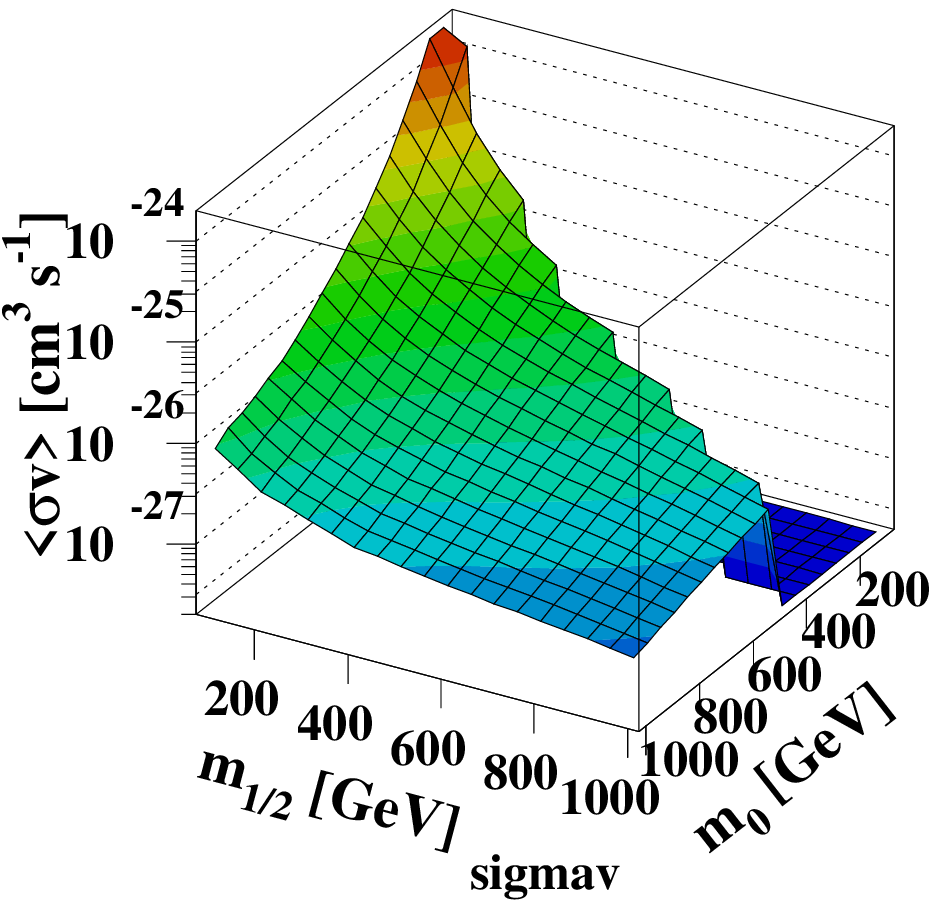}
 \caption[]{\label{sigmav} \it
 The thermally averaged total annihilation cross section times velocity
 for neutralino annihilation as function of $m_0$ and $m_{1/2}$ for
 $\tan\beta$= 5 (left) and 50 (right).)
 The neutralino mass equals $\approx 0.4 m_{1/2}$ in the CMSSM, so
 in the plots the neutralino varies from 40 to 400 GeV along the
 front axis. Note the strong decrease of the cross section for
 heavier SUSY mass scales and the different vertical scales due to
 the strong increase of the cross section with \tb.}
\end{center}
\end{figure}
\section{Annihilation Cross section Constraints from WMAP}
In the early universe all particles were produced abundantly and
were in thermal equilibrium through annihilation and production
processes. The time evolution of the number density of the
particles is given by the Boltzmann equation, which can be written
for neutralinos as: \bq \frac{dn_\chi}{dt}+3Hn_\chi=-<\sigma
v>(n_\chi^2-n_\chi^{eq 2}), \eq where H is the Hubble expansion
rate, $n_\chi$ is the actual number density, $n_\chi^{eq}$ is the
thermal equilibrium number density (before freeze-out), $<\sigma
v>$ is thermally averaged value of the total annihilation cross
section times the relative velocity of the annihilating
neutralinos. The Hubble term takes care of the decrease in number
density because of the expansion, while the first term on the
right hand side represents the decrease due to annihilation and
the second term represents the increase through creation by the
inverse reactions.

At temperatures below the mass of the neutralinos the number
density drops exponentially. The annihilation rate $\Gamma=<\sigma
v> n_\chi$ drops exponentially as well, and if it drops below the
expansion rate, the neutralinos cease to annihilate. They fall out
of equilibrium (freeze-out) at a temperature of about $m_\chi/25$
~\cite{kolb} and a relic cosmic abundance remains.

For the case that $<\sigma v>$ is energy independent, which is a
good approximation in case there is no coannihilation, the present
mass density in units of the critical density is given
by~\cite{jungman}: \bq \Omega_\chi h^2=\frac{m_\chi
n_\chi}{\rho_c}\approx (\frac{3\cdot 10^{-27} cm^3 s^{-1}}{<\sigma
v>})\label{wmap}.\eq One observes that the present relic density
is inversely proportional to the annihilation cross section at the
time of freeze out, a result independent of the neutralino mass
(except for logarithmic corrections). For the present value of
$\Omega_\chi h^2=0.1$ the thermally averaged total cross section
at the freeze-out temperature of $m_\chi/25$ must have been
$3\cdot 10^{-27}cm^3 s^{-1}$. This can be achieved only for
restricted regions of parameter space in the MSSM, as will be
discussed in the next section.
\section{Dark Matter Predictions from Supersymmetry}
Supersymmetry~\cite{susyrev} presupposes a symmetry between
fermions and bosons, which can be realized in nature only if one
assumes each particle with spin j has a supersymmetric partner
with spin $\vert j-1/2\vert$ ($\vert j-1/2\vert$ for the Higgs
bosons). This leads to a doubling of the particle spectrum.
Unfortunately the supersymmetric particles or ``sparticles'' have
not been observed so far, so the sparticle masses must be above
the limits set by searches at present accelerators. Obviously SUSY
cannot be an exact symmetry of nature; or else the supersymmetric
partners would have the same mass as the normal particles. The
mSUGRA model, i.e. the Minimal Supersymmetric Standard Model
(MSSM) with supergravity inspired breaking terms, is characterized
by only 5 parameters: $m_0,~m_{1/2},~\tb,~\mbox{sign}(\mu), ~A_0$.
Here $m_0$ and $m_{1/2}$ are the common masses for the gauginos
and scalars at the GUT scale, which is determined by the
unification of the gauge couplings. Gauge unification is still
possible with the precisely measured couplings at LEP~\cite{bs}.
The ratio of the vacuum expectation values of the neutral
components of the two Higgs doublets in Supersymmetry is called
\tb ~ and $A_0$ is the trilinear coupling at the GUT scale. We
only consider the dominant trilinear couplings of the third
generation of quarks and leptons and assume also $A_0$ to be
unified at the GUT scale. The constraints on the supersymmetric
parameters space are practically independent of $A_0$ due to a
coincidence from the constraints from the \bsg~ rate and the lower
limit on the Higgs mass of 114 GeV~\cite{bs}. The absolute value
of the Higgs mixing parameter $\mu$ is determined by electroweak
symmetry breaking, while its sign is taken to be positive, as
preferred by the anomalous magnetic moment of the muon~\cite{bs}.

The GUT scale masses are connected to low energy masses by the
Renormalization Group Equations (RGE), as shown in Fig. \ref{evol}.
The running masses of the gauginos at low energy
obey the simple solutions of the RGE:

\bq M_i(t)=\frac{\tilde{\alpha}_i(t)}{\tilde{\alpha}_i(0)}m_{1/2}. \eq
Numerically at the weak scale $(t=2\ln(\mgut/\mz)=66)$ one finds (see fig.
\ref{evol}):
\bqa
 M_3(\tilde{g})&\approx& 2.7m_{1/2},\\
 M_2(M_Z)&\approx& 0.8m_{1/2},\\
 M_1(M_Z)&\approx& 0.4m_{1/2}.\label{gaugi}
\eqa

The gluinos obtain corrections from the strong coupling constant
$\alpha_3$; therefore they grow heavier than the gauginos of the
$SU(2)_L\otimes U(1)_Y$ group.
Since the Higgsinos and gauginos are all spin 1/2 particles
and are equal in all other quantum nubers,
the mass eigenstates are in general mixtures of them,
which are called generically charginos (neutralinos)
for the mixture of the supersymmetric partners of the charged (neutral)
 gauge bosons and charged (neutral) Higgs bosons.
The Majorana neutralino and Dirac chargino fields can be written as:
$$ \chi = \left(
\begin{array}{c}\tilde{B} \\ \tilde{W}^3 \\
 \tilde{H}^0_1 \\ \tilde{H}^0_2
 \end{array}\right), \ \ \ \psi = \left( \begin{array}{c}
 \tilde{W}^{+} \\ \tilde{H}^{+}
\end{array}\right),$$
while the mass matrices can be written as~\cite{susyrev}:

\bq
M^{(0)}=\left(
\begin{array}{cccc}
 M_1 & 0 & -M_Z\cos\beta \sw & M_Z\sin\beta \sw \\
 0 & M_2 & M_Z\cos\beta \cw  & -M_Z\sin\beta \cw \\
 -M_Z\cos\beta \sw & M_Z\cos\beta \cw & 0 & -\mu \\
 M_Z\sin\beta \sw & -M_Z\sin\beta \cw & -\mu & 0
\end{array} \right)\label{neutmat}
\eq

\bq
 M^{(c)}=\left(
 \begin{array}{cc}
  M_2 & \sqrt{2}M_W\sin\beta \\ \sqrt{2}M_W\cos\beta & \mu
 \end{array} \right)\label{charmat}
\eq

The last matrix leads to two chargino eigenstates $\tilde{\chi}_{1,2}^{\pm}$.
The dependence on the parameters at the GUT scale can
be estimated by substituting for $M_2$ and $\mu$ their values at the weak
scale: $M_1(M_Z)=2 M_2(M_Z)\approx 0.4m_{1/2}$ and $\mu(M_Z) \approx 0.63\mu(0)$.

From Fig. \ref{evol} it can be seen that the mass parameters in
the Higgs potential, $m_1$ and $m_2$, are driven negative, largely
because of the large Yukawa couplings of the third generation of
quarks and leptons. This leads to radiative electroweak symmetry
breaking (EWSB), so the Higgs mechanism in Supersymmetry needs not
to be introduced {\it ad hoc}, as in the Standard Model, but is
caused by radiative corrections. The running is only strong enough
for top masses between 140 and 200 GeV and if the starting value
$\sqrt{\mu^2+m_0^2}$ at the GUT scale is large enough, which in
practice implies $\mu>M_2$. From the mass matrices \ref{neutmat}
and \ref{charmat} it is clear that for $M_{1,2}<\mu$ the lightest
chargino is wino-like with a mass given by $M_2$ if the mixing is
neglected and similarly the lightest neutralino is bino like with
a mass given by $M_1\approx 0.5M_2$.

In practice, there is some mixing and the neutralino mass eigenstates
are linear combinations of the weak eigenstates, i.e.
$$\chi_i^0=N_1|\tilde{B}> +N_2|\tilde{W}_3>+N_3|\tilde{H}_1^0>+
N_4|\tilde{H}_2^0>.$$ The gaugino fraction $N_1^2+N_2^2$ is
nevertheless close to one, especially if the diagonal elements are
large compared with the off-diagonal elements proportional to
$M_Z$. This is demonstrated in Fig. \ref{gaugino}.

The interaction of the sparticles with normal matter is governed by a new
{\it multiplicative} quantum number called R-parity, which is needed
in order to prevent baryon- and lepton number violation. In GUT theories
quarks, leptons and Higgses are all contained in the same supermultiplet,
which allows couplings between quarks and leptons. Such transitions,
which could lead to rapid proton decay, are not observed in
nature. Therefore, the SM particles are assigned a positive
R-parity and the supersymmetric partners have a negative one, which can be
related to the known conserved quantum numbers of spin S, baryon
number B and lepton number L by $R=(-1)^{3B+l+2S}$. Requiring
R-parity conservation implies that at each vertex one needs
{\it two} supersymmetric particles, from which it follows that:

\begin{itemize}
\item The rapid proton decays involving vertices with only one sparticle do
 not occur.
\item Sparticles can be produced only in pairs, e.g. $\overline{p} p\ra
 \tilde{q}\tilde{g}X$ or $e^+ +e^-\ra \tilde{\mu}^+\tilde{\mu}^-$.
\item The heavier sparticles can decay to lighter ones, like $\tilde{e}\ra
 e\tilde{\gamma}$ or $\tilde{q}\ra q\tilde{g}$, but the Lightest
 Supersymmetric Particle (LSP) is stable, since its decay into normal matter
 would change R-parity.
\item The LSP has to be neutral to be a good candidate for Dark Matter.
\item The interactions of particles and sparticles can be different. For
 example, the photon couples to electron-positron pairs, but the photino does
 not couple to electron-positron - or selectron-spositron pairs, since in
 these cases the R-parity would change from -1 to +1.
\item The LSP is weakly interacting with normal matter, since the final state
 has to contain the LSP again, so its interaction with quarks would only be
 elastic scattering by e.g. Z-, Higgs or squark exchange.
\end{itemize}

Consequently, the LSP is an ideal candidate for Dark Matter, since
it has all the properties of a Weakly Interacting Massive Particle
(WIMP), namely it is neutral, heavy and weakly interacting, so it
will form galactic haloes.

The neutralinos can annihilate through the diagrams, shown in Fig.
\ref{anni}. The main features of the amplitudes have been
indicated below the diagrams:

\begin{itemize}
\item The annihilation into fermion-antifermion pairs is
proportional to the fermion mass in the limit v$\to$0, which is
the important case in the present universe at a temperature of a
few Kelvin. This can be understood as follows: The neutralino is a
Majorana particle, so it is its own antiparticle. In addition it
has spin 1/2, thus obeying Fermi statistics, which implies it
cannot have identical quantum numbers. Furthermore at low velocity
it annihilates into an s-wave state, which implies the spins have
to be antiparallel (just like for electrons in the hydrogen
s-wave). Therefore also the spins in the final state have to be
antiparallel, which leads to an amplitude proportional to the
fermion mass to account for the required helicity
flip\cite{goldberg}. Consequently heavy final states are enhanced
at low momenta, as demonstrated in Fig. \ref{sigmap}. Note that at
higher momenta not only s-waves contribute and the helicity
suppression disappears, so during the time of freeze-out all final
states were produced. Note that these arguments are only valid for
the diagrams with sfermion and Z-exchange. For the Higgs exchange
the proportionality to the final state fermion mass  arises from
the Yukawa coupling. These cross sections were calculated with the
program package CalcHep\cite{calchep}. \item The second important
point concerns the \tb~ dependence: the diagram via pseudoscalar
Higgs exchange is proportional to \tb ~ for down-type quarks and
1/\tb~ for up-type quarks. This implies that at large \tb~
($\tb>5$) the b-quark final states are enhanced over t-quark final
states, as shown in Fig. \ref{sigmaA}. The amplitudes of the Higgs
exchange and the Z-exchange have an opposite sign. Since the top
final states have a large amplitude for Z-exchange (amplitude
$\propto $ mass), they are additionally suppressed by the negative
interference with the t-channel amplitudes. The cross sections for
various final states are shown as function of $m_0, m_{1/2}$ in
Fig. \ref{sigmavall}. The strong increase of the total
annihilation cross section as function of \tb~ is demonstrated in
Fig. \ref{sigmav} in the $m_0, m_{1/2}$ plane, which also shows
the strong dependence on the SUSY masses: only low SUSY scales and
large \tb~lead to cross sections of the order of magnitude
required by the WMAP data (Eq. \ref{wmap}). \item As shown in Fig.
\ref{anni}, the amplitude for pseudoscalar Higgs exchange is
proportional to $N_1 N_{3,4}$, i.e. it requires that the lightest
neutralino has both bino- and Higgsino components, which implies
that the diagonal elements in the mass matrix \ref{neutmat} should
not be too large compared with the off-diagonal elements
proportional to $M_Z$. So unless one tunes \tb~ and the SUSY
masses such that one hits the resonance ($m_A\approx 2m_\chi$), in
which case very small Higgsino admixtures are enough, one needs
relatively light neutralino masses.
\end{itemize}
\begin{figure}[t]
\begin{center}
 \includegraphics[angle=0,width=0.6\textwidth]{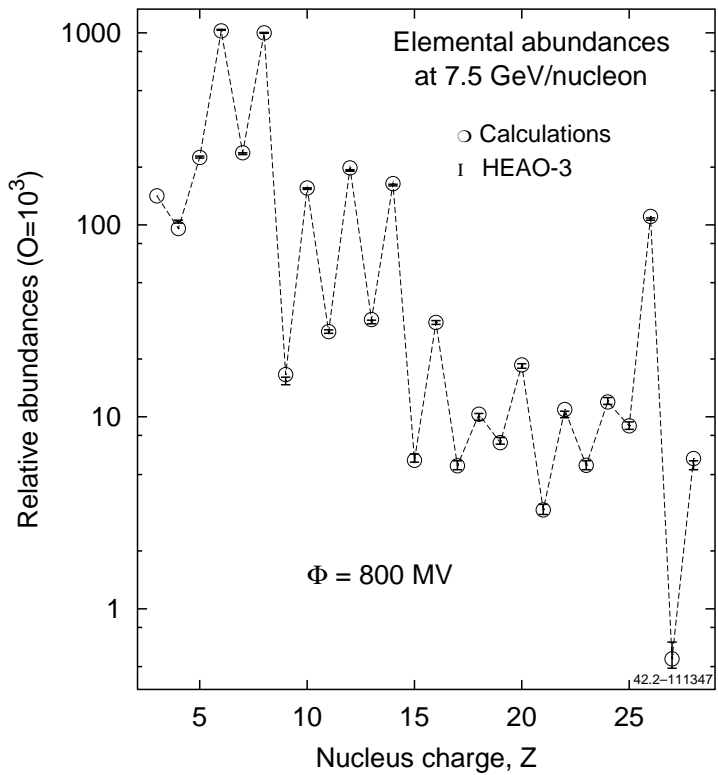}
 \caption[]{\it
 Propagated abundances at 7.5 GeV/nucleon, as calculated
 by Galprop in comparison with data. From Ref.~\cite{ms_nucl}.}
 \label{ms_nucl}
\end{center}
\end{figure}
\begin{figure}[t]
\begin{center}
 \includegraphics[angle=0,width=0.45\textwidth]{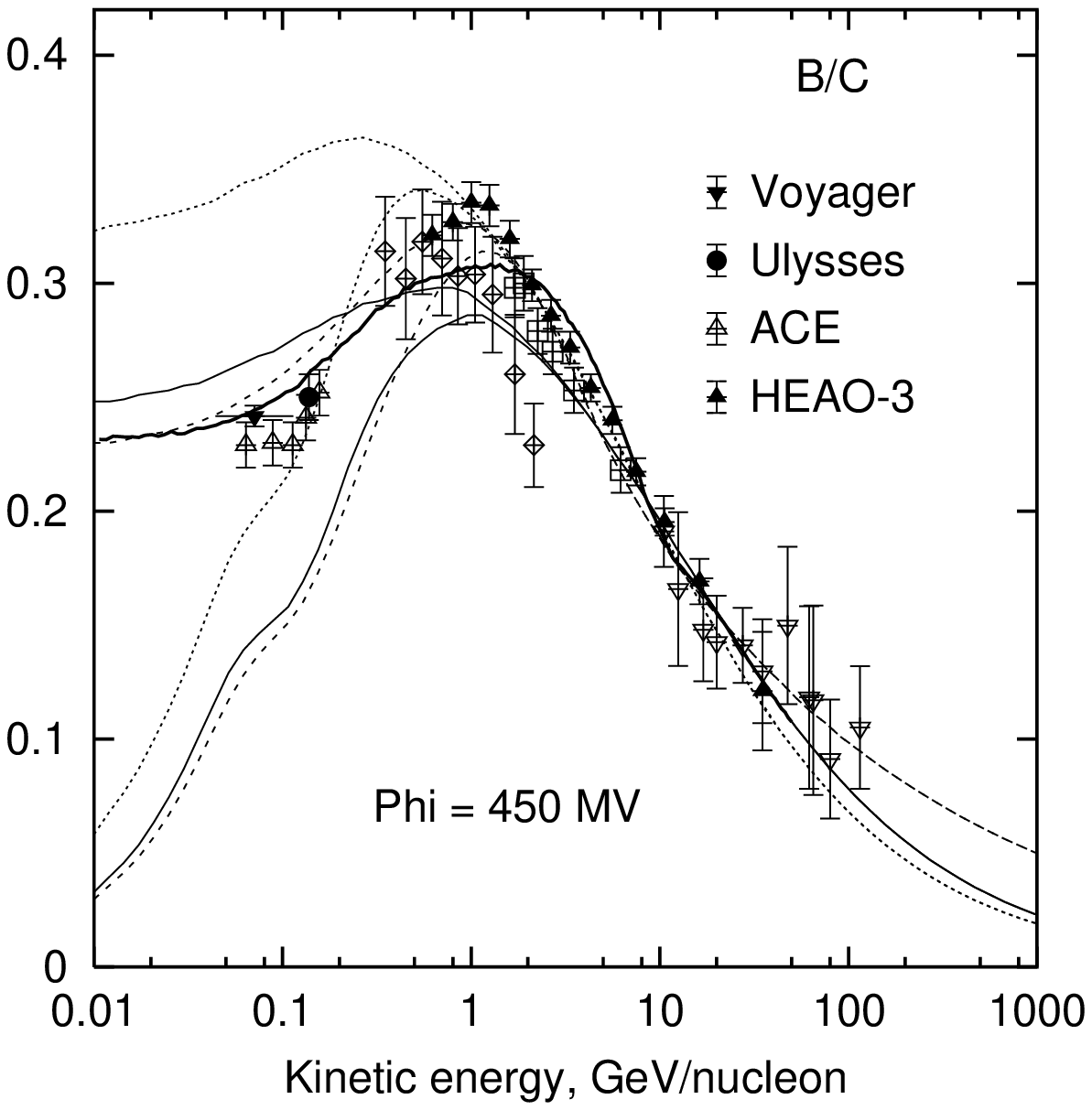}
 \includegraphics[angle=0,width=0.45\textwidth]{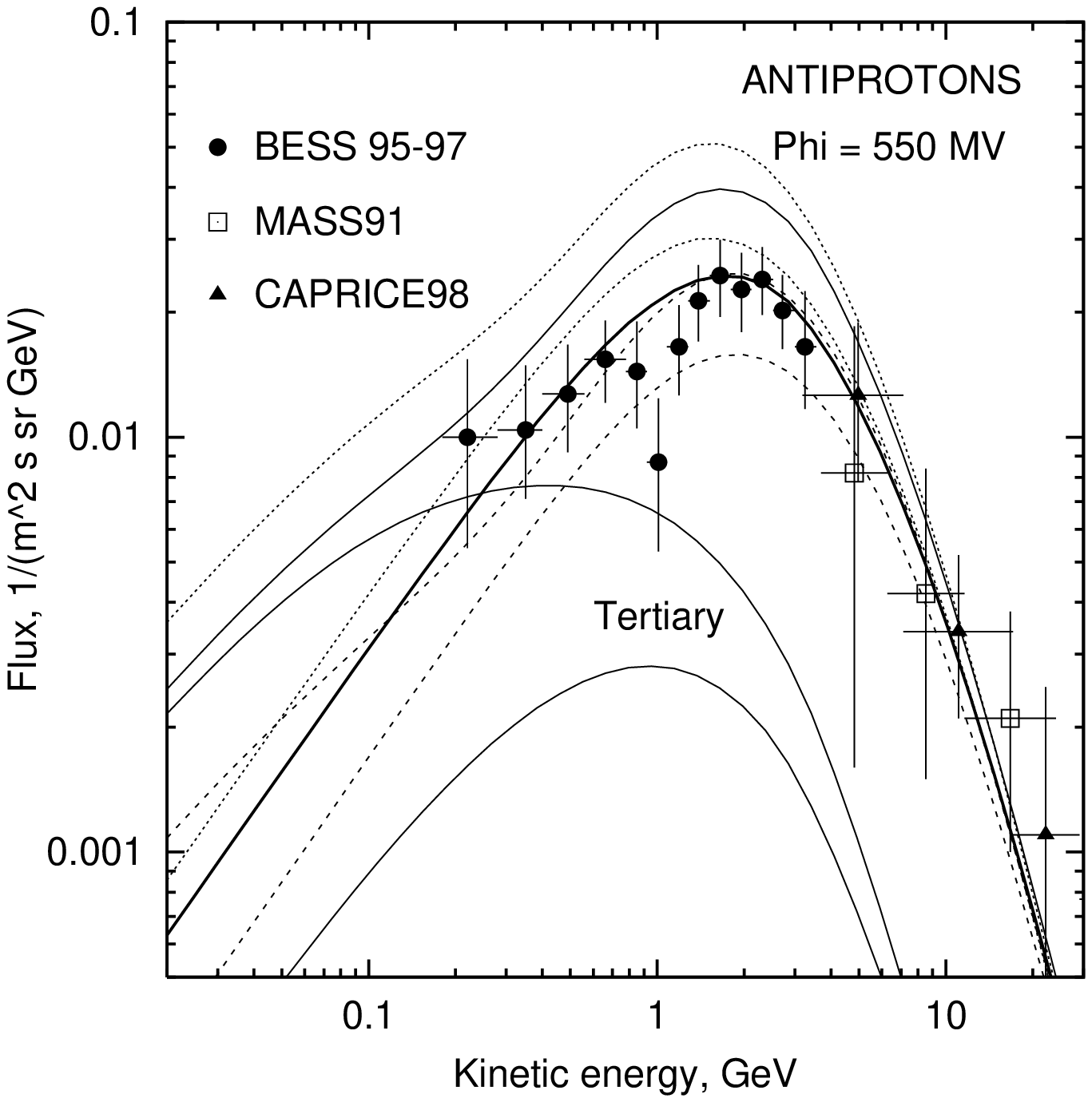}
 \caption[]{\it
 The B/C ratio as an example of secondary/primary nucleon ratios for
 various Galprop models in comparison with data. The
 dashed lines are the model with diffusive reacceleration, the solid ones
 for diffusion plus convection, and the dotted ones for plain diffusion.
 The lower (upper) curves of each kind are the interstellar (solar modulated)
 ones.
 The diffusive reacceleration curve (dashed)
 provides the best fit to the B/C ratio, but due to the large diffusion coefficient
 required, this leads to a deficiency in the flux of antiprotons
 (right). From Ref.~\cite{ms_problems}.}
 \label{ms_bc}
\end{center}
\end{figure}
\begin{figure}
\begin{center}
 \includegraphics[angle=0,width=0.5\textwidth]{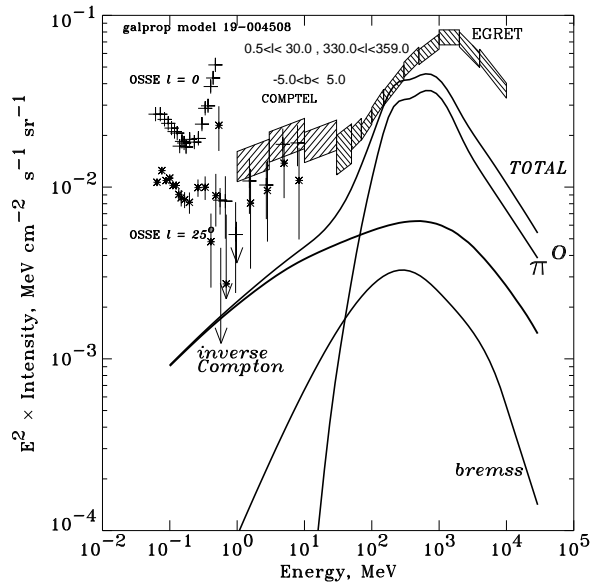}
 \caption[]{\it
 The gamma-ray energy spectrum of the inner Galaxy as calculated
 by Galprop~\cite{ms_gamma} in comparison with EGRET data.
 Clearly, there is an access of data above 1 GeV.}
 \label{ms_gamma}
\end{center}
\end{figure}

In summary, the annihilation cross section becomes large for large
\tb~ and is dominated for $\tb >5$ by the s-channel pseudoscalar
Higgs exchange into $b\overline{b}$ quark pairs. Fig. \ref{sigmav}
shows that values of \tb~ around 50 yield the  annihilation cross
sections  required by WMAP, given in  Eq. \ref{wmap}. Regions of
coannihilation\cite{griest}  at smaller \tb ~ are allowed by WMAP
data, but have too small annihilation cross sections to explain
the deficiencies in the positron, antiproton and gamma ray fluxes,
as will be shown in the last section. Before discussing the global
fits, the cosmic ray fluxes from nuclear interactions are
discussed.

\section{Cosmic Rays generated by Nuclear Interactions}

The sources of charged and neutral cosmic rays are believed to be
supernovae and their remnants, pulsars, stellar winds and binary
systems\cite{ms_1998_nucl,ms_1998_pos}. Observations of X-ray and
$\gamma$-ray emissions from these objects reveal the acceleration
of charged particles near them. Particles accelerated near the
sources propagate tens of millions of years in the interstellar
medium where they can loose or gain energy and produce secondary
particles and $\gamma$-rays. The spallation of primary nuclei into
secondary nuclei gives rise to rare isotopes. Nuclear interactions
produce not only matter, but also antimatter, like antiprotons and
positrons. The latter originate mainly from the decay of charged
pions and kaons.

The detailed studies of cosmic rays teach us about the production
and propagation in the universe. The gammas can deliver
information over intergalactic distances, while the charged
particles pro\-pagate mainly on galactic distances. Secondary
nuclei are produced in the galactic disc, from where they escape
into the halo by diffusion and Galactic winds (convection). They
may gain energy by ``diffusive'' reacceleration in the
interstellar medium by the 2nd order Fermi acceleration mechanism,
i.e. on average more interstellar gas clouds from opposite
directions are hitting a given nuclei than ``comoving'' gas
clouds. In elastic collisions this leads on average to an energy
gain, thus depleting the low energy part of the source spectrum.
Long-living radioactive secondaries  tell how long they survive in
the halo before interacting in the disc again, thus determining
the size of the halo. The gas density and acceleration time scale
can be probed by the abundances of the K-capture isotopes, which
would decay via electron K-capture in the interstellar gas.

A global fit to all this information allows one to build a model
of our galaxy. Analytical and semi-analytical models often fail
when compared with all data. Therefore advanced models
incorporating nuclear reaction networks, cross sections for
production of antiprotons, positrons, $\gamma$-rays and
synchrotron radiation, energy losses, convection, diffusive
reacceleration, distribution of sources, gas and radiation field
etc. are needed.

In addition, the
distributions of matter and antimatter in the interstellar medium (ISM)
are modified locally by affects of the solar activity and magnetic
fields inside our solar system, e.g. from the planets. Gleeson \&
Axford~\cite{solar} modelled the periodically varying solar
activity with a typical half cycle of 11 years\footnote{A cycle
can vary between 8 and 14 yr.} by a radial solar wind in which
the charged particles loose kinetic energy depending on their
rigidity R and on the distance $r$ from the sun. On this time
scale the incoming flux from the Galaxy does not vary and the
problem reduces to an adiabatical deceleration by the solar wind
with a dependence only on the radial coordinate. This can be
solved analytically:
\bq
 J(r,E,t)=\frac{R^2}{R^{\prime 2}}
 J(\inf,E^\prime),\label{solmod}
\eq
where $J(r,E,t)$ is the
measured differential flux at a distance $r$ from the sun for
particles with energy $E$ and mass $E_0^2/c^2$, $J(\inf,E^\prime)$
 is the incoming flux to the solar system and $R^2=(E^2-E_0^2)/(Ze)^2$ is the
rigidity with $R^\prime=R(E^\prime)$. The energy loss can be
parametrized by the solar modulation parameter $\Phi(t)$ as
$E=E^\prime-|Ze|\Phi(t)$, where $|Ze|$ is the absolute charge of
the particle; $\Phi(t)$ varies
 between 350 and 1500 MeV depending on the solar cycle.
The solar modulation shifts the particle spectrum
to lower energies, but the effect is only noticeable for rigidities
below 10 GV. Recent determinations of
the local interstellar flux (LIS) from the modulated (=measured) ones
for electrons, positrons and protons can be found in Ref.~\cite{casadei}.

The most complete and publicly available code for the
production and propagation of particles in our galaxy is the
Galprop code~\cite{ms_1998_nucl,ms_1998_pos}.
It provides a numerical solution to
 the transport equation including a cross section database with more than
2000 points, source functions, density distributions, etc. The
cross section tables include all possible cross sections: $p+p$,
$p+He$, $p+N$, $He+N$, $N+N$, where all nuclei up to the heaviest
ones (Ni) are considered. Fig. \ref{ms_nucl} shows the composition
of the primary and secondary nuclei, as calculated by Galprop in
comparison with data. Clearly, the production of secondary nuclei
is well described.

Fig. \ref{ms_bc} shows the spectrum of the Boron over Carbon (B/C)
ratio, which shows a characteristic depletion at low energies.
Since Boron is a purely secondary produced nuclei, while Carbon is
primarily produced, the depletion at low energy is a sensitive
handle on the question of diffusive reacceleration and solar
modulation. As shown, the modulation effects the spectra mainly at
kinetic energies below 10 GeV/nucleon. In order to reproduce the
sharp peak in the ratios of secondary to primary nuclei without
any unphysical breaks in the energy dependence of the diffusion
coefficients and/or the injection spectrum, the diffuse
reacceleration with a rather large diffusion coefficient is
needed~\cite{ms_problems}. But this leads to too few antiprotons,
as shown by the dashed line on the right hand side in Fig.
\ref{ms_bc}. The possible way out of the discrepancy between the
B/C ratio and too few antiprotons was suggested by Strong and
Moskalenko: a ``fresh'' local
 unprocessed component at low energies of primary nuclei, thus decreasing
the B/C ratio and allowing for a smaller diffusion coefficient.
Also too few gammas are generated by Galprop, as shown in Fig.
\ref{ms_gamma}, which would also need either a harder nucleon
spectrum or a harder electron spectrum, but this would need
spatial variations which make the spectrum in our local region
unrepresentative of the large scale average~\cite{ms_gamma}.

However, an alternative explanation may be the annihilation of
neutralinos, which increases the yield of gammas, antiprotons, and
high energy positrons, but does NOT effect the B/C ratio. This
goes exactly in the direction of solving these discrepancies
between Galprop and present data {\it simultaneously}, as will be
shown in the next section by a global fit to all data.

\begin{figure}
\begin{center}
 \includegraphics [width=0.4\textwidth,clip]{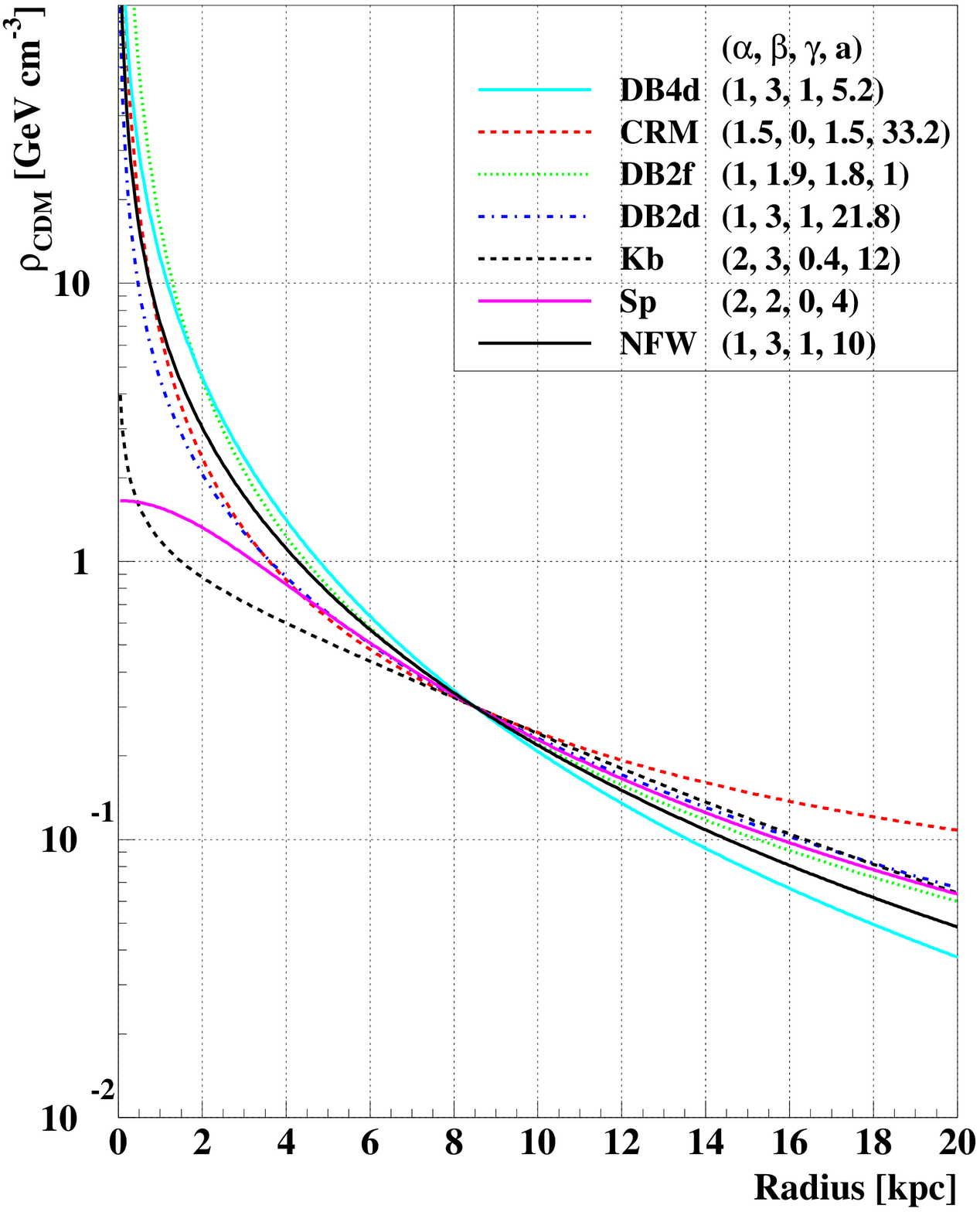}
 \includegraphics [width=0.4\textwidth,clip]{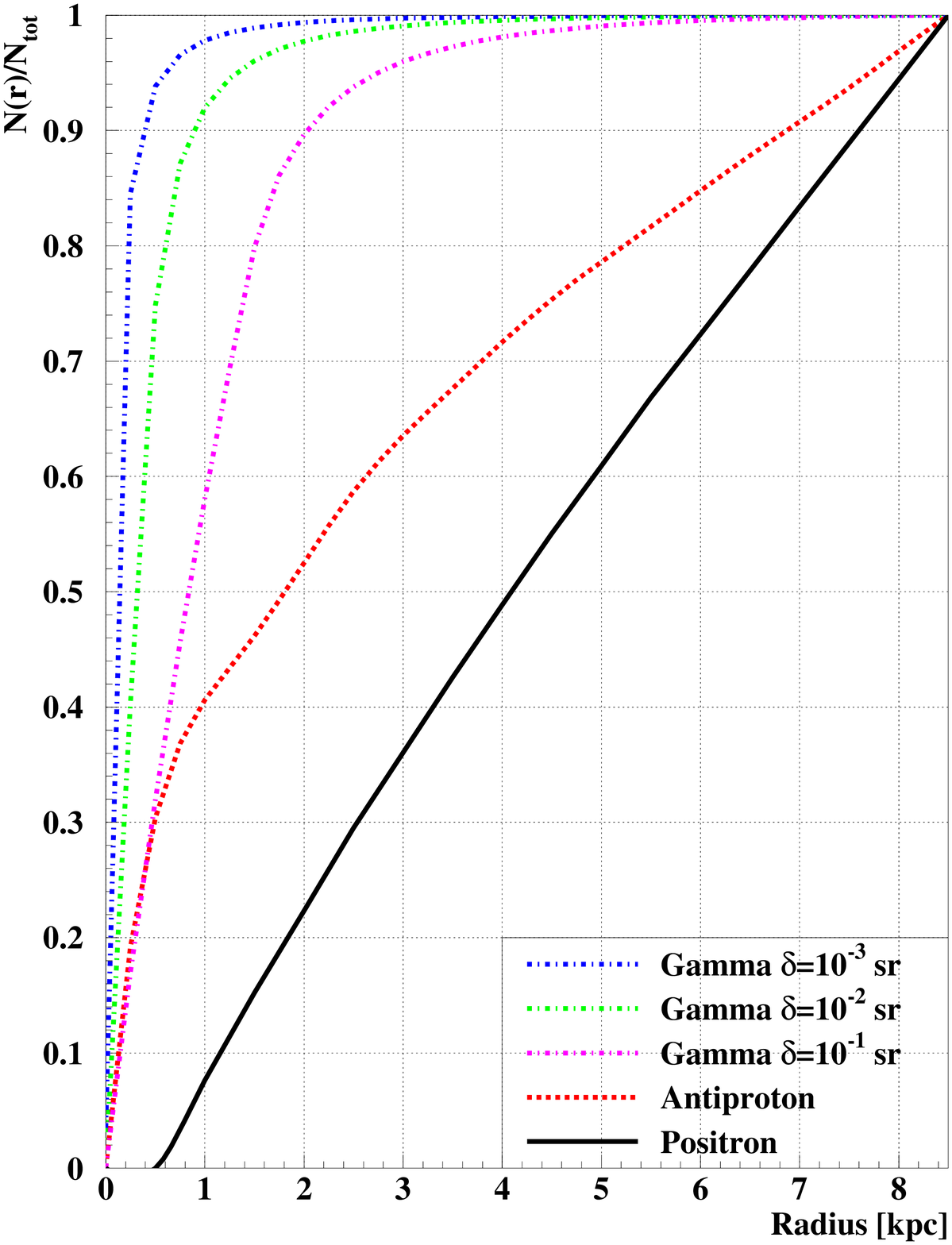}
 \caption[]{\it
 The halomodel used in this paper in comparison with other
 halomodels in the literature (left) and the integrated fraction of the fluxes
 as function of the
 distance from the center of the galaxy (right). One observes
 that more than 50\% of the gamma rays originate from a region of
 less than 0.2 (1.5) kpc from
 the center for a detector subtending a solid angle of
 $10^{-3}$ ($10^{-1}$) sr, while
 the antiprotons and positrons reach the 50\% at 1.9 and 4.1 kpc,
 respectively.}
 \label{halo}
\end{center}
\end{figure}

\begin{figure}
\begin{center}
 \includegraphics [width=0.9\textwidth,clip]{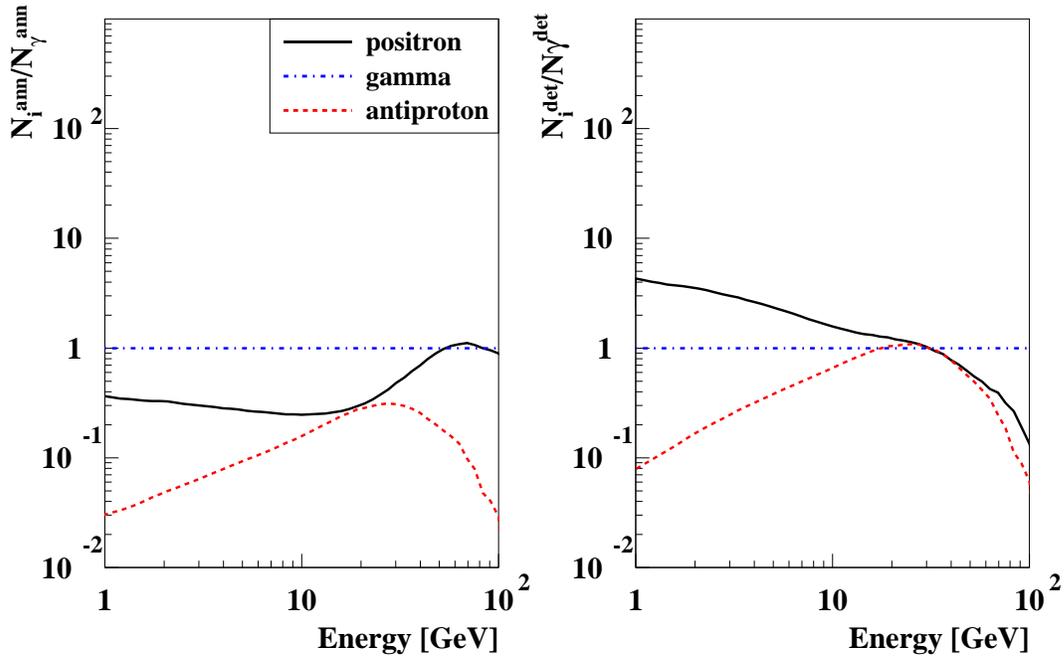}
 \caption[]{\it
 The flux of antiprotons and positrons
 normalized to the gamma flux from  neutralino annihilation before
 (left) and after (right) diffusion with the halomodel and
 diffusion parameters described in the text.}
 \label{yield}
\end{center}
\end{figure}
\section{Global Fits to positrons, antiprotons and gamma rays}

Trying to disentangle the contributions from nuclear interactions
and neutralino annihilation to the antimatter fluxes and gamma
rays is in practice not easy. Ideally one would like to implement
the neutralino annihilation as a source function in the Galprop
code, so the antimatter and gamma rays from nuclear interactions
and annihilation would be transported through the galaxy in an
identical way. However, this numerical code is too slow to be used
in a fit program. Therefore, we used the second best possibility,
namely using the publicly available code DarkSusy~\cite{darksusy}
for neutralino annihilation, which has semi-analytical solutions
to the diffusion equation and includes the important energy losses
for positrons. We changed the diffusion parameters and code in
DarkSusy in such a way, that the results resembled as closely as possible
the Galprop results. The main difference in the diffusion
parameters between DarkSusy and Galprop lies in the fact that
Galprop uses diffusive reacceleration, while DarkSusy does not.
The diffusive reacceleration is needed to fit the B/C ratio and
results in an almost order of magnitude larger diffusion constant
with a much smaller energy dependence. The diffusion parameters
and energy losses in DarkSusy were changed as follows:

\begin{figure}
\begin{center}
 \includegraphics [width=0.3\textwidth,clip]{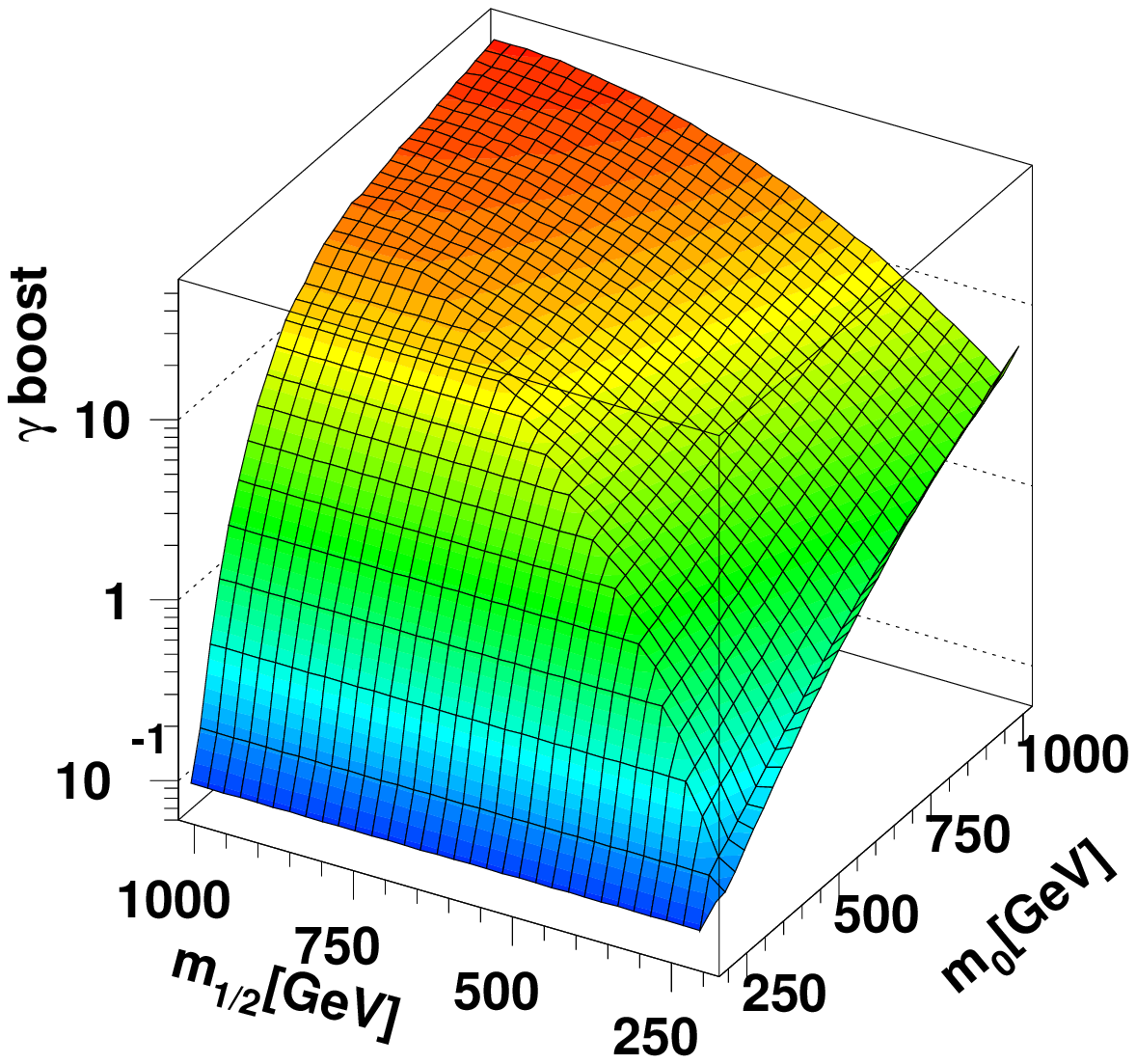}
 \includegraphics [width=0.3\textwidth,clip]{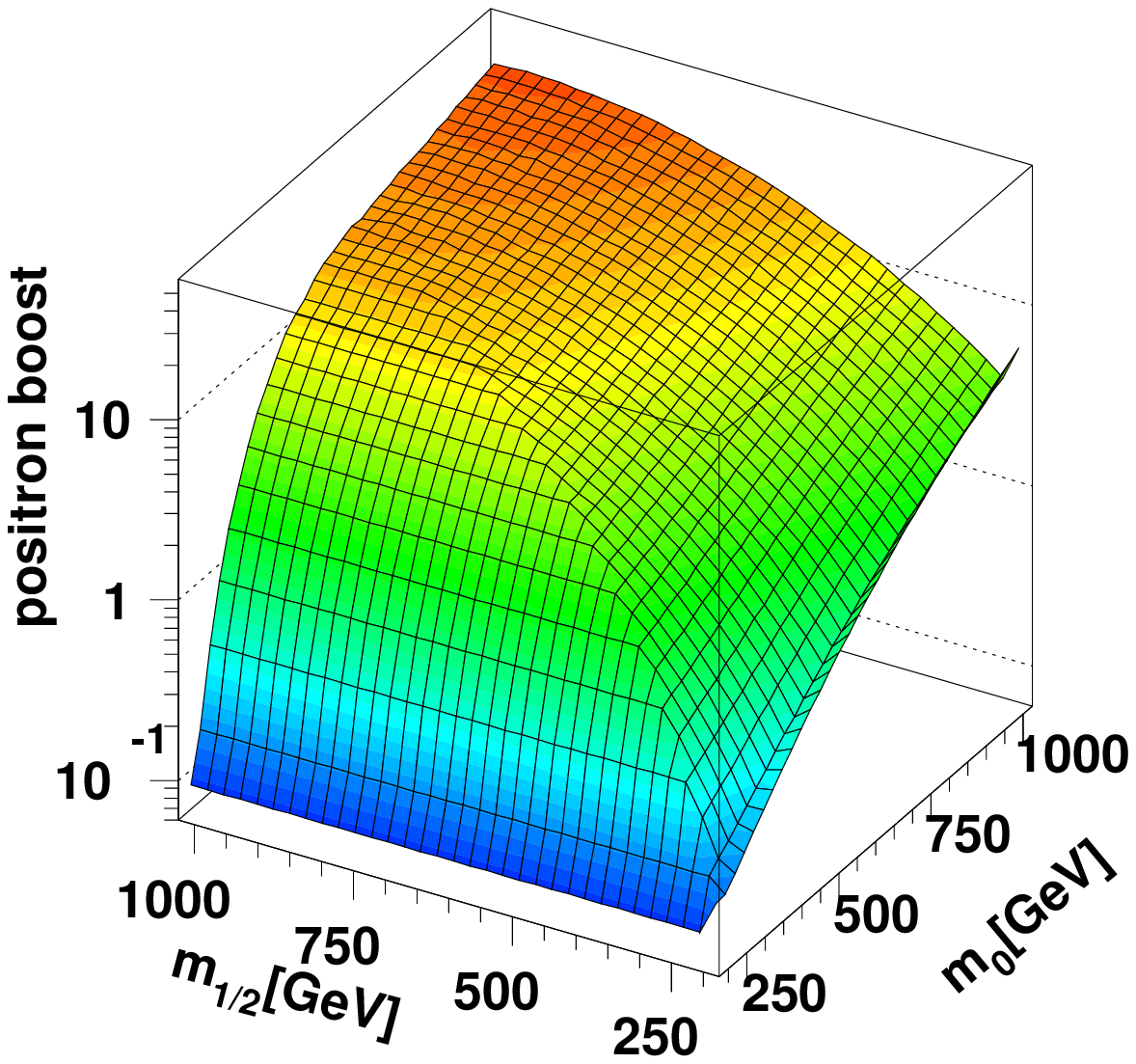}
 \includegraphics [width=0.3\textwidth,clip]{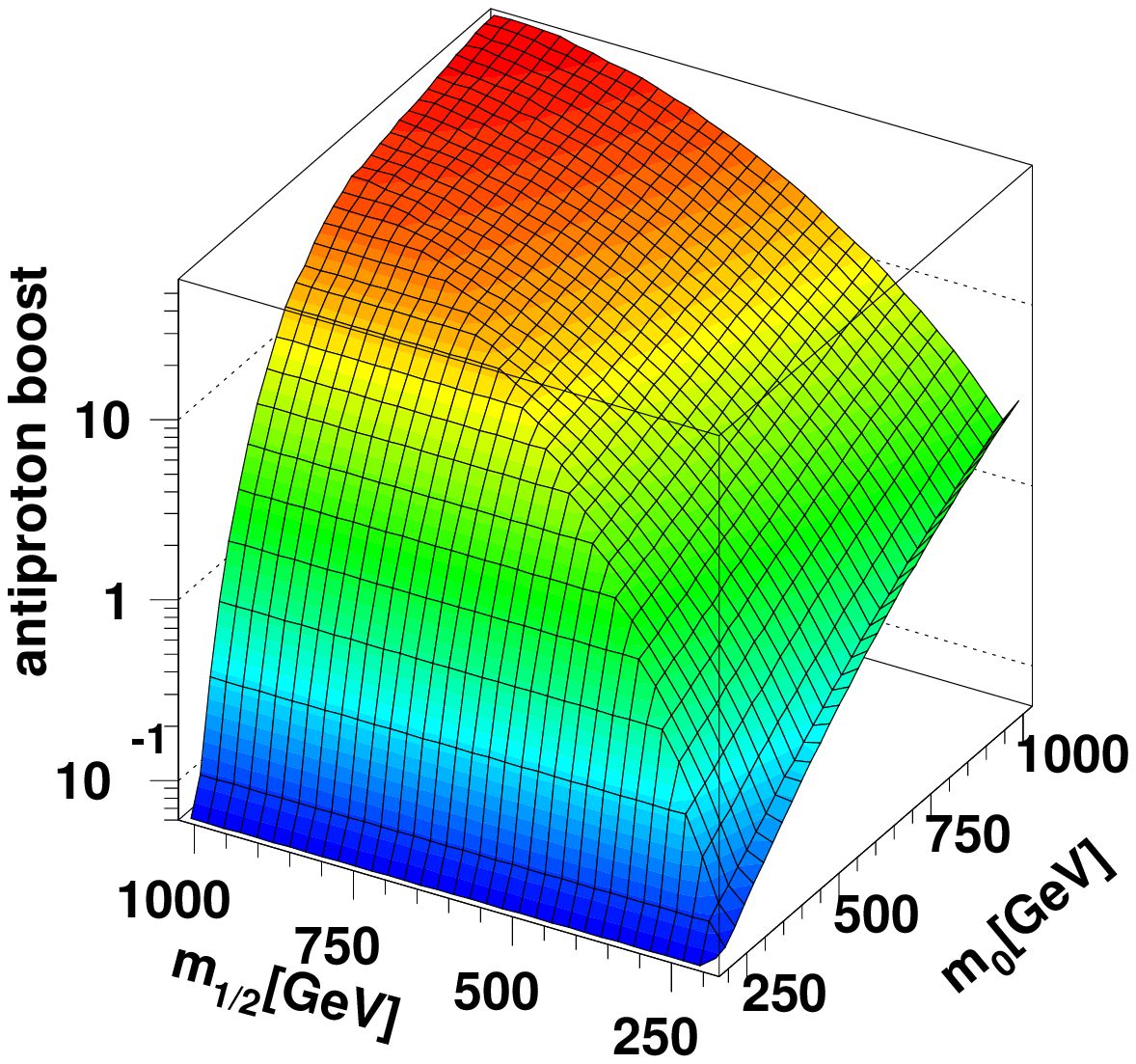}
 \caption[]{\it
 Boost factors for gamma rays,
 positrons and antiprotons (from left to right) as function of
 $m_0$ and $m_{1/2}$ for \tb=51. Note that the boost factors are
 similar for all three fluxes, which is a strong constraint for any
 model.}
 \label{boost}
\end{center}
\end{figure}

\begin{figure}
\begin{center}
 \includegraphics [width=0.35\textwidth,clip]{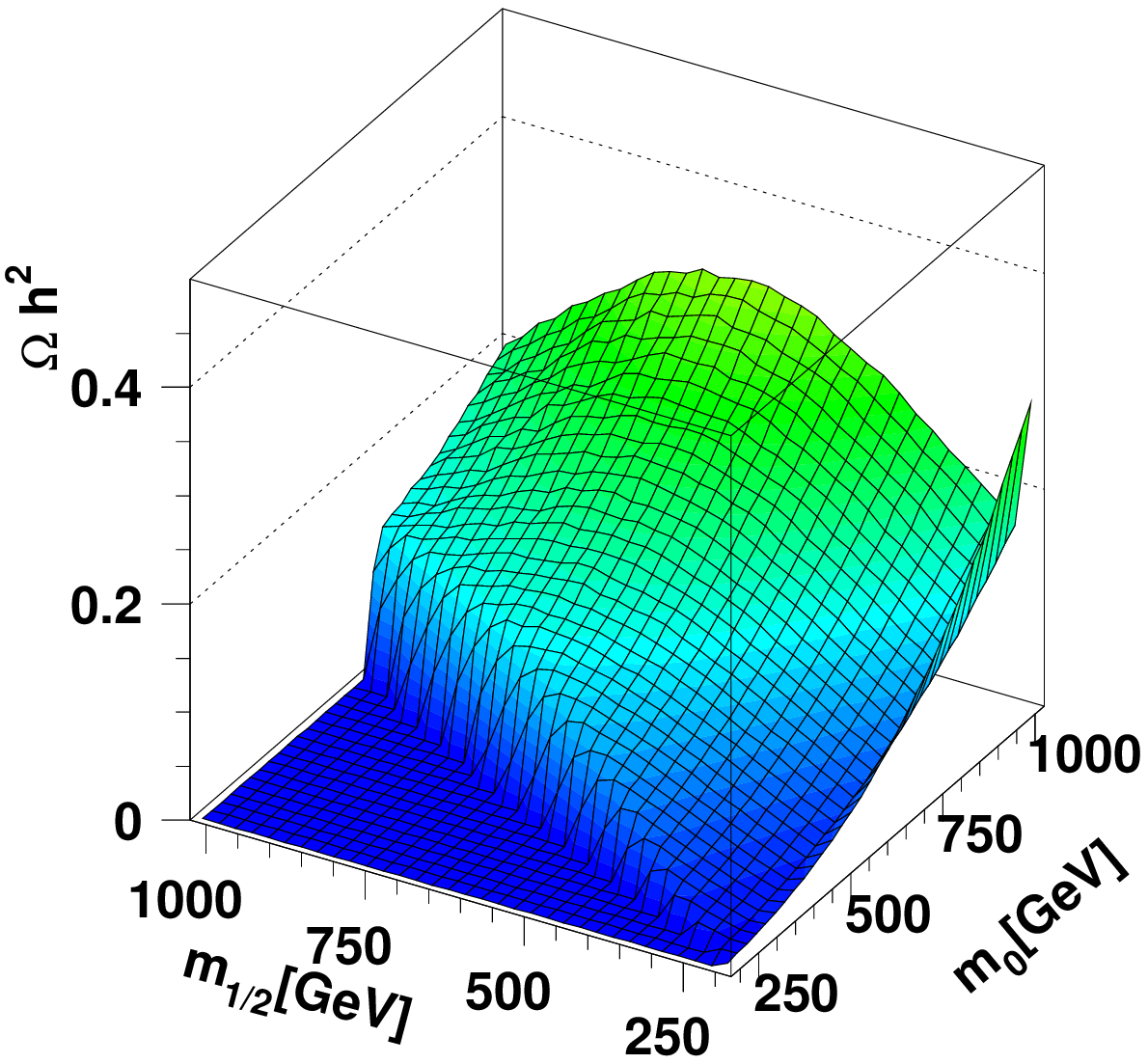}
 \includegraphics [width=0.35\textwidth,clip]{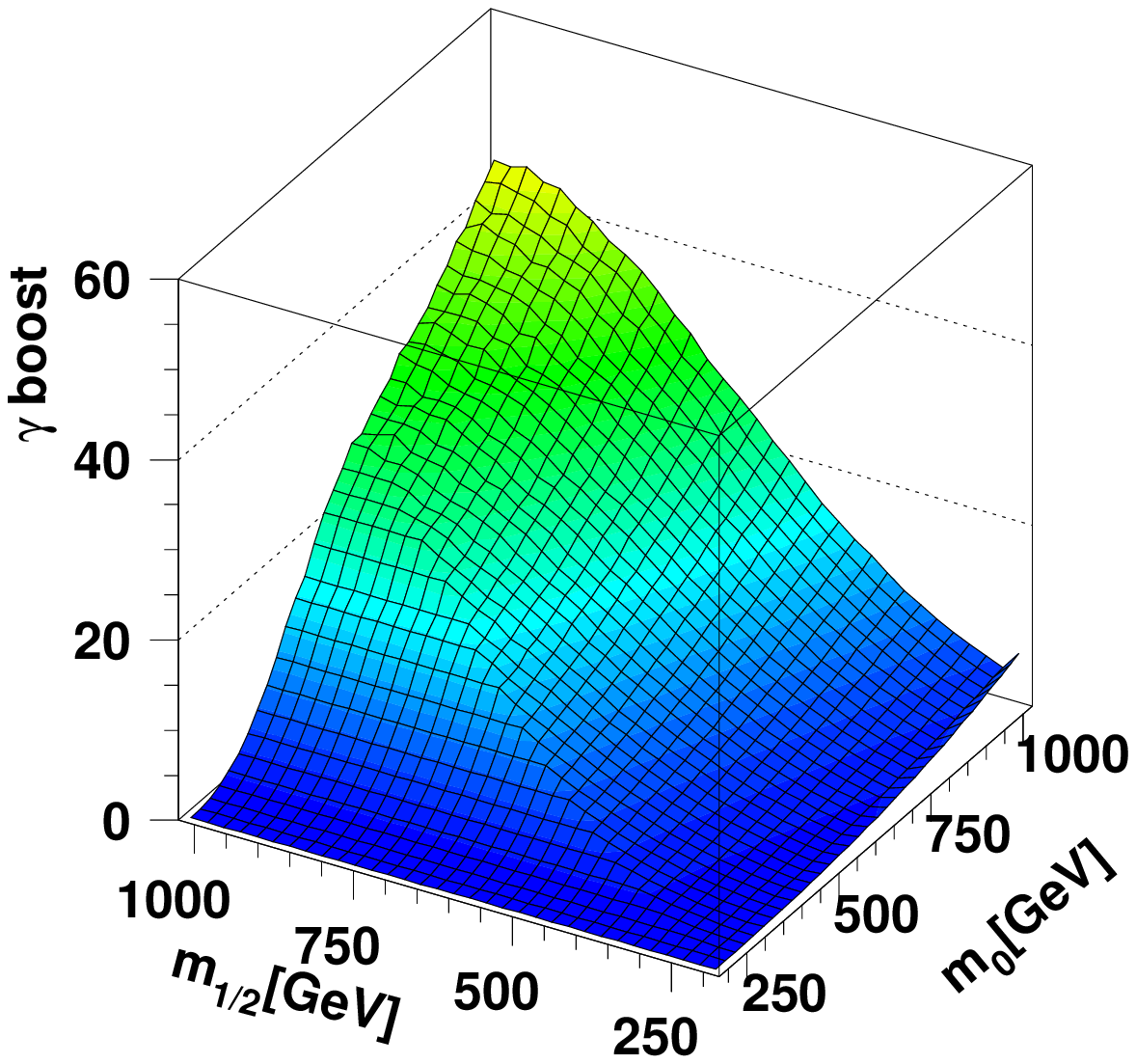}
 \caption[]{\it
 The relic density (left) and boost factor for the gamma rays
 as function of $m_0$ and $m_{1/2}$.
 Note that for large mass scales the relic density is reduced due to
 significant coannihilation cross sections, which
 are not operative in the present universe, thus causing large boost factors.
 For small mass scales the boost factor is practically proportional to the
 relic density, since both are inversely proportional to the annihilation
 cross section.}
 \label{relic3d}
\end{center}
\end{figure}

\begin{figure}[t]
\begin{center}
 \includegraphics [width=0.4\textwidth,clip]{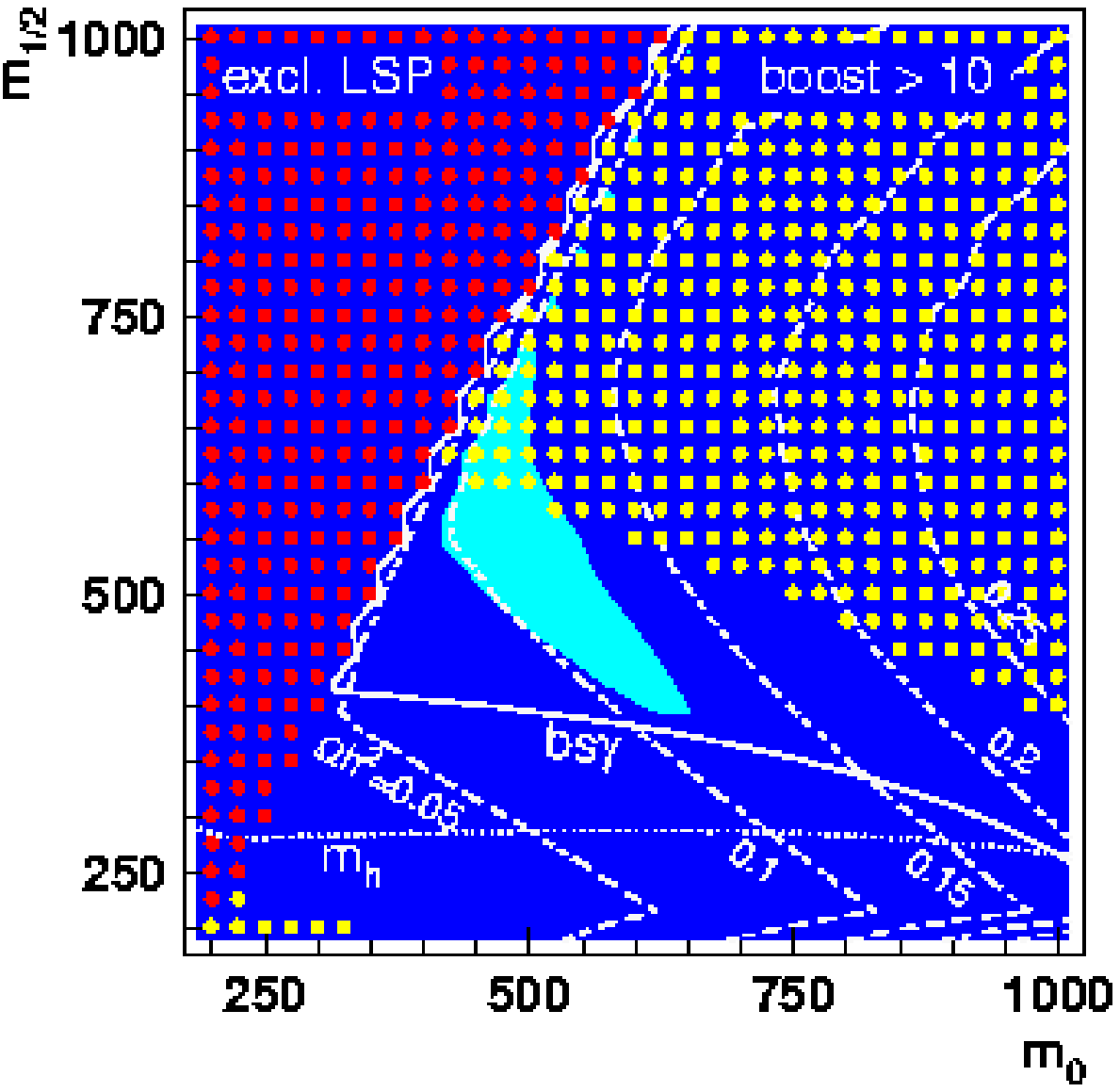}
 \includegraphics [width=0.4\textwidth,clip]{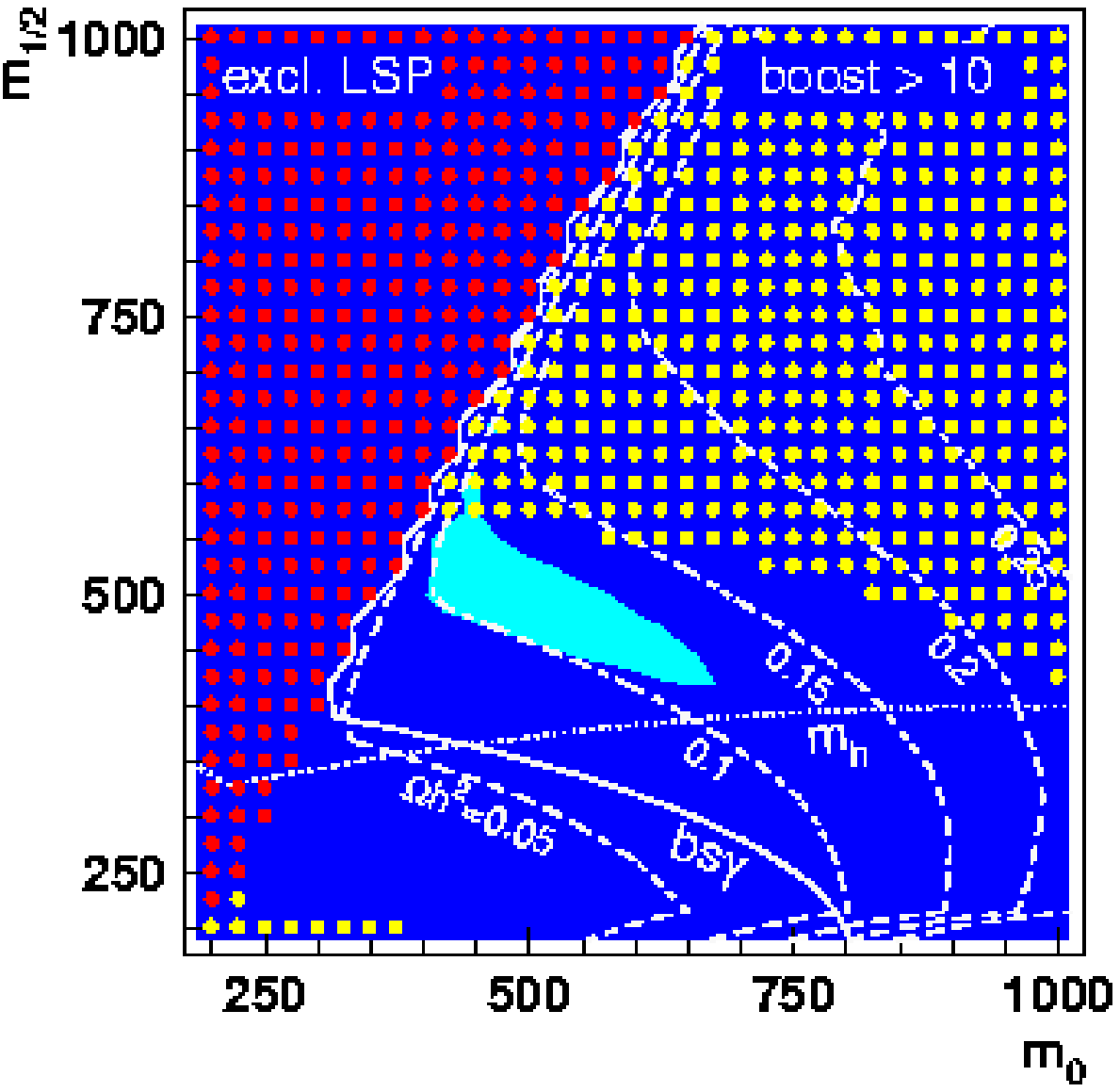}
 \includegraphics [width=0.4\textwidth,clip]{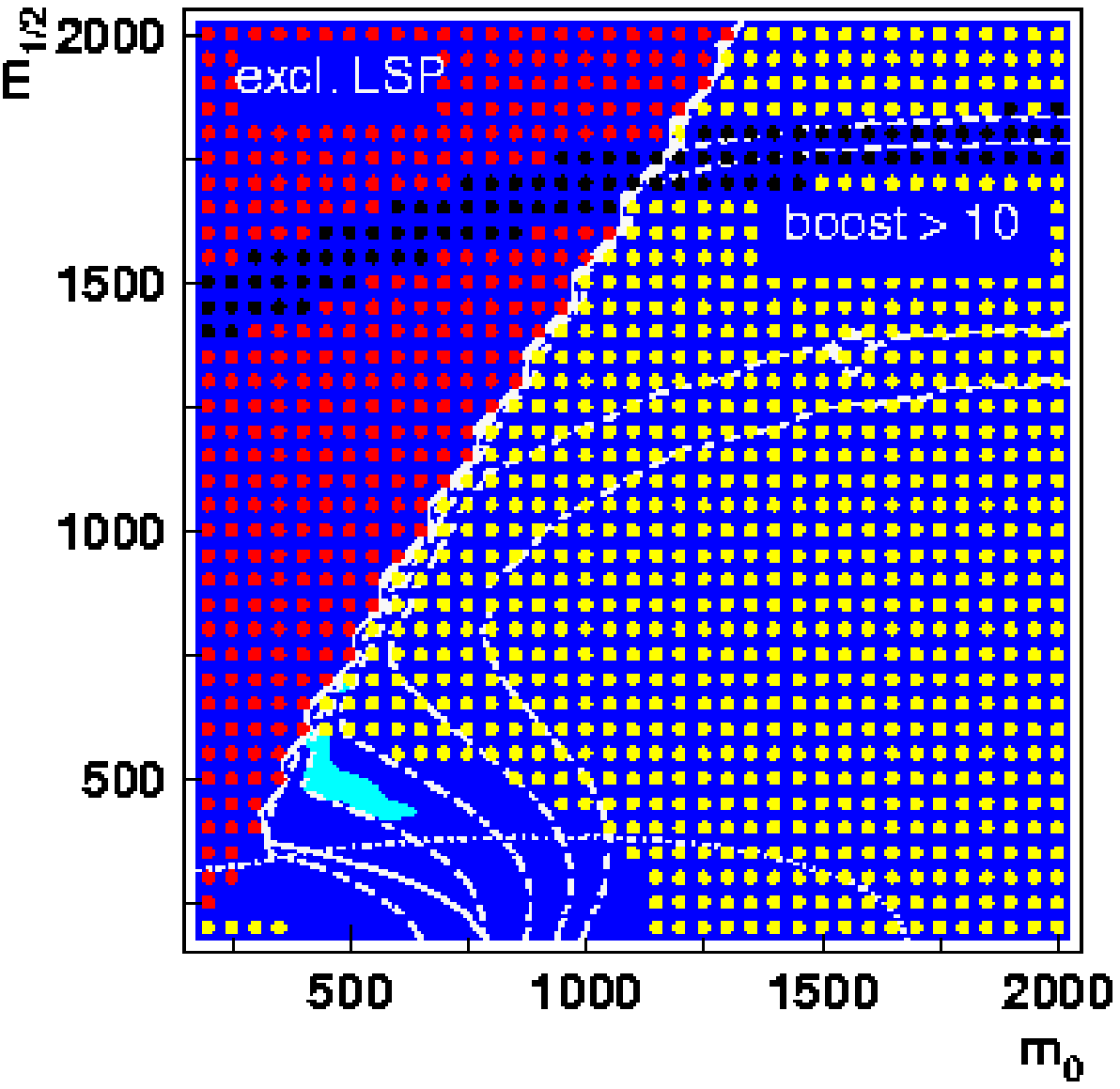}
 \includegraphics [width=0.4\textwidth,clip]{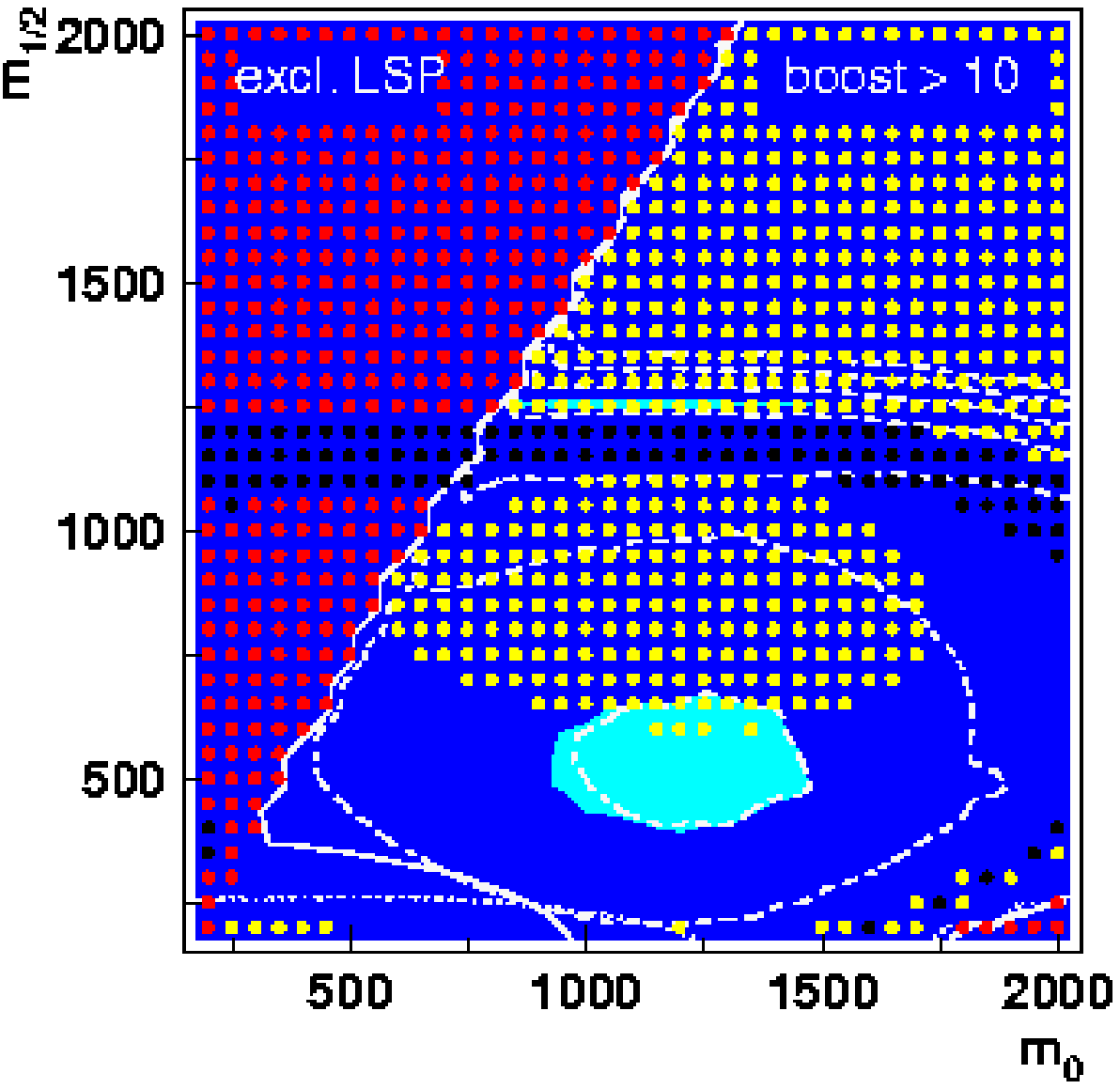}
 \includegraphics [width=0.4\textwidth,clip]{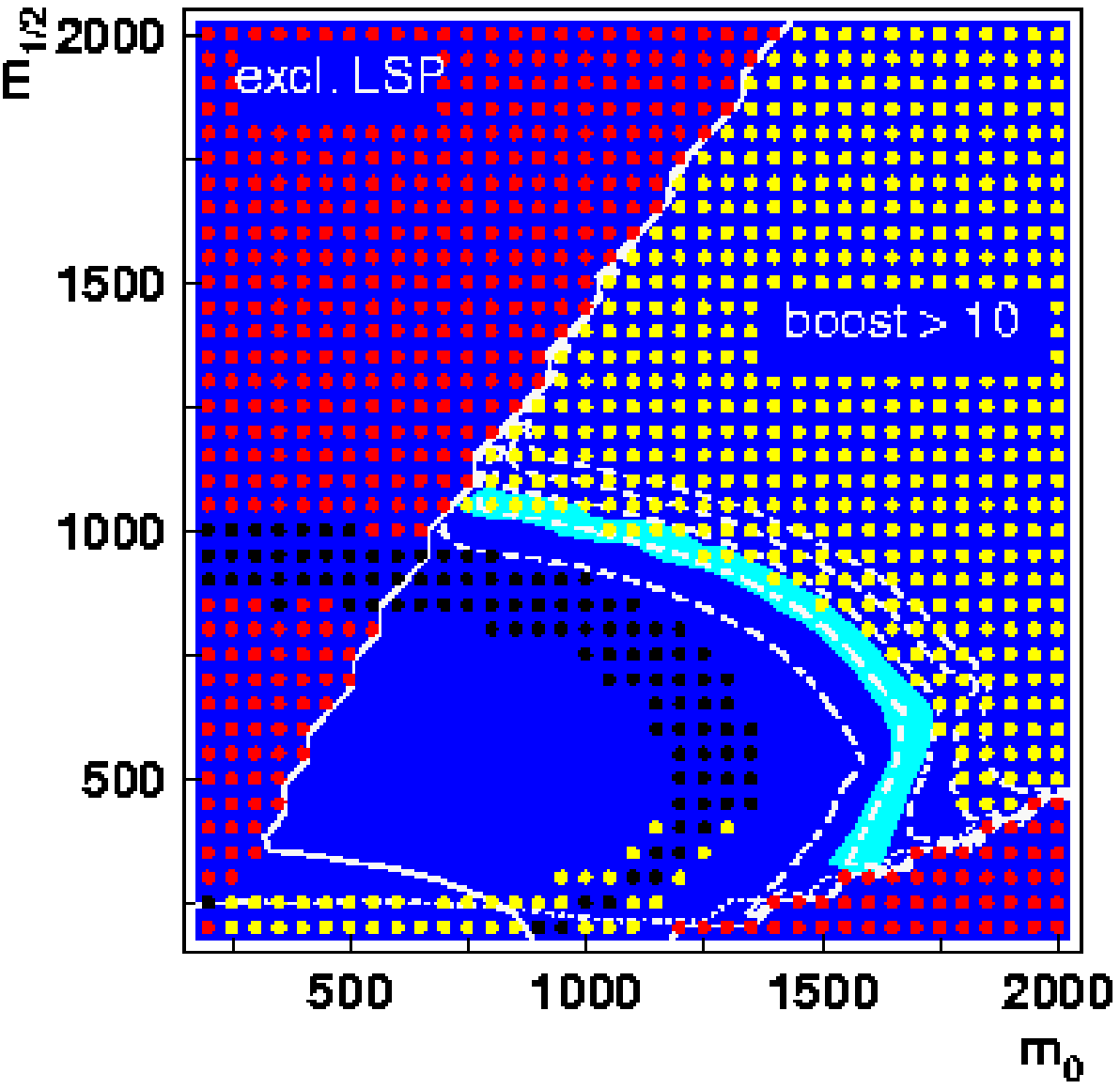}
 \includegraphics [width=0.4\textwidth,clip]{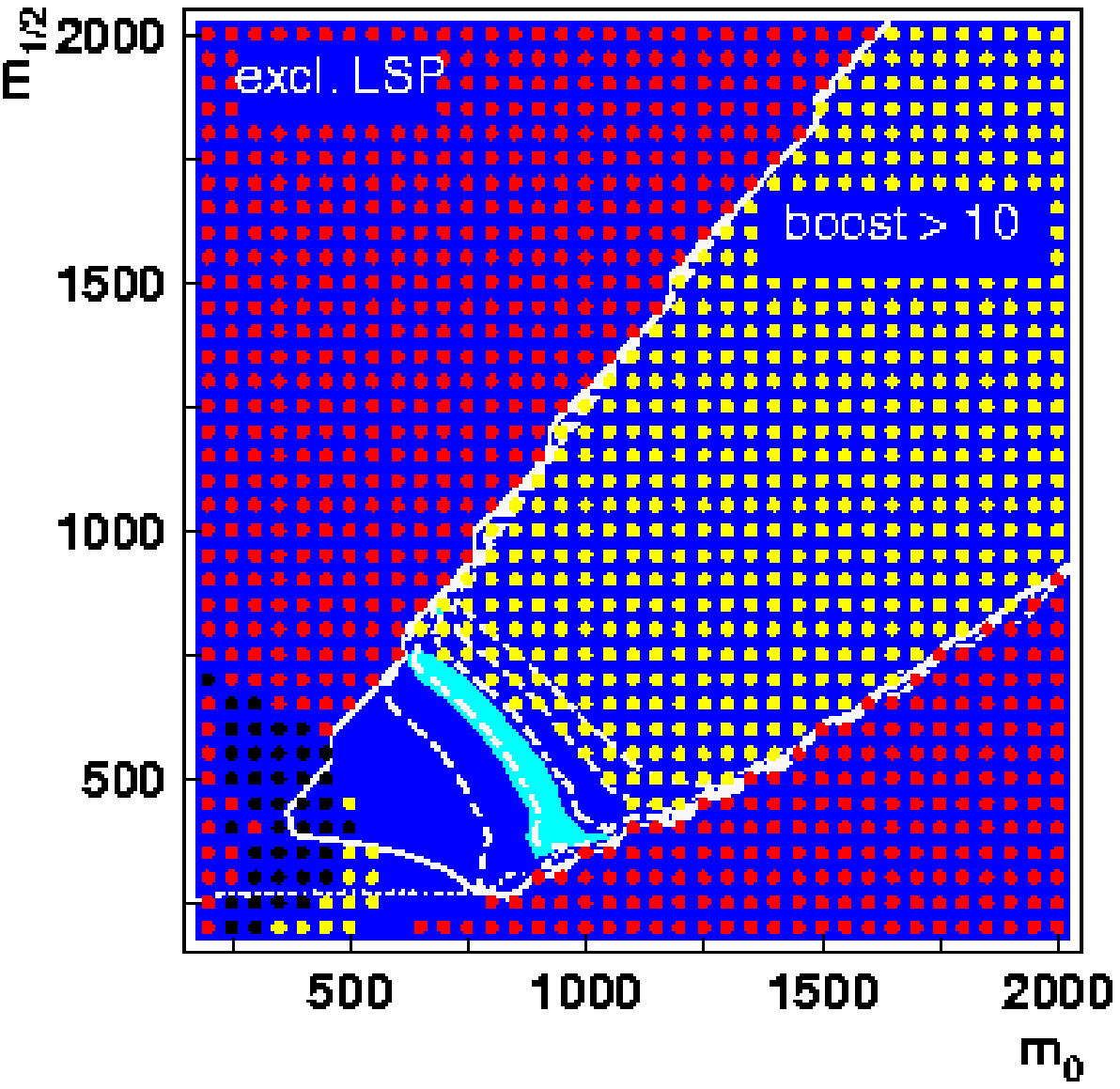}
 \caption[]{\it
 The region of relic density allowed by the WMAP data.
 The upper row is for \tb=51 and $A_0=0$ (left) and $A_0=m_0$ (right), which
 shows that the role of Higgs constraint (dotted line)
 and \bsg ~ constraint (solid line) are interchanged
 for the different values of the trilinear coupling, but the lower limit
 on $m_{1/2}$ is not very sensitive to $A_0$.
 The light shaded (blue) area is the region allowed by WMAP and the
 contours of larger $\Omega h^2$ are indicated by the dashed lines
 in steps of 0.05. The second (third) row show the same information
 for a larger region
for \tb=51, 52 (l. and r.) ( 53 and 55 (l. and r.)).
 The excluded regions, where the stau would be the LSP or EWSB fails or
 the boost factors are above 10 are indicated by the dots.
 The black dots indicate the resonance region,
 where $|m_A-2m_{\chi_0}| \le 10$ GeV. For $\tb>52$ the acceptable
 values for the relic densities are for $m_{1/2}$ values above the
 resonance region.
 }
 \label{relic}
\end{center}
\end{figure}

\begin{itemize}
\item
 The size of the galaxy is determined by a cylindrical box
 with half height h=4 kpc and radius r=30 kpc, as in Galprop.
 The solar system
 is at a distance of 8.5 kpc from the center.
 Outside this cylinder the diffusion stops and the particles escape.
\item
 The energy dependence of the diffusion constant was taken
 to be the same as in Galprop and given by:
 $$D=\beta D_0\cdot (p/p_0)^{\alpha},$$
 where $\beta=v/c$ is the velocity, $D_0=42\cdot 10^{-27}~cm^2/s$
 is the diffusion coefficient at momentum $p_0=1$ GeV and
 $\alpha=0.33$. This diffusion coefficient is at least a factor 7
 larger than the default one in DarkSusy and has a much weaker
 energy dependence, which is  of importance mainly
 for  antiprotons.
\item
 The large diffusion constant implies that the galactic center starts to
 contribute significantly to the antiproton flux in the solar
 system. Therefore the effective thickness of the disk was
 increased from 0.1 to 0.5 kpc and the density of hydrogen in the
 disc was increased from 1 to 3 GeV/cm$^{-3}$, thus increasing the
 interactions between antiprotons and protons in the disc. The
 density of hydrogen in the halo was neglected. \item The energy
 loss $dE/dt=\zeta E^2$ of the positrons (mainly by synchrotron
 radiation and inverse Compton scattering on star light) was
 doubled using $\zeta=1.52\cdot 10^{-9}~yr^{-1}~GeV^{-1}$,
 following Galprop~\cite{positrons} and~\cite{kt}. \item The Dark
 Matter halo profile is usually assumed to be of the Navarro, Frenk
 \& White (NFW) type~\cite{nfw}, as supported by numerical
 simulations of galaxy formation~\cite{galsim,galsim1}. Effectively
 we have chosen for the Dark Matter density distribution an
 isothermal spherical symmetric profile:
 $$\rho (r)=\rho_0\cdot (\frac{r}{a})^{-\gamma}\left[1+
 (\frac{r}{a})^\alpha\right] ^{\frac{\gamma-\beta}{\alpha}},$$
 where $a$ is a scale radius and the slopes $\alpha$, $\beta$ and
 $\gamma$ can be thought of as the radial dependence at $r\approx
 a$, $r>>a$ and $r<<a$, respectively. At large distances we expect
 a $1/r^2$ dependence for a flat rotation curve, while at small
 distances more a $1/r$ dependence is needed. For definiteness we
 use the default $(\alpha, \beta, \gamma)$ =(1,3,1) for a scale
 $a=10$ kpc, but e.g. (1.5,2,1) for a=2 kpc yields practically
 identical results. At a distance less than $10^{-5}$ kpc from the center
 the profile is kept constant to avoid any singularity.
 The density $\rho_0$ is adjusted such that the
 local halo density is in the range 0.2-0.8 GeV/cm$^3$, as required
 by the rotation curve of our galaxy~\cite{bergstrom1}. The halo
 density increases quite steeply towards the center, so most
 annihilations will take place in the center of the galaxy, thus
 producing there gamma rays, positrons and antiprotons.
Different halo profiles are compared in Fig. \ref{halo}.
 The percentage of gamma rays coming from the center depends
 on the halo profile and on the solid angle, as
 shown on the right hand side of Fig. \ref{halo}. The positrons
 only reach the detector from a much smaller region due to the
 higher energy losses, while the antiprotons come from an
 intermediate region, as shown in Fig. \ref{halo} as well. These
 curves were obtained by calculating the flux for a halo profile
 truncated after a given radius.

 The gamma rays can travel over large distances without loosing
 energy, but they will arrive at the detector only, if they were
 emitted along the line of sight. Antiprotons and positrons on the
 other hand change direction during the propagation along the
 magnetic field lines and by collisions, so they can
 arrive at the detector even if they were not emitted along the
 line of sight. This causes a larger acceptance for the antiprotons
 and positrons in comparison with the gamma rays,
 as shown in Fig. \ref{yield}.
 Here the gamma flux at the Earth - $n_\gamma$ per unit of time,
 surface and solid angle - from neutralino annihilation in the
 galactic halo can be written as:

 \bq \frac{dn_\gamma}{dt~dS~d\Omega}=
  \frac{1}{4\pi}\frac{N_\gamma<\sigma v>}{m_\chi^2}\int_{l.o.s.} \rho_\chi^2 ds =
  \frac{1}{4\pi}\frac{N_\gamma<\sigma v>}{m_\chi^2}<J>\label{gamma},
 \eq

 where $N_\gamma$ is the number
 of photons per annihilation (about 38 for b-quark final states)
 and $<J>$ is the averaged value of the integral of the neutralino
 density squared along the line of sight. Since the density peaks
 at the center, the averaged value depends on the solid angle over
 which one averages. For the chosen profile $<J>$ is about 20000 for a
 solid angle of $10^{-3}$ sr towards the center of the galaxy
 and 1000 for $d\Omega=0.17$ sr, which
 is the solid angle used for the EGRET data.

\item
 Diffusive reacceleration is an essential ingredient of Galprop, of special
 importance for the antiprotons. DarkSusy has neither the antiproton background
 spectrum from nuclear interactions nor the possibility of
 reacceleration. Therefore we used the parametrization of the antiproton
 spectrum from Simon, Molnar and R\"osler~\cite{simon}.
 The effect of the reacceleration is a flattening of the antiproton
 spectrum, which was simulated in DarkSusy by a larger value
 of the solar modulation constant $\Phi$. With the parametrization
 from Simon et al., the large diffusion coefficient and $\Phi=1200$ MeV we could
 reproduce the Galprop spectra
 for data from the years 1997 and 1998 sufficiently well. We will use only data
 from this part of the solar cycle.
\end{itemize}

The annihilation cross sections in DarkSusy differs considerably
from the one in Micromegas~\cite{micro} for large \tb.
This was traced back
to a different width of the pseudoscalar Higgs boson.
Micromegas follows the usual
procedure of minimizing the Higgs potential at a scale given by
$\sqrt{m_{\tilde{t}_1} m_{\tilde{t}_2}}$, which minimizes the
higher order corrections. Furthermore, the
pseudoscalar Higgs width is determined by the Yukawa couplings, of which
the most important one in our case is the one given by
 $m_b$. However, the QCD corrections to the width
lead to expressions of the type $m_b(\mu)(1+c_1\alpha_s(\mu)/\pi
\ln(\mu^2/m_A^2)+ ...$; the large logarithms are minimized by
choosing for the renormalization scale $\mu=m_A$. After using the
running b-mass in DarkSusy to calculate the width, good agreement
in the cross sections was found for the large \tb~ region, where
pseudoscalar Higgs exchange dominates.

The relic density was calculated with Micromegas as well,
since this program is particularly suited for large \tb,
where the Higgs exchange and its loop corrected width are
important and
it furthermore incorporates all coannihilation channels.
Micromegas was interfaced to the program package Suspect, which
calculates the low energy SUSY masses from the GUT scale
parameters~\cite{suspect}.
The problem of calculating the relic density at large \tb~\cite{baer}
is caused by the fact that both the pseudoscalar Higgs mass and its
width have large corrections: the mass at tree level is determined by
$\sqrt{m_1^2+m_2^2}$ evaluated at the electroweak scale, but as
can be seen from Fig. \ref{evol}, the running of $m_2^2$ becomes
steep at low energies. Since $m_2^2$ becomes negative for large
\tb, the tree level mass becomes small or even negative and a
positive final mass is obtained by radiative corrections.
Therefore the different calculations give quite a spread in
masses~\cite{kraml} and depend sensitively on the scale, where the
Higgs potential is minimized.
\begin{table}[t]
 \begin{center}
 \begin{tabular}{|c|c|c|}
  \hline
  Parameter & Value & Value \\
  \hline
  $m_0$ & 500 GeV & 1000 GeV \\
  $m_{1/2}$ & 500 GeV & 1000 GeV \\
  $A_0$ & 500 GeV & 0 GeV \\
  $\tan\beta$ & 51 & 53 \\
  $\mbox{sgn}~\mu$ & + & + \\
  \hline
  Particle & Mass [GeV] & Mass [GeV] \\
  \hline
  $\chi^0_{0,1,2,3}$ & 207,375,568,583 & 427, 783, 1119, 1127 \\
   $N^{\chi^0}_{0,1,2,3}$ & 0.995, 0.017, 0.093, 0.036 &
  0.999, 0.004, 0.047, 0.019 \\
  $\chi^\pm_{0,1}$ & 375, 584 & 783, 1128\\
  $\tilde{g}$ & 1121 & 2154 \\
  $\tilde{t}_{1,2}$ & 852, 1010 & 1616, 1842 \\
  $\tilde{b}_{1,2}$ & 909, 998 & 1719, 1831\\
  $\tilde{\tau}_{1,2}$ & 292, 535 & 591, 1037 \\
  $h, H$ & 116, 504 & 121, 823 \\
  $A, H^\pm$ & 506, 515 & 832, 838 \\
  \hline
  Observable & Value & Value \\
  \hline
  $Br(b\to X_s\gamma)$~\cite{micro} & $2.73 \cdot 10^{-4}$ & $3.45 \cdot 10^{-4}$ \\
  $a_\mu$~\cite{micro} & $258 \cdot 10^{-11}$ & $67 \cdot 10^{-11}$\\
  $\Omega h^2$~\cite{micro} & 0.114 & 0.095 \\
  $<\sigma v( {\chi+\chi\rightarrow anything})>$ & $1.6\cdot 10^{-26}
  cm^3/s$& $4.5\cdot 10^{-26} cm^3/s $\\
  $<\sigma v ({\chi+\chi\rightarrow b\overline{b}})>$ & $1.4\cdot 10^{-26}
  cm^3/s$& $4.0\cdot 10^{-26} cm^3/s$\\
  $<\sigma v ({\chi+\chi\rightarrow \tau\overline{\tau}})>$ &
  $0.2\cdot 10^{-26}cm^3/s$& $0.5\cdot 10^{-26} cm^3/s$\\
  \hline
 \end{tabular}
 \caption[]{\it
 mSUGRA parameters with the
 corresponding supersymmetric
 particle spectrum and some predicted observables for an LSP mass of 207
 and 427 GeV. The cross sections were calculated
 by DarkSusy~\cite{darksusy}, which agree with the cross sections
 from Micromegas~\cite{micro} for these parameters after the modifications
 to DarkSusy mentioned in the text. The particle spectrum was calculated
 by Suspect~\cite{suspect} and Feynhiggsfast~\cite{fhf}.}
 \label{t1}
 \end{center}
\end{table}
\begin{table}[b]
\begin{center}
 \begin{tabular}{|c||c|c|c|c|c|c|}\cline{2-7}
 \multicolumn{1}{ c }{ } &
 \multicolumn{2}{|c|}{Background}&\multicolumn{2}{|c|}{$m_\chi=207$ GeV }&\multicolumn{2}{|c|}{$m_\chi=427$ GeV}\\
 \cline{2-7}
 \multicolumn{1}{ c }{   } & \multicolumn{1}{|c| }{$\chi^2_b/d.o.f.$} &
 \multicolumn{1}{|c| }{Prob.} &
 \multicolumn{1}{|c|}{$\chi^2_{b+s}/d.o.f.$} &
 \multicolumn{1}{|c|}{Prob.}&
 \multicolumn{1}{|c|}{$\chi^2_{b+s}/d.o.f.$} & \multicolumn{1}{|c|}{Prob.}
 \\ \hline
 Positrons &59.4/18 & 2.5$\cdot 10^{-6}$& 16.7/17 & 0.48&
  14.3/17 & 0.59\\ \hline
 Antiprotons &15.9/13 &2.6$\cdot 10^{-1}$& 6.6/12 & 0.89&
  6.7/12 & 0.88 \\ \hline
 Gammas&34.5/7&1.4$\cdot 10^{-5}$& 9.5/6 & 0.15&
  12.3/6 & 0.06\\ \hline \hline
  Total &110/38 &6.7$\cdot 10^{-9}$& 33.7/35 & 0.53&
  33.3/35 & 0.55  \\ \hline
 \end{tabular}
 \end{center}
 \caption[]{The $\chi^2$ and probabilities of the fits
 for fluxes from background only, i.e. from nuclear interactions
 (labelled ``b'' ) and for fluxes including ``signal'' contributions from
 neutralino annihilation (labelled ``b+s'') for two neutralino masses.}
 \label{t2}
\end{table}

\begin{figure}[tbh]
\begin{center}
 \includegraphics [width=0.45\textwidth,clip]{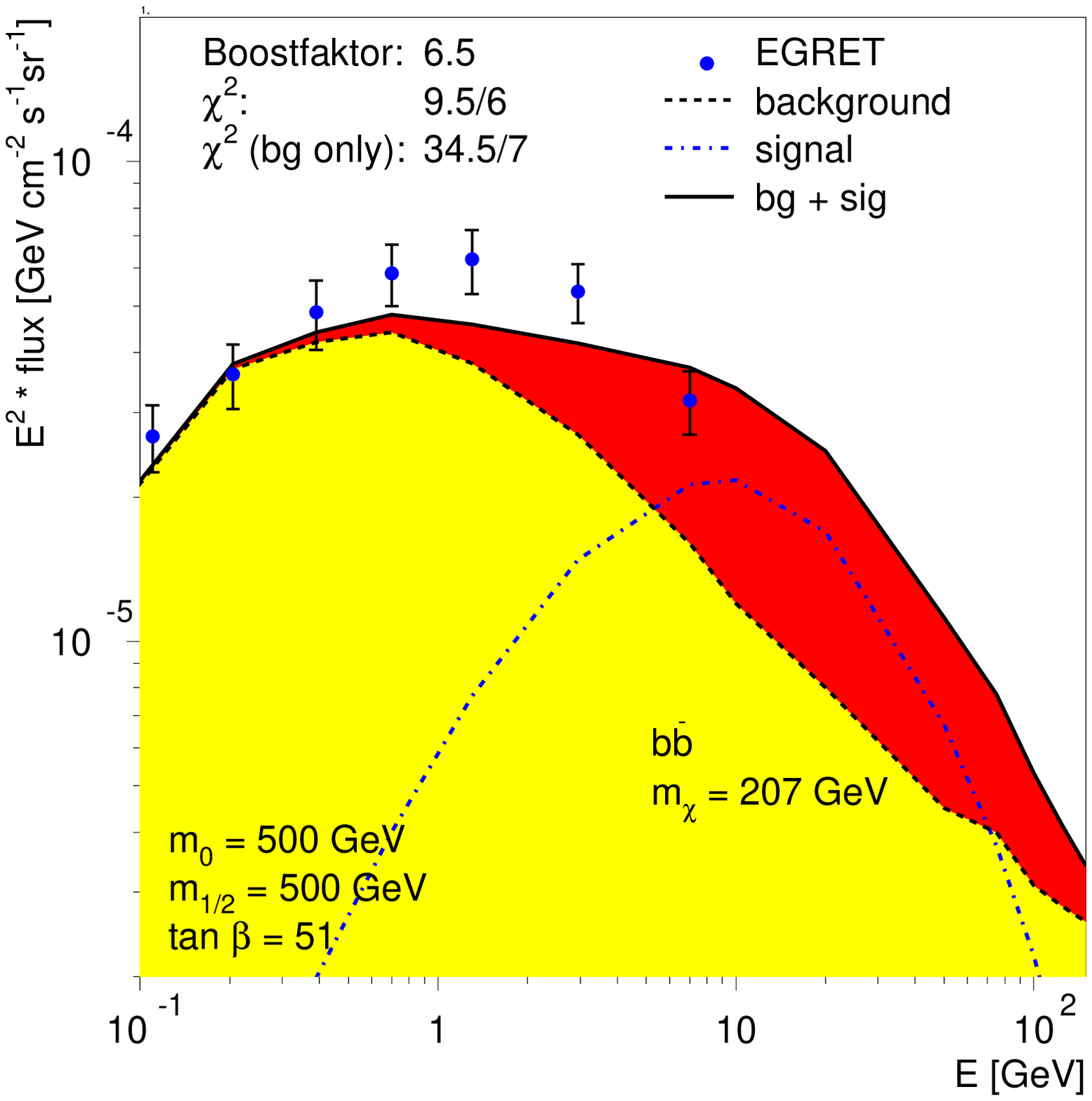}
 \includegraphics [width=0.45\textwidth,clip]{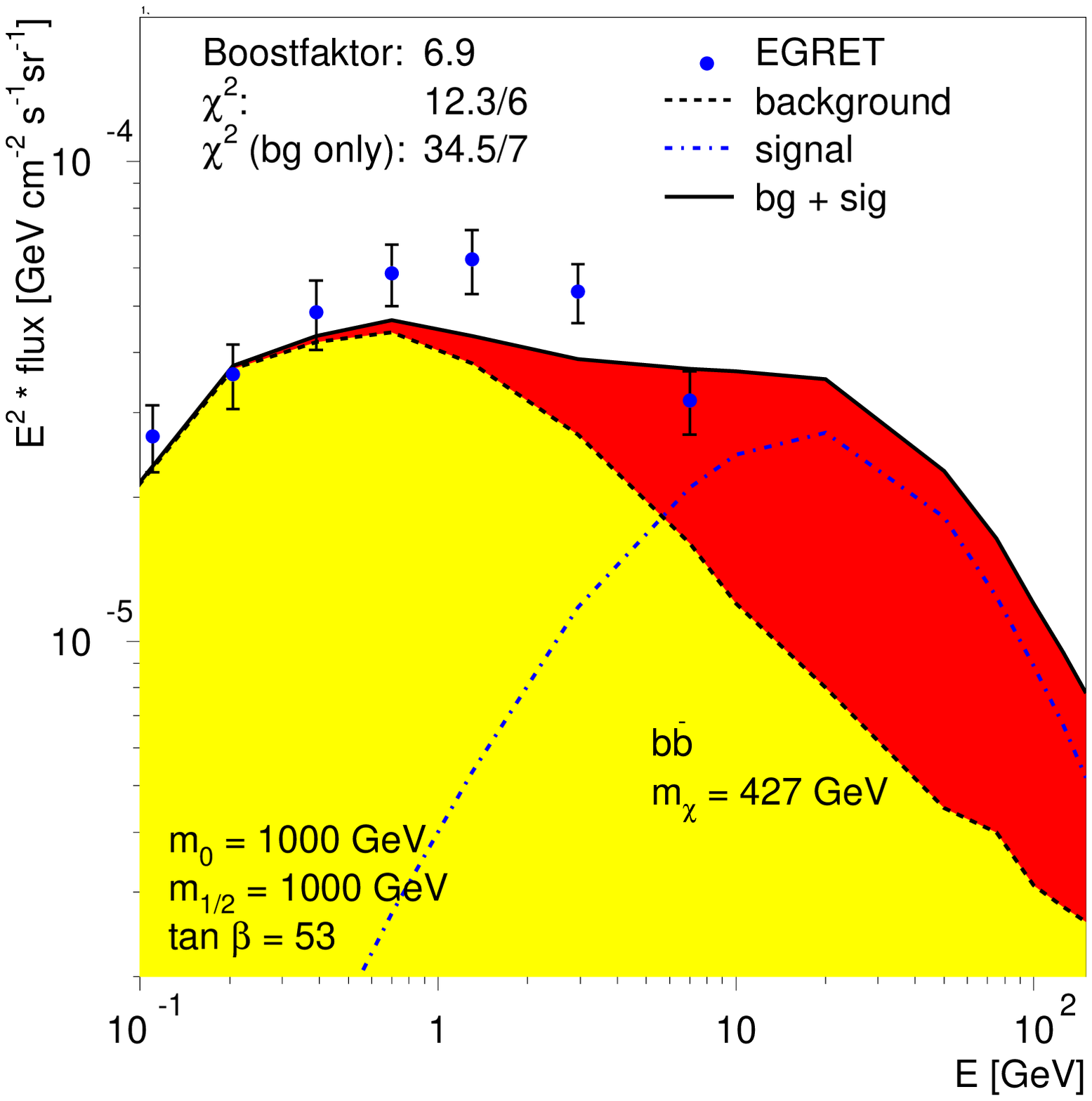}
 \caption[]{\it
 Gamma ray spectrum with contributions from nuclear interactions (grey/yellow)
 and neutralino annihilation (dark/red) for a neutralino mass of
 207 (left) and 427 GeV (right).}
 \label{fit_gamma}
\end{center}
\end{figure}

 \begin{figure}
\begin{center}
 \includegraphics [width=0.45\textwidth,clip]{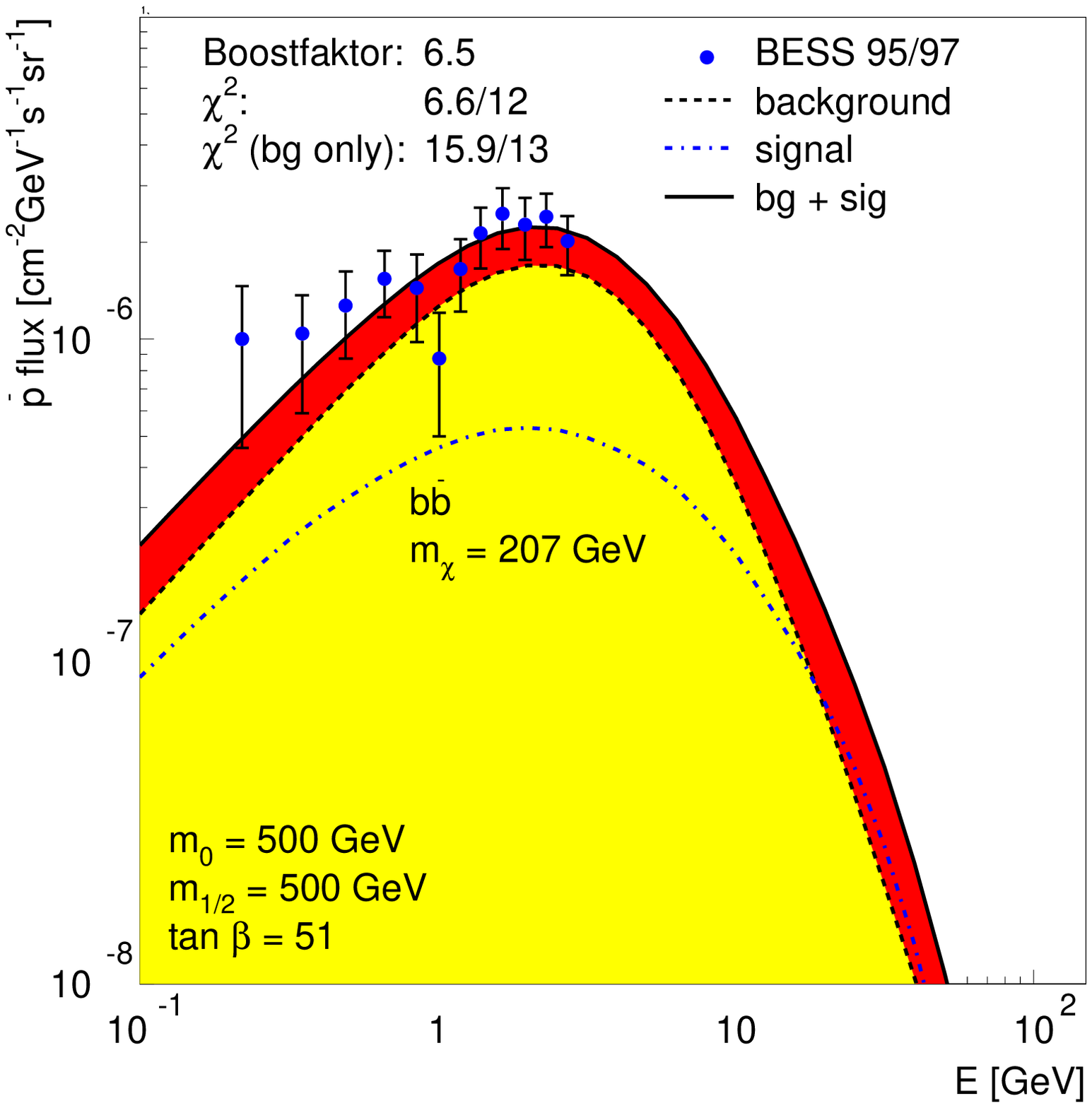}
 \includegraphics [width=0.45\textwidth,clip]{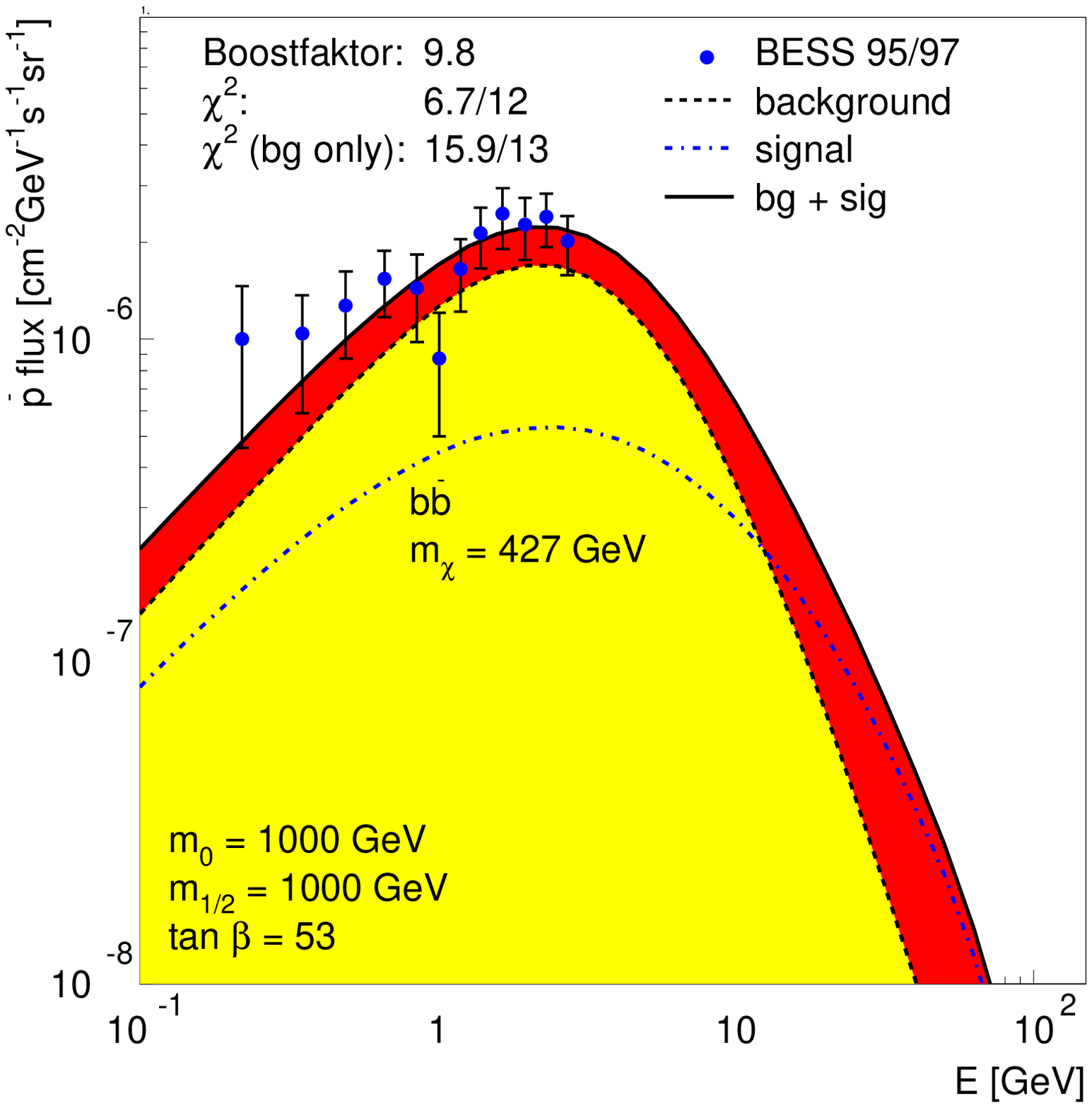}
 \caption[]{\it
 Antiproton spectrum with contributions from nuclear interactions (grey/yellow)
 and neutralino annihilation (dark/red) for a neutralino mass of
 207 (left) and 427 GeV (right).}
 \label{fit_pb}
\end{center}
\end{figure}

\begin{figure}
\begin{center}
 \includegraphics [width=0.45\textwidth,clip]{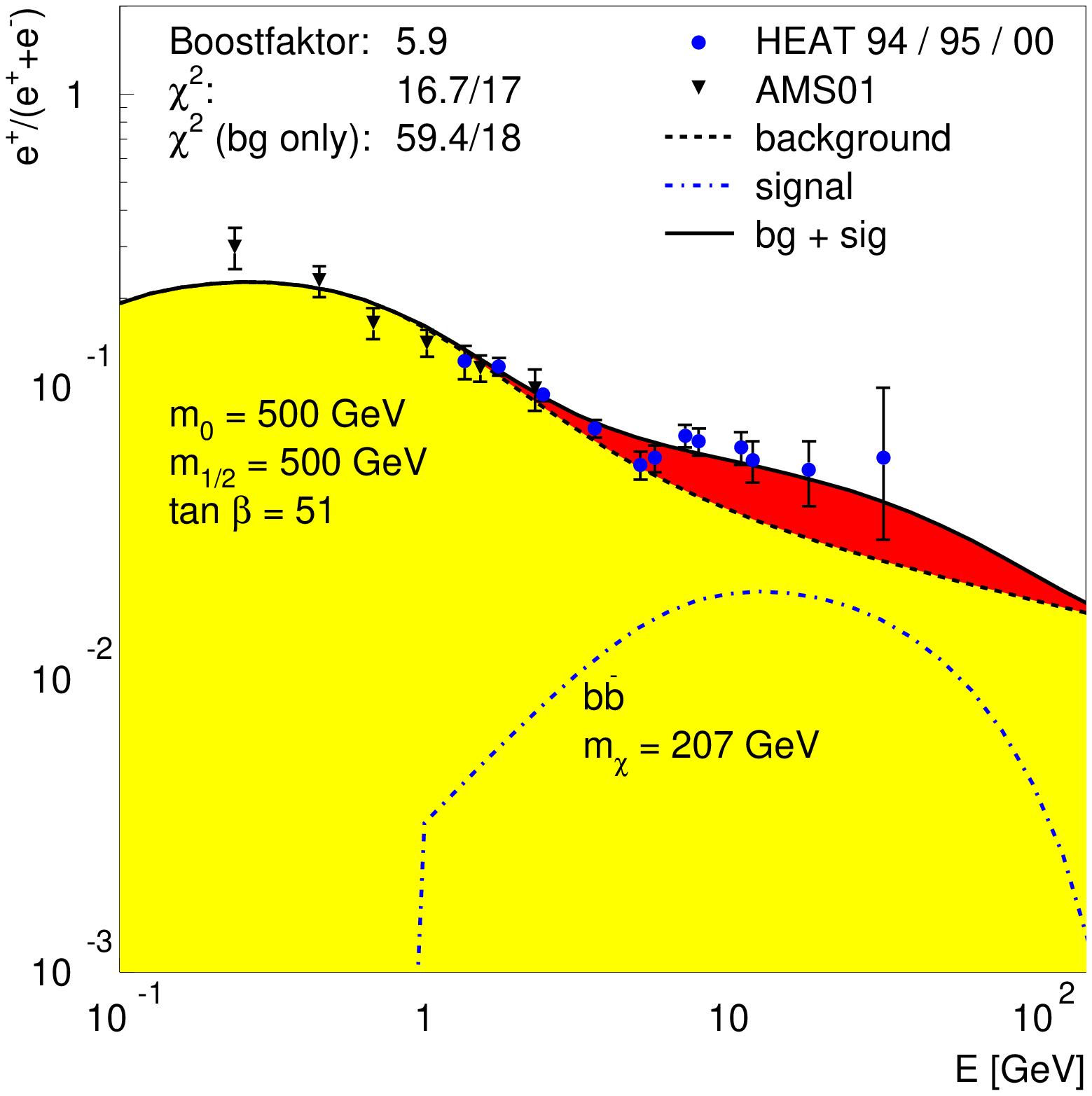}
 \includegraphics [width=0.45\textwidth,clip]{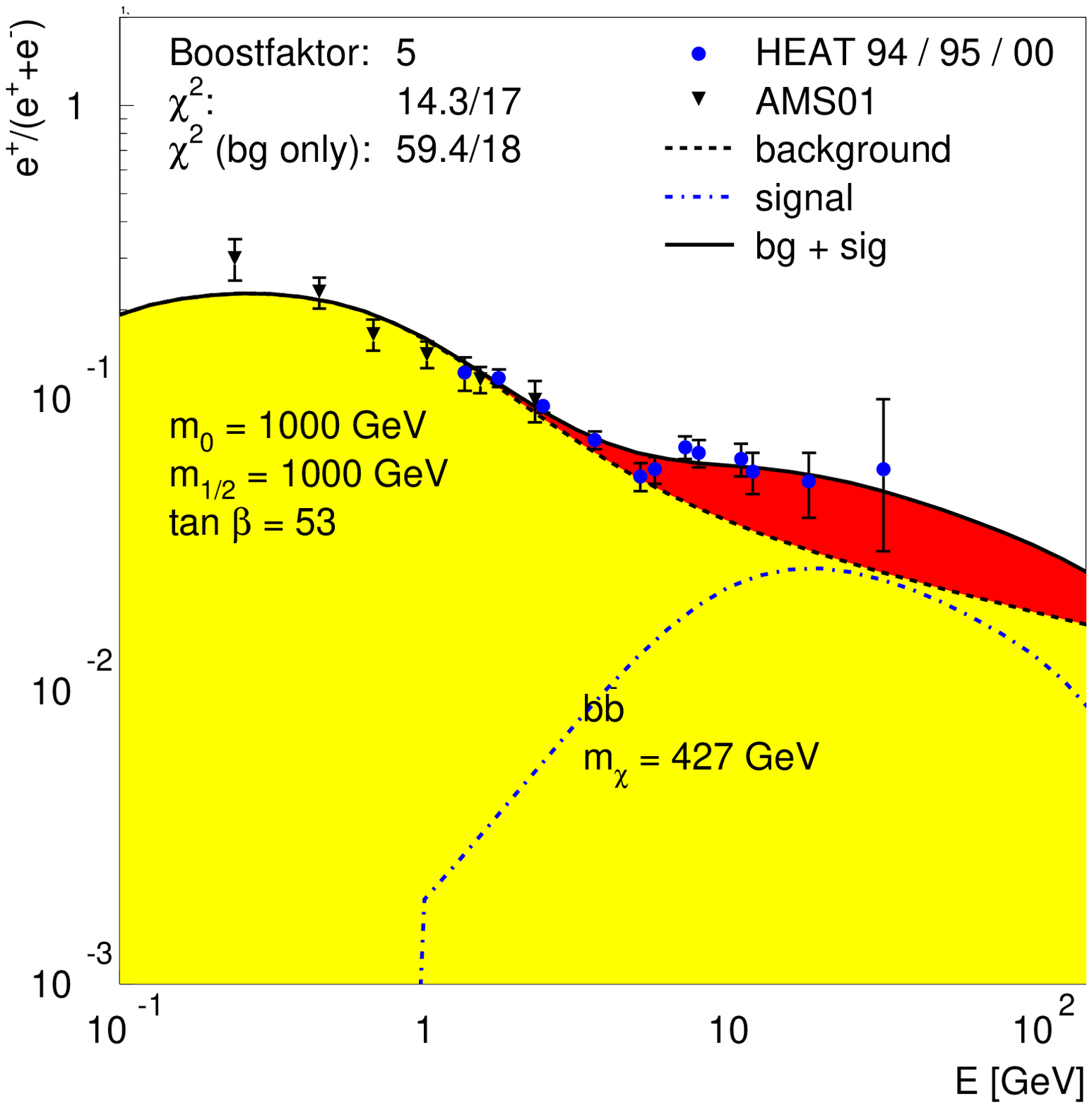}
 \caption[]{\it
 Positron spectrum with contributions from nuclear interactions (grey/yellow)
 and neutralino annihilation (dark/red) for a neutralino mass of
 207 (left) and 427 GeV (right).}
 \label{fit_ep}
\end{center}
\end{figure}

\begin{figure}[tbh]
\begin{center}
 \includegraphics [width=0.75\textwidth,clip]{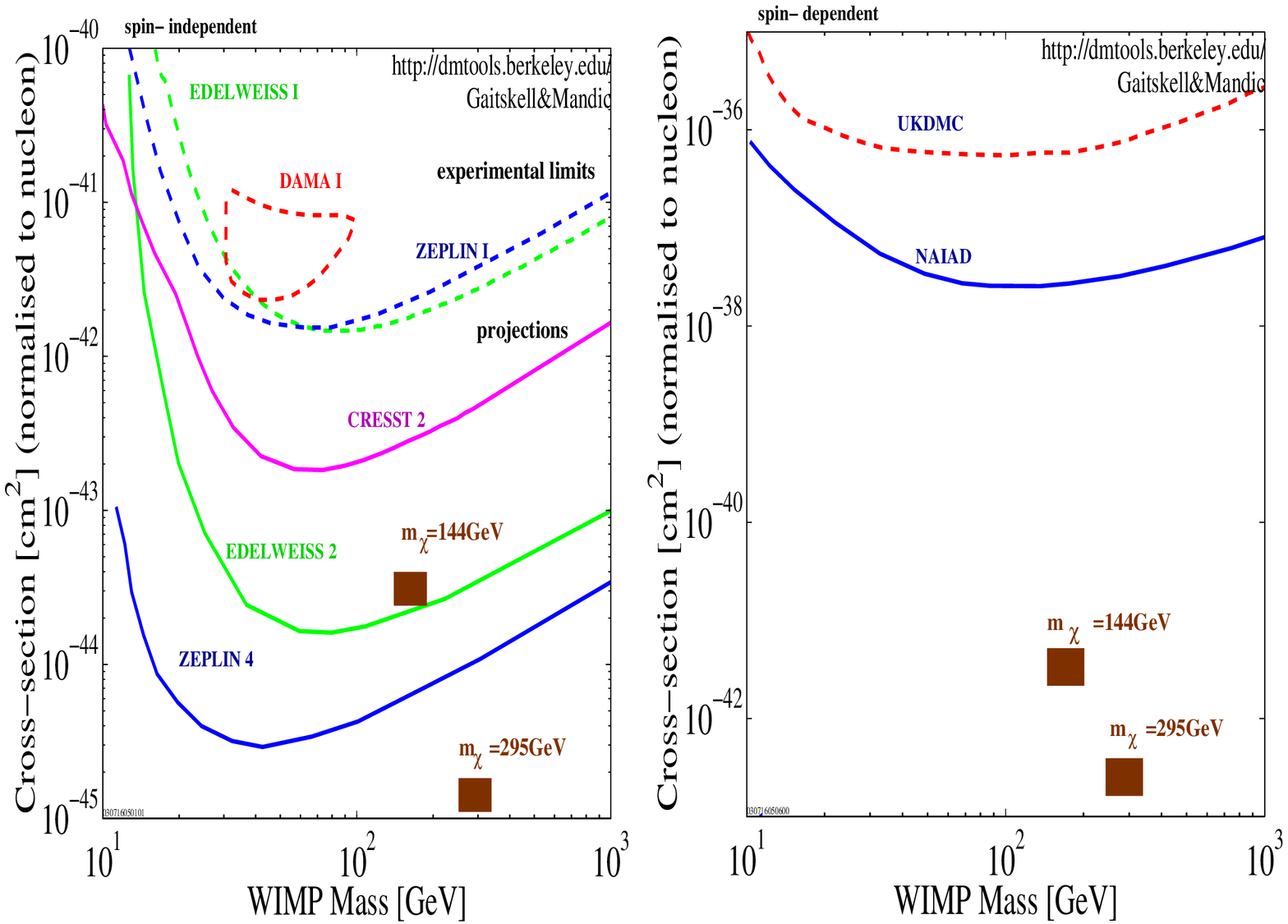}
 \caption[]{\it
 The spin independent (left) and spin
 dependent (right) cross section for direct WIMP detection on
 protons in comparison with existing (dashed lines) and projected
 (solid lines) limits and the cross section expected for neutralino
 masses of 144 and 295 GeV (squares).  A neutralino mass as low as 144 GeV
 is at the border of compatibility with the constraints from the Higgs mass
 and \bsg, so a mass between the
 two masses indicated is more likely.}
 \label{direct}
\end{center}
\end{figure}

The following data were used in the fit:
\begin{itemize}
\item
 Gamma ray data from the galactic center in the angular range
 $330^\circ<\ell<30^\circ$ and $-5^\circ<b<5^\circ$
 from the EGRET space telescope, which has
 been taking data for about 9 years on the NASA Compton Gamma Ray
 Observatory (CGRO).
 We use the data as presented in Ref.~\cite{egret}.
\item
 Positron data from AMS~\cite{ams01} and HEAT~\cite{heat}.
\item
 Antiproton data from BESS in the years 1997 and 1998~\cite{bess}
\end{itemize}

In order to see if the deficiencies in the galactic data on
positron, antiprotons and gamma rays can be fitted by the
contributions from neutralino annihilation, the
following strategy was persued. A $\chi^2$ minimization was performed between
the combined data and the sum of the annihilation signal and
background. The normalizations of the fluxes of positrons,
antiprotons and gamma rays from neutralino annihilation, the
so-called boost factors, were taken as free parameters in the fit
for a given neutralino mass, since the annihilation rate is
proportional to the Dark Matter density squared, so any
 clustering of dark matter can increase
the absolute normalization.
 The normalization of the background was not varied, since it is well
determined by other data, which Galprop used to fit the diffusion
parameters. The Galprop data describes well the positrons and
gamma rays in the region where the neutralino annihilation does
not contribute, but fails for the high energy parts of the
spectra. The antiprotons are too low over the whole spectrum for
the diffusion parameters determined from the B/C fraction, as
mentioned in the previous section.

The $\chi^2$ function for gamma rays was defined as $\sum_i
(f_\gamma D_i-T)^2/(f_\gamma \sigma_i)^2+
(1-f_\gamma)^2/\sigma_n^2$, where the sum runs over all data bins
with errors $\sigma_i$ and $T$ is the sum of the parametrized
background from nuclear interactions, as calculated by Galprop and
the contribution from neutralino annihilation, as calculated by
DarkSusy with the modifications mentioned above. $f_\gamma$ is a
common normalization factor for all data points with a systematic
error given by $\sigma_n$. Normally $f_\gamma$ was set to one, but
sometimes it was left free in the fit in order  to study the
effect of possible correlations between the data points. Similar
$\chi^2$ functions were defined for the antiprotons and positrons
and the total $\chi^2$ was simply the sum of the $\chi^2$
contributions for gamma rays, antiprotons and positrons, since no
correlations exist.

The $\chi^2$ function was minimized by the Minuit package\cite{minuit}
and the fit was repeated for all values of the SUSY parameters $m_0,
m_{1/2}, \tb$, where the mass scales were varied between 200 and
1000 GeV and \tb~ was varied between 50 and 55. The sign of the $\mu $
parameter was chosen positive, as preferred by the data on
$Br(b\to X_s\gamma)$ and the anomalous magnetic moment of the muon
$g_\mu-2$~\cite{bs}.

The boost factors for positrons, antiprotons and gamma rays are
shown as function of $m_0$ and $m_{1/2}$ for \tb=51 in Fig.
\ref{boost}. They were all close to each other over most of the
parameter space, which is absolutely non-trivial, if one considers
the large enhancements for positrons and antiprotons from the
diffusion as compared to gamma rays, as shown before in Fig.
\ref{yield}. This is a strong indication that the deficiencies in
the prediction of the antiprotons, positrons and gamma rays are
indeed due to the missing dark matter annihilation with a
standard dark matter halo profile. One expects a similar boost,
since the fluxes all originate from quite a large region
of the galaxy, as shown before in Fig. \ref{halo}, so the
averaging over the clumpiness should be similar for all. The boost
factors are determined by the { present} annihilation cross
section at a temperature of a {\it few Kelvin}, while the relic
density is determined by the cross section { during freeze out},
which is about $m_\chi/25$ or { $10^{14}$ K}. At the high
temperatures coannihilations can become important, especially at
high mass scales, so the annihilation cross sections can become
large, but the boost factors can be large as well due to the small
annihilation cross section in the present universe. This is
demonstrated by a comparison of the relic density and the boost
factor in Fig. \ref{relic3d}, where at large $m_0$ and $m_{1/2}$
the relic density is small due to coannihilation, but the boost
factors are large. Small boost factors are only obtained at small
mass scales for which the neutralino annihilation cross sections
are large (see Fig. \ref{sigmav}).

The regions of parameter space allowed by the WMAP data are
plotted in Fig. \ref{relic} for different values of \tb. It is
clear that for $\tb\approx 50$ only a small region is allowed, if
in addition the electroweak constraints from the Higgs mass and
\bsg, and the requirement that the LSP is a neutral particle have
to be fulfilled. Values of \tb~ below 50 are excluded completely,
if one wants to be consistent with all constraints and if one
requires in addition that the boost factors are below 10. The last
requirement implies that clumpiness can enhance the annihilation
signal by at most a factor of 10, which is the value suggested by
simulations of galaxy formation and it excludes the regions of
coannihilations between staus and neutralinos. The boost factors
are strongly correlated with the value of the local halo density
$\rho_\chi$. To obtain a conservative excluded region by the
requirement that the boost factor is below 10, $\rho_\chi$  was
set to its maximum allowed value  of 0.2-0.8 GeV/cm$^3$.
Coannihilations occur if the staus are nearly degenerate in mass
with the neutralinos, which happens next to the region labelled
``excl. LSP'' in fig. \ref{relic}. In this region the stau is the
Lightest Supersymmetric Particle (LSP) and next to it the stau is
almost degenerate with the neutralino, so it cannot decay into a
neutralino and tau. In this case  the stau is practically stable
and can annihilate with a neutralino  into a tau plus photon. This
coannihilation reduces the relic density to values required by the
WMAP data. For values of \tb~ above 50 there are quite a range of
neutralino masses allowed, since for larger values of \tb~ one
hits the resonance of pseudoscalar Higgs exchange, in which case
much heavier neutralinos are allowed, as shown in Fig. \ref{relic}
(bottom left).

The results of two good fits with two different neutralino masses
(207 and 427 GeV) are shown in Figs. \ref{fit_gamma}, \ref{fit_pb}
and \ref{fit_ep}. The parameters are summarized in Table \ref{t1}.
Fig. \ref{relic} shows that these parameters are close to  masses
at the lower and upper range of possible masses. As indicated in
the figures the boost factors for antiprotons, positrons and gamma
rays are all similar for the NRW halo profile discussed before and
the $\chi^2$ improves significantly with the inclusion of Dark
Matter in the fits. The $\chi^2/d.o.f.$ is reduced from 110/38 for
the ``background-only'' fit to 34/35 for the fit including
neutralino annihilation. This corresponds to an increase in
probability from about 10$^{-8}$ to 0.5, as shown in Table
\ref{t2} together with the results for the individual spectra. For
the antiprotons the increase in
 probability is the  least significant, as expected, since the shape
 of background and signal are similar. This causes also
 a strong correlation between the normalization of the data and
 the boost factor. For the gamma rays and positrons this
 correlation is less severe because of the different shapes of
 signal and background. Leaving all the normalizations free in the fit
 does not change the results significantly for normalization errors
 below 10 \%, since mainly the antiprotons are affected, but they anyhow
 do not have a strong statistical significance. They are nevertheless
 important for the analysis, since they check the expectation of
 similar boost factors for all species
 and constrain the analysis by the requirement of not overproducing
 the antiprotons.

It is interesting to see what are the prospects to observe the
neutralinos in other channels or by direct detection for the
points of parameter space selected by the present analysis. For a
neutralino mass of 207(427) GeV the flux of muons with energy
above 1 GeV from neutralino annihilation inside the sun are only 8
(0.3) $km^{-2}~yr^{-1}$, while the flux from the earth is even
smaller for the SUSY and halo parameters considered in this
analysis. These fluxes were calculated with
DarkSusy~\cite{darksusy}.

The cross section for {\it direct} Dark Matter detection was
calculated with DarkSusy~\cite{darksusy} as well. For the heavier
masses the cross section rapidly decreases, but the projected
sensitivity of future direct detection experiments may be
sufficient, if the neutralino mass happens to be at the lower
range allowed by the present analysis, as demonstrated in Fig.
\ref{direct}.

Unfortunately, the high cross section deduced from the 6.3$\sigma$
 solar modulation signal
observed in the DAMA experiment~\cite{dama} seems difficult to
reconcile with the present analysis, since the diagrams for
neutralino-nucleon scattering are by crossing symmetry similar to
the ones for neutralino annihilation, so such a large cross
section in direct detection would produce an excessive amount of
positrons, antiprotons and gamma rays.
\clearpage
\section{Conclusion}

The galactic models can very well describe the nuclear reactions,
especially the ratio of secondary to primary produced nuclei and
the small concentration of long-lived radioactive nuclei. These
observations fix the diffusion parameters, especially the need for
diffusive reacceleration
 as an additional term in the diffusion equation was emphasized by Moskalenko
 and Strong. However, the diffusion parameters from such a global view
produce too few antiprotons, too few hard gammas and to a lesser
extent to few hard positrons. Up to now many investigations on
the fluxes of positrons, antiprotons and gamma rays have been
performed, either to study the contributions from neutralino
annihilation or to study ``Local Bubble'' type contributions to
defeat the deficiencies of the standard propagation models.
However, it is shown that Dark Matter annihilation within the
Constrained Minimal Supersymmetric Model can solve these
deficiencies {\it simultaneously}.

This is the first time that these fluxes are studied
in a global fit using the best available diffusion
parameters without artificial constructs to remedy the
deficiencies of the galactic models and add to the contributions
from nuclear interactions the ones from neutralino annihilation in
the Constrained Minimal Supersymmetric Model (CMSSM) with all the
known constraints from electroweak precision data and the relic
density from the WMAP satellite. Actually, all we need from the
supersymmetric model to fill up the deficiencies is a neutralino
with a thermally averaged annihilation cross section times
relative neutralino velocities of about $10^{-26}~cm^3/s$ for the
present universe, i.e. without coannihilation. This cross section
can be smaller, if the halo distribution of the neutralinos is not
smooth, but clumpy, thus enhancing the annihilation rate by a
certain ``boost factor'', which simulations of galaxy formation
show to be  of ``the order of a few''. The boost factors for
antiprotons, positrons and gamma rays were left as independent
free parameters in the fit, but they all come out to be similar for a
given halo profile and given neutralino mass. This result is
non-trivial given the very different mean free paths of positrons,
antiprotons and gamma rays and provides a strong support for
neutralino annihilation as a common origin of the deficiencies. The
needed large cross sections are obtained in the CMSSM only for
relatively light neutralinos in the mass range of 150 - 400 GeV
and $\tb>50$. At large \tb~ the annihilation is dominated by
pseudoscalar Higgs exchange into $b\overline{b}$ quark pairs. The
upper neutralino mass is limited by the rather arbitrary
requirement that the boost factors are below 10 and the lower mass
limit is given by the electroweak constraints at large \tb. The
probability of the global fit to galactic fluxes of positrons,
antiprotons and gamma rays increases from 10$^{-8}$ to 0.5 by
including the contribution from neutralino annihilation,
which can be interpreted in case of gaussian errors as
an improvement by about 6 standard deviations.
These facts, statistical significance of the global fit combined
with similar boost factors for positrons, antiprotons and gamma
rays, provide strong experimental evidence for the supersymmetric
nature of Dark Matter.

\section{Acknowledgements}
We thank V. Moskalenko and A. Strong for sharing with us all
their knowledge about our galaxy, O. Reimers to provide us with
the EGRET data, L.Bergstr\"om, J. Edsj\"o and P. Ullio for useful
discussions about DarkSusy and neutralino annihilation, F. Stoehr,
V. Springel and G. B\"orner for discussions on haloes, A. Belyaev
and G. B\'elanger for discussions about relic density, A. Pukhov
for discussions about CalcHEP, J. K\"uhn for a
discussion on radiative corrections to the Higgs boson widths, and
last but not least D. Kazakov and his group for early contributions
to this analysis.

This work was supported by the DLR (Deutsches Zentrum f\"ur
Luft- und Raumfahrt) and
the BMBF (Bundesministerium f\"ur Bildung und Forschung).


\vspace*{-0.5cm}

\begin{thebibliography}{99}

\bibitem{wmap} The results of the first year of operation
 of the WMAP satellite can be found on the Web:
 http://map.gsfc.nasa.gov/m\_mm/pub\_papers/firstyear.html

\bibitem{bennett} C.L. Bennett, et al., astro-ph/0302207.

\bibitem{spergel} D.N. Spergel, et al., astro-ph/0302209.

\bibitem{verde} L. Verde et al., astro-ph/0302218

\bibitem{lspdm}
{\rm J. Ellis et al., Nucl. Phys. {\bf B238} (1984) 453.}


\bibitem{jungman}
G.~Jungman, M.~Kamionkowski and K.~Griest,
Phys.\ Rep.\ {\bf 267} (1996) 195.

\bibitem{ms_problems}
 I.V. Moskalenko, A.W. Strong, J.F. Ormes and M.S. Potgieter,
 Astrophys. J. {\bf 565} (2002) 280.

\bibitem{ms_gamma} A.W. Strong, I.V. Moskalenko and O. Reimer,
Astrophys. J. {\bf 537} (2000) 763.

\bibitem{ms_nucl} A.W. Strong, I.V. Moskalenko, S.G. Mashnik and J.F. Ormes,
Astrophys. J. {\bf 586} (2003) 1050.

\bibitem{bergstrom}
L.~Bergstrom,
Rept.\ Prog.\ Phys.\ {\bf 63} (2000) 793
[arXiv:hep-ph/0002126];

\bibitem{Rudaz:1987ry}
S.~Rudaz and F.~W.~Stecker,
Astrophys.\ J.\ {\bf 325} (1988) 16.

\bibitem{Chardin:2003ci}
G.~Chardin,
DAPNIA-02-413
{\it Prepared for 37th Rencontres de Moriond on the Cosmological
Model, Les Arcs, France, 16-23 Mar 2002}

\bibitem{Orloff:2003jg}
J.~Orloff, E.~Nezri and V.~Bertin,
arXiv:hep-ph/0301215.

\bibitem{Nezri:2002ay}
E.~Nezri,
arXiv:hep-ph/0211082.

\bibitem{Bertin:2002ky}
V.~Bertin, E.~Nezri and J.~Orloff,
Eur.\ Phys.\ J.\ C {\bf 26} (2002) 111 [arXiv:hep-ph/0204135].


\bibitem{Barger:2001ur}
V.~D.~Barger, F.~Halzen, D.~Hooper and C.~Kao,
Phys.\ Rev.\ D {\bf 65} (2002) 075022 [arXiv:hep-ph/0105182].

\bibitem{Feng:2000zu}
J.~L.~Feng, K.~T.~Matchev and F.~Wilczek,
Phys.\ Rev.\ D {\bf 63} (2001) 045024 [arXiv:astro-ph/0008115].

\bibitem{Halzen:1995xn}
F.~Halzen,
arXiv:astro-ph/9508020.

\bibitem{Kamionkowski:1994dp}
M.~Kamionkowski, K.~Griest, G.~Jungman and B.~Sadoulet,
Phys.\ Rev.\ Lett.\ {\bf 74} (1995) 5174 [arXiv:hep-ph/9412213].

\bibitem{Morselli:cc}
A.~Morselli,
Int.\ J.\ Mod.\ Phys.\ A {\bf 17} (2002) 1829.

\bibitem{Donato:2003xg}
F.~Donato, N.~Fornengo, D.~Maurin, P.~Salati and R.~Taillet,
arXiv:astro-ph/0306207.

\bibitem{Edsjo:2002wj}
J.~Edsjo,
arXiv:astro-ph/0211354.

\bibitem{Birkedal-Hansen:2002am}
A.~Birkedal-Hansen and B.~D.~Nelson,
Phys.\ Rev.\ D {\bf 67} (2003) 095006 [arXiv:hep-ph/0211071].

\bibitem{Bertin:2002sq}
V.~Bertin, E.~Nezri and J.~Orloff,
JHEP {\bf 0302} (2003) 046 [arXiv:hep-ph/0210034].

\bibitem{Baltz:2001ir}
E.~A.~Baltz, J.~Edsjo, K.~Freese and P.~Gondolo,
Phys.\ Rev.\ D {\bf 65} (2002) 063511 [arXiv:astro-ph/0109318].

\bibitem{Lahanas:2001yr}
A.~B.~Lahanas and V.~C.~Spanos,
Eur.\ Phys.\ J.\ C {\bf 23} (2002) 185 [arXiv:hep-ph/0106345].

\bibitem{Nihei:2001qs}
T.~Nihei, L.~Roszkowski and R.~Ruiz de Austri,
JHEP {\bf 0105} (2001) 063 [arXiv:hep-ph/0102308].

\bibitem{Bergstrom:2000bk}
L.~Bergstrom, J.~Edsjo and C.~Gunnarsson,
Phys.\ Rev.\ D {\bf 63} (2001) 083515 [arXiv:astro-ph/0012346].

\bibitem{Ullio:2002pj}
P.~Ullio, L.~Bergstrom, J.~Edsjo and C.~Lacey,
Phys.\ Rev.\ D {\bf 66} (2002) 123502 [arXiv:astro-ph/0207125].

\bibitem{Cesarini:2003nr}
A.~Cesarini, F.~Fucito, A.~Lionetto, A.~Morselli and P.~Ullio,
arXiv:astro-ph/0305075.

\bibitem{Ullio:by}
P.~Ullio,
Int.\ J.\ Mod.\ Phys.\ A {\bf 17} (2002) 1777.


\bibitem{Baltz:2002ua}
E.~A.~Baltz, J.~Edsjo, K.~Freese and P.~Gondolo,
arXiv:astro-ph/0211239.

\bibitem{Ellis:2003ch}
J.~R.~Ellis,
arXiv:astro-ph/0305038;

\bibitem{Ellis:2003ug}
arXiv:astro-ph/0304183.

\bibitem{Ellis:2001hv}
J.~R.~Ellis, J.~L.~Feng, A.~Ferstl, K.~T.~Matchev and K.~A.~Olive,
Eur.\ Phys.\ J.\ C {\bf 24} (2002) 311 [arXiv:astro-ph/0110225].

\bibitem{Kane:2002nm}
G.~L.~Kane, L.~T.~Wang and T.~T.~Wang,
Phys.\ Lett.\ B {\bf 536} (2002) 263 [arXiv:hep-ph/0202156].

\bibitem{kolb}
E.~W.~Kolb and M.~S.~Turner, ``The Early Universe,'' Frontiers in
Physics, Addison-Wesley, 1990.




\bibitem{Calcaneo-Roldan:2000yt}
C.~Calcaneo-Roldan and B.~Moore,
Phys.\ Rev.\ D {\bf 62} (2000) 123005
[arXiv:astro-ph/0010056].
\bibitem{Gondolo:2000pn}
P.~Gondolo,
Phys.\ Lett.\ B {\bf 494} (2000) 181
[arXiv:hep-ph/0002226].
\bibitem{Baltz:1999ra}
E.~A.~Baltz, C.~Briot, P.~Salati, R.~Taillet and J.~Silk,
Phys.\ Rev.\ D {\bf 61} (2000) 023514
[arXiv:astro-ph/9909112].

\bibitem{Corsetti:1999ma}
A.~Corsetti and P.~Nath,
Int.\ J.\ Mod.\ Phys.\ A {\bf 15} (2000) 905
[arXiv:hep-ph/9904497].




\bibitem{Bergstrom:1998jj}
L.~Bergstrom, J.~Edsjo, P.~Gondolo and P.~Ullio,
Phys.\ Rev.\ D {\bf 59} (1999) 043506
[arXiv:astro-ph/9806072].

\bibitem{Bottino:1998tw}
A.~Bottino, F.~Donato, N.~Fornengo and P.~Salati,
Phys.\ Rev.\ D {\bf 58} (1998) 123503
[arXiv:astro-ph/9804137].



\bibitem{Ullio:1997ke}
P.~Ullio and L.~Bergstrom,
Phys.\ Rev.\ D {\bf 57} (1998) 1962
[arXiv:hep-ph/9707333].



\bibitem{Bern:1997ng}
Z.~Bern, P.~Gondolo and M.~Perelstein,
Phys.\ Lett.\ B {\bf 411} (1997) 86
[arXiv:hep-ph/9706538].

\bibitem{Bergstrom:1996kp}
L.~Bergstrom, J.~Edsjo and P.~Gondolo,
Phys.\ Rev.\ D {\bf 55} (1997) 1765
[arXiv:hep-ph/9607237].


\bibitem{Jungman:1994cg}
G.~Jungman and M.~Kamionkowski,
Phys.\ Rev.\ D {\bf 51} (1995) 3121
[arXiv:hep-ph/9501365].
\bibitem{Bottino:1994xp}
A.~Bottino, N.~Fornengo, G.~Mignola and L.~Moscoso,
Astropart.\ Phys.\  {\bf 3} (1995) 65
[arXiv:hep-ph/9408391].

\bibitem{Jungman:1993yn}
G.~Jungman and M.~Kamionkowski,
Phys.\ Rev.\ D {\bf 49} (1994) 2316
[arXiv:astro-ph/9310032].
\bibitem{Bottino:1991ep}
A.~Bottino, V.~de Alfaro, N.~Fornengo, A.~Morales, J.~Puimedon and S.~Scopel,
Mod.\ Phys.\ Lett.\ A {\bf 7} (1992) 733.






\bibitem{bergstrom1}
L.~Bergstrom, P.~Ullio and J.~H.~Buckley,
Astropart.\ Phys.\ {\bf 9} (1998) 137
[arXiv:astro-ph/9712318].
\bibitem{susyrev}
{\rm Reviews and original references can be found in:
 W.~de Boer,
 Prog.\ Part.\ Nucl.\ Phys.\ {\bf 33} (1994) 201 [arXiv:hep-ph/9402266];
 \\
 A.B. Lahanus and
 D.V. Nanopoulos, Phys. Rep. {\bf 145} (1987) 1;\\ H.E. Haber and G.L. Kane,
 Phys. Rep. {\bf 117} (1985) 75;\\ M.F. Sohnius, Phys. Rep. {\bf 128} (1985)
 39;\\ H.P. Nilles, Phys. Rep. {\bf 110} (1984) 1;\\ P. Fayet and S. Ferrara,
 Phys. Rep. {\bf 32} (1977) 249.}
\bibitem{ms_1998_nucl} A.W. Strong and I.V. Moskalenko,
Astrophys. J. {\bf 509} (1998) 212.

\bibitem{ms_1998_pos} I.V. Moskalenko and A.W. Strong,
Astrophys. J. {\bf 493} (1998) 694.

\bibitem{darksusy}
 DarkSUSY, P.~Gondolo,
J.~Edsjo, L.~Bergstrom, P.~Ullio and E.~A.~Baltz,
 arXiv:astro-ph/0012234 and http://www.physto.se/\~edsjo/darksusy/.

\bibitem{suspect}
A.~Djouadi, J.~L.~Kneur and G.~Moultaka,
arXiv:hep-ph/0211331.

\bibitem{fhf}
S.~Heinemeyer, W.~Hollik and G.~Weiglein,
arXiv:hep-ph/0002213.

\bibitem{micro}
G.~B\'elanger, F.~Boudjema, A.~Pukhov and A.~Semenov,
arXiv:hep-ph/0210327 and http://wwwlap.in2p3.fr/lapth/micromegas.
%



\bibitem{bs}
W.~de Boer and C.~Sander,
arXiv:hep-ph/0307049 and references therein.

%
\bibitem{goldberg}
 H. Goldberg, Phys. Rev. lett. {\bf 50} (1983) 1419.

\bibitem{calchep}
A.~Pukhov
 {\it et al.},
 arXiv:hep-ph/9908288.
\bibitem{griest} 
K. Griest and D. Seckel, Phys. Rev. {\bf D43} (1991) 3191

\bibitem{solar}
 L.J. Gleeson and W.I. Axford, ApJ {\bf 149} (1967) L115; ApJ {\bf 154} (1968) 1011.

\bibitem{casadei} D. Casadei and V. Bindi, astro-ph/0302307.

\bibitem{positrons}
I.~V.~Moskalenko and A.~W.~Strong,
Phys.\ Rev.\ D {\bf 60} (1999) 063003
[arXiv:astro-ph/9905283].

\bibitem{kt} M. Kamionkowski and M.S. Turner, Phys. Rev. {\b D 43} (1991) 1774.

\bibitem{nfw} J.F. Navarro, C.S. Frank and S.D.M. WHite, ApJ {\bf 490} (1997) 493.

\bibitem{galsim}
F.~Stoehr, S.~D.~White, V.~Springel, G.~Tormen and N.~Yoshida,
arXiv:astro-ph/0307026;\\

\bibitem{galsim1}
D.~Zhao, H.~Mo, Y.~Jing and G.~Boerner,
Mon.\ Not.\ Roy.\ Astron.\ Soc.\ {\bf 339} (2003) 12
[arXiv:astro-ph/0204108].



\bibitem{simon} M. Simon, A. Molnar and S. R\"osler, ApJ {\bf 499} (1998) 250;
 Numerical details of the background parametrization can be found in
 M. Horn, Diplomarbeit IEKP-KA/2002-17,
 http://www-ekp.physik.uni-karlsruhe.de/theses/dipl/2002.html, p. 69.

%
%
\bibitem{baer}
H.~Baer, C.~Balazs and A.~Belyaev,
JHEP {\bf 0203} (2002) 042
[arXiv:hep-ph/0202076].

\bibitem{kraml}
B. Allanach, S. Kraml, and W. Porod,
[arXiv:hep-ph/0207314].
\bibitem{egret}
A.~W.~Strong, I.~V.~Moskalenko and O.~Reimer,
arXiv:astro-ph/0306345.
%
\bibitem{ams01} J.~Alcaraz {\it et al.} [AMS Collaboration],
 Phys.\ Lett.\ B {\bf 484} (2000) 10 [Erratum-ibid.\ B {\bf 495} (2000) 440].
%
\bibitem{heat}
 S.~W.~Barwick {\it et al.} [HEAT Collaboration],
 Astrophys.\ J.\ {\bf 482} (1997) L191
 [arXiv:astro-ph/9703192].\\
 M.~A.~DuVernois {\it et al.},
 Astrophys.\ J.\ {\bf 559} (2001) 296.
\bibitem{bess} BESS Coll. S. Orito et al., Phys. Rev. Lett {\bf 84} (2000) 1078. T. Maeno et al., Atrop. Phys. {\bf 16} (2001) 121;
 astro-ph/0010381.
\bibitem{minuit}
{\rm F. James, M. Roos, {\rm MINUIT Function Minimization and Error
 Analysis\/}, CERN Program Library Long Writeup D506; Release 92.1, March
 1992;}
\bibitem{dama}
R.~Bernabei {\it et al.},
Riv.\ Nuovo Cim.\  {\bf 26} (2003) 1
[arXiv:astro-ph/0307403].

\end{thebibliography}
\end{document}